\documentclass{aa}  

\usepackage{natbib}
\usepackage{xcolor}

\usepackage{csquotes}
\usepackage{graphicx}
\usepackage{txfonts}
\usepackage{lipsum}
\usepackage{subfigure}
\usepackage{caption}
\usepackage{subcaption}         % necessary for continued figures, example in section 3
\usepackage{lscape}             % to rotate a single page table, example in appendix.
\usepackage{placeins}           % useful with \FloatBarrier, to keep 
\newcommand{\mgii}{\mbox{Mg\,{\sc ii}}}
\newcommand{\civ}{\mbox{C\,{\sc iv}}}
\newcommand{\hi}{\mbox{H\,{\sc i}}}
\newcommand{\ovi}{\mbox{O\,{\sc vi}}}

\newcommand{\ciii}{\mbox{C\,{\sc iii}}}
\newcommand{\siiv}{\mbox{Si\,{\sc iv}}}

\begin{document}

   \title{The environmental dependence of the circumgalactic medium in a high-resolution cosmological simulation}
   
%%%%%%%%%%%%%%%%%%%%%%%%%%%%%%%%%%%%%%%%
% Please do not include ORCIDs next to author names.
% Only ORCIDs authenticated by individual authors in EDP Sciences editorial system will be taken into account.
% ORCIDs included here will be removed.
%%%%%%%%%%%%%%%%%%%%%%%%%%%%%%%%%%%%%%%%

   \author{Georg Herzog\inst{1}\corrauth{georg.herzog@inaf.it}
        \and Rajeshwari Dutta \inst{2}\email{rajeshwari.dutta@iucaa.in}
        \and Michele Fumagalli \inst{3, 4}\email{michele.fumagalli@unimib.it}
        }

   \institute{INAF-IASF Milano, via A. Corti 12, I-20133 Milano, Italy.
    \and IUCAA, Postbag 4, Ganeshkhind, Pune 411007, India. 
    \and Dipartimento di Fisica \enquote*{G. Occhialini}, Universit\`{a} degli Studi di Milano-Bicocca, Piazza della Scienza 3, 20126 Milano, Italy.
    \and INAF - Osservatorio Astronomico di Trieste, via G.B. Tiepolo 11, I-34143 Trieste, Italy.}

   \date{Received September 30, 20XX}

% \abstract{}{}{}{}{}
% 5 {} token are mandatory
 
  \abstract
  % context heading (optional)
  % {} leave it empty if necessary  
   {There is increasing evidence from observations that the circumgalactic  medium (CGM) of galaxies depends on the large-scale structure in which they are embedded.
   When probing the CGM in absorption using quasar sightlines, studies find an enhanced sky coverage in the CGM of galaxies in overdensities compared to galaxies in isolation.
   However, the exact reason for this environmental dependence is still unclear.}
  % aims heading (mandatory)
   {In this work we aim to model, for the first time, the influence of the large-scale structure on the cool and warm ($T\sim10^{4-5}$\,K) gas phases of the CGM.
   } 
  % methods heading (mandatory)
   {We use a high-resolution ($m_{gas}\approx 4.5\times 10^4$ M$_{\odot}$, $m_{dm}\approx 2.4\times 10^5$ M$_{\odot}$) cosmological simulation based on the \texttt{EAGLE} model of galaxy formation.
   We select all galaxies at $z=0$ with stellar mass $M_*>10^8$ M$_\odot$ and split them into galaxies in overdensities (group galaxies) and galaxies in isolation using a Friends-of-Friends algorithm.
   For these two samples, we investigate how the large-scale structure influences the physical properties of the CGM and the measured covering fractions of the cool and warm gas phases.}
  % results heading (mandatory)
   {When the two samples of group and isolated galaxies are matched in stellar mass, halo mass, and we use only central galaxies, we do not find any significant difference in the physical properties of the CGM and the measured covering fractions.
   However, when satellite galaxies are included, we recover the observational trends in the difference of covering fractions with the environment.
   }
  % conclusions heading (optional), leave it empty if necessary
   {The difficulty of recovering the observational trends shows the complexity of capturing the multiphase CGM in simulations.
   However, since our results concerning the admixture of satellites are independent of the employed subgrid physics, this work shows that central galaxies and satellites need to be disentangled in observational studies to clearly discern the role of the environment on the CGM.
   }


   \maketitle
\nolinenumbers

%%%%%%%%%%%%%%%%% Paper %%%%%%%%%%%%%%%%%%%%%%%%%%%%%%%%

\section{Introduction}
The circumgalactic medium (CGM) is a multiphase gaseous halo around galaxies that connects the interstellar medium (ISM) with the intergalactic medium (IGM) and the filaments of the cosmic web.
Galaxies sustain their star formation due to gas accreted from the CGM and cosmic web filaments.
Feedback processes such as supernovae and stellar winds as well as feedback from active galactic nuclei (AGNs) then redeposit metal-enriched gas back into the CGM.
Thus, the properties of the CGM are set by a balance of inflows and outflows \citep{Tumlinson2017CGMreview, Peroux2020CosmicBaryonCycle, Fumagalli2024CGMreview, ChenZahedy2026CGMreview}.

However, recent observations point in the direction that the CGM is also affected by the large-scale environment in which a galaxy resides in.
While it has long been established that the environment influences the properties of galaxies \citep{Boselli2022RamPressureReview, GunnAndGott1972, Dressler1980GalaxyMorphology, BlantonAndMoustakas2009, NaabAndOstriker2017, Applebaum2021, Benavides2021, Herzog2023}, its influence on the CGM is a novel result.
When comparing galaxies in groups and galaxies in isolation, studies have found that their CGM is different.
\footnote{In this context, the term \textit{group} refers to an association of two or more galaxies by means of a distance criterion and is independent of the group halo mass.
These structures represent overdensities in the universe, but are not necessarily virialized.}
For example, \cite{Bordoloi2011} found a more extended \mgii~absorption around group galaxies compared to isolated galaxies at $0.5<z<0.9$, but argued that this can be explained by the superposition of the CGM of several galaxies.
The more extended \mgii~absorption around group galaxies was confirmed by \cite{Nielsen2018ApJ} for galaxies at $0.1<z<0.9$.
They also found an increased \mgii~covering fraction in group galaxies and attributed this to the intragroup medium, i.e., gas that is bound to the group instead of individual galaxies.
Finally, \cite{Burchett+2016} found an environmental dependence of the CGM using \civ~absorption at $z<0.015$.
For galaxies with stellar mass, $M_*>10^{9.5}$ M$_\odot$, more than half of the isolated galaxies show \civ~absorption, while \civ~is suppressed in the richest groups probed.
However, the results of this study are limited by the small sample size.

With the sensitive integral field unit (IFU) capabilities of the Multi Unit Spectroscopic Explorer (MUSE) at the VLT \citep{MUSEinstrument2006, MUSEinstrument2010}, complete studies of galaxies and their environment down to low galaxy masses became possible.
\cite{Fossati2019MUDF2} found higher \mgii~absorption in group galaxies at $z\approx 1$ compared to isolated galaxies, attributing it to gravitational interactions within the group that strip gas from galaxies and distribute it in the intergroup medium.
Similar results were found by \cite{Dutta2020}, who found a higher covering fraction of \mgii~absorption in group galaxies compared to isolated galaxies.
\cite{Dutta2021} confirmed the previous results for \mgii~and additionally found an environmental dependence of \civ~absorption.
They showed that the \civ~covering fraction in groups was higher compared to isolated galaxies, although the trend for \civ~was not very significant.
Similar results at $z\sim 1$ have also been found by \cite{Qu2023CUBS} and \cite{Cherrey2024MNRAS.528..481C}.
Finally, the environmental dependence of the CGM has been observed out to redshifts $z\sim 3-4$, with higher covering fractions of \hi, \mgii~and \civ~absorption in group environments \citep{Muzahid2021MUSEQuBES, Lofthouse2023, Galbiati2023, Galbiati+2024, Banerjee2023MUSEQuBESCIV, Banerjee2025MUSEQuBES}.
Thus, the observations converge on the fact that the environment a galaxy is embedded in influences the abundance of the ions that trace the different gas phases.

Although the CGM and its properties have been extensively studied in simulations and compared with observations \citep{FaucherGiguere2011, Fumagalli+2011, VanDeVoort2012, Suresh+2015, Turner+2017, Hafen2017, Oppenheimer+2018, Fielding2020}, until now there is almost no theoretical basis to guide the interpretation of the observational results concerning the environmental dependence of the CGM.
Recently, \cite{ManamiRoy2024} undertook a first attempt to model the influence of ram pressure stripping on the CGM and found that satellites can add cold gas to the host CGM through ram pressure stripping, induced cooling, and gas removal due to feedback.
However, their experiment was set up in an idealized way and several open questions connected to the influence of the environment on the CGM persist.
Most prominently, the exact physical mechanism that is causing the higher covering fraction in groups is still unclear.

What has been studied in more detail is the methodology of quasar sightline studies and how the practice of assigning gas to galaxies within a certain line-of-sight velocity window influences the results.
For example, using the EAGLE simulation \cite{Ho2020} found that for galaxies with stellar mass around $10^9-10^{9.5}$ M$_{\odot}$, up to 80\% of the \mgii\ gas selected within a typical velocity window of 500 km\,s$^{-1}$ at an impact parameter of 100 pkpc is outside the virial radius of the galaxy, which influences the detectability of co-rotating gas.
\cite{WengPeroux2024} use the TNG50 simulation to investigate the origin of \hi~absorption lines within a velocity window of 500 km\,s$^{-1}$ around a galaxy and find that, depending on the halo mass and impact parameter, the contribution of gas from satellites, other haloes, and the intergalactic medium can be substantial.
Such a bias when selecting the gas in velocity space could be responsible for the environmental dependence of the CGM.
Since with simulations we have the power to look into the gas distribution in real space, we can test whether this is indeed the case.

When using simulations to investigate the CGM, one needs to take into account the importance of spatial resolution on the cold gas content in the simulated CGM.
Several works converge on the fact that low resolution simulations might underestimate the amount of cold gas in the CGM \citep{VanDeVoort2019CGMHighResolution, Hummels2019CGMandResolution, Peeples2019Foggie, Ramesh2024GIBLESmallScaleGas}.
However, these studies are either based on only a single halo that is resimulated at high resolution \citep[c.f.][]{VanDeVoort2019CGMHighResolution, Hummels2019CGMandResolution, Peeples2019Foggie} or on a small sample of resimulated haloes \citep[8 in case of][]{Ramesh2024GIBLESmallScaleGas}.
For a statistical analysis of the CGM in different environments, one needs a much larger sample of galaxies, which currently one can only get from cosmological boxes.

Therefore, any study investigating the influence of the environment on the CGM in a statistical manner needs to balance two aspects in the simulation.
On the one hand, it needs to use simulations that provide a resolution that is high enough to resolve the small scale structures of cold gas in the CGM.
On the other hand, it needs to balance this high resolution with a galaxy sample that is both cosmologically realistic and statistically meaningful.
Since currently no cosmological simulation fulfills both requirements, this leads necessarily to a trade off between resolution and having a statistically meaningful sample when choosing the simulation.

In this paper, we undertake a first attempt to model the effect of the environment on the cool and warm phases of the multiphase CGM.
We use a high-resolution ($m_{gas}\approx 4.5\times 10^4$ M$_{\odot}$, $m_{dm}\approx 2.4\times 10^5$ M$_{\odot}$) cosmological simulation based on the \texttt{EAGLE} model of galaxy formation \citep{Schaye2015, Crain2015}.
Although we are still forced to trade resolution in the CGM for having a cosmologically realistic and statistically viable sample, the simulation we use is one of the highest-resolution cosmological boxes currently available and thus one of the most suitable simulations for the questions we want to answer.
Furthermore, in a previous work we were able to quantify the number of field galaxies that undergo ram pressure stripping in galaxy haloes in this simulation \citep{Herzog2023}.
Thus, using the same simulation allows us to directly connect our results on ram pressure stripping with the environmental dependence of the CGM.

We structure the paper as follows: In Sec. \ref{Sec:Methods} we describe the simulation, the method to split our sample into group and isolated galaxies, the selection of gas tracing different gas phases, and the calculation of covering fractions.
In Sec. \ref{Sec:Results} we present our results of the dependence of the covering fractions and the physical properties of the CGM on the environment.
In Sec. \ref{Sec:Discussion} we discuss the implications of these results, compare them to the literature and mention potential caveats and weaknesses of our model.
Finally, we present our conclusions in Sec. \ref{Sec:Conclusion}.

\section{Methods}
\label{Sec:Methods}

\subsection{Simulation}
\label{sec:simulation}

We use the high-resolution Smoothed-Particle-Hydrodynamics (SPH) cosmological simulation first presented in \cite{BLF2020} for our investigation of the influence of the environment on the CGM.
Here we only give a short overview of the simulation and present those parts of the simulation in more detail that are relevant for the analysis.
For a thorough description of this simulation, we refer the reader to the paper of \cite{BLF2020} where it was initially presented.

The simulation is based on the \texttt{EAGLE} model of galaxy formation \citep{Schaye2015, Crain2015} and evolves a 20 cMpc cube with periodic boundary conditions from $z=127$ to $z=0$.
The initial conditions have been carried out with the publicly available code \texttt{MUSIC} \citep{HahnAbel2011}, which fills the cube with a random realization of 1024$^3$ gas and dark matter particles.
Gas particles have a mass of $m_{gas}\approx 4.5\times 10^4$ M$_{\odot}$, while dark matter particles have a mass of $m_{dm}\approx 2.4\times 10^5$ M$_{\odot}$.
Following the initialization, the simulation is then evolved using \texttt{P-gadget3}, an adapted version of the \texttt{gadget2} code \citep{Springel2005}, adopting a Plummer-equivalent softening length of $\epsilon \approx 195$ comoving pc.
At $z=11.5$ the UV background of \cite{HaardtMadau2001} is turned on by instantaneously injecting 2 eV per proton mass.
Throughout, the simulation assumes the cosmology of early Planck results \citep{Planck2014}.

For star formation to proceed, the gas has to cool down and exceed a density threshold of $\rho_{th}=1.0$ cm$^{-3}$.
Gas cooling and heating are implemented following \cite{Wiersma2009}.
When $\rho_{th}$ is exceeded, gas particles are stochastically transformed into star particles of a mass of $M_*\approx 4.5 \times 10^4$ M$_{\odot}$ following a pressure law that reproduces the star formation rate of the Kennicutt-Schmidt law \citep{Schaye2015}.
Star-forming gas is put on an artificial equation of state $T(\rho)=T_0(\rho / \rho_{th})^{\gamma-1}$, where $\gamma=4/3$ and $T_0=8000$ K, to prevent the formation of extremely high density gas.
Since this affects only star-forming gas found in the center of galaxies, this does not directly affect our analysis of the CGM.
Finally, metal enrichment follows the \texttt{EAGLE} model, in which the mass of each element lost during the evolution of a stellar population is then added to the neighboring gas particles \citep{Schaye2015}.
Throughout the simulation, AGN feedback was turned off.
Since AGN feedback could have a significant impact on the metal distribution in the CGM \citep{Suresh+2015}, the distribution of metals in this simulation may not be realistic.
Thus, instead of using ion tracers, we decided on a different approach in testing the multiphase CGM.
We describe our approach in Sec. \ref{Sec:SelectionGasPhases}.

\subsection{Halo Finder and Group Finder}
For our analysis of the environmental dependence of the CGM, we need to divide our sample of galaxies into group and isolated galaxies.
This is a two step process.
First, we need to identify galaxies in the simulation data and classify them as central and satellite galaxies.
Second, we then divide this sample of galaxies into group and isolated galaxies using a Friends-of-Friends (FoF) algorithm.
We will describe both of these steps in more detail.

In the simulation, gravitationally bound structures are found using the \texttt{HBT+} halo finder of \cite{Han2018}.
\texttt{HBT+} takes a catalogue of FoF halos as well as the associated particles of each FoF halo as input, where the standard linking length of $b=0.2$ was used in constructing the FoF catalogue.
To determine the bound particles of each halo, \texttt{HBT+} carries out an unbinding procedure, discarding each particle with a kinetic energy bigger than its potential energy.
Furthermore, \texttt{HBT+} also constructs merger trees of the dark matter (DM) halos, setting the most massive halo as the central halo.
At each snapshot \texttt{HBT+}, therefore, returns a catalogue of central and subhalos with a list of particles (DM, stars, gas) that are bound to these halos as well as several halo properties such as virial mass, bound stellar mass, bound gas mass, virial radius, half mass radius, and the position and average physical velocity of the halo.

Using the \texttt{HBT+} classification of central and subhalos, we identify central galaxies as the most massive galaxies of a given DM halo, and satellite galaxies as galaxies inhabiting subhaloes contained within the main DM halo.
As the stellar mass of a galaxy, we use the total bound stellar mass of a DM halo.
The virial radius $R_{vir}$ is taken to be $R_{200}$, which is the radius of a sphere with a mean density 200 times the critical density of the universe.
For satellites, $R_{200}$ depends on the location within the parent halo.
Closer to the center of the halo, the ambient density is higher and thus a sphere with a mean density enclosing 200 times the critical density will be smaller.
To avoid using a quantity in our analysis that is not well defined for satellites, we decided to use the half mass radius R$_{1/2}$ instead, which is the radius that includes half the bound mass of a halo and for satellites is independent of the location inside the parent halo.
To guide the reader, in our sample on average, $R_{vir}\approx 2.4\times R_{1/2}$.

We selected all galaxies with a stellar mass $M_*>10^8$ M$_\odot$, irrespective of whether they are central galaxies or satellite galaxies.
This mass cut-off corresponds to the typical stellar mass completeness limit at $z\lesssim 1$ in CGM surveys using MUSE \citep[e.g.,][]{Lofthouse2020MAGG_I, Dutta2020}.
We then run our FoF group finder with a linking length of $r=500$ kpc on this sample of galaxies, which is the linking length commonly used to define groups in observations \citep{Knobel2009LinkingLength, Diener2013LinkingLength, Fossati2019MUDF2}.
For each galaxy in our sample, we therefore check whether there is another galaxy with $M_*>10^8$ M$_\odot$ within a radius of $r=500$ kpc to classify whether this galaxy is part of a group or is isolated.

Applying the above selection criterion, we get a sample of 289 galaxies with $M_*>10^8$ M$_\odot$ at $z=0$.
Using our FoF algorithm with linking length $r=500$ kpc, we obtain 158 group galaxies (36 being central galaxies) and 131 isolated galaxies (127 being central galaxies).
At $z=1$, the same selection criterion gives 288 galaxies.
Of these, 156 are group galaxies (48 being central galaxies) and 132 are isolated galaxies (130 being central galaxies).
We performed the analysis at both $z=0$ and $z=1$.
Since the results between the two redshifts are consistent, we will focus below only on the results at $z=0$ and show the results at $z=1$ in Appendix \ref{Appendix:Z1Results}.
In Appendix \ref{Sec:MassCutFoF}, we show that the mass-cut of $10^8$ M$_\odot$ in the FoF algorithm does not influence our results on the environmental dependence of the CGM.

\begin{figure}
    \centering
    \includegraphics[width=\columnwidth]{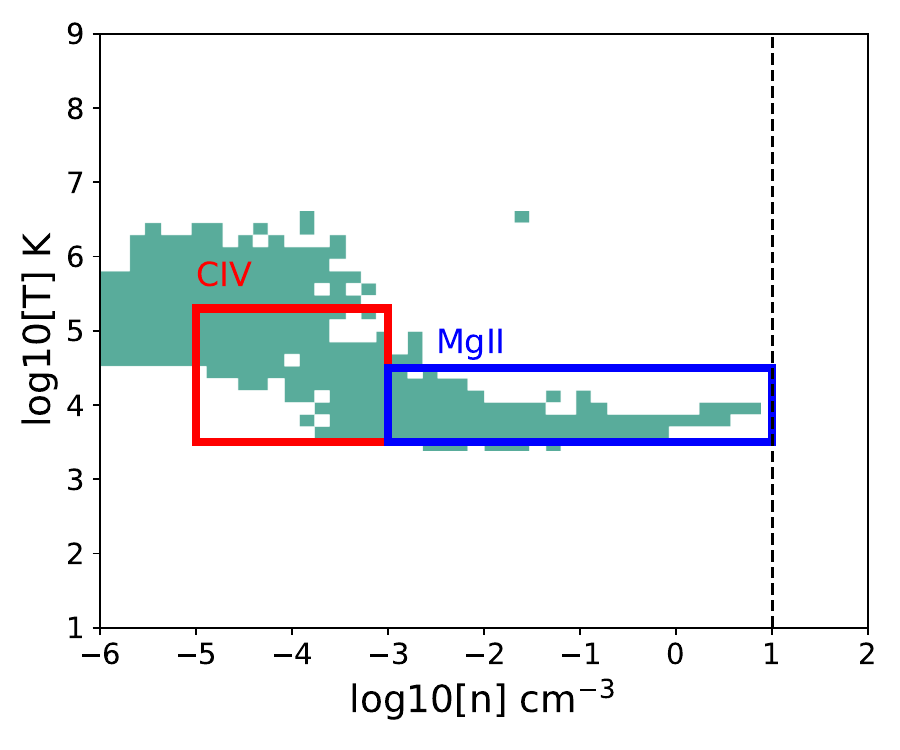}
    \caption{Phase space diagram of gas around a galaxy of mass M$_{200}\approx 2\times 10^{11}$ M$_\odot$.
    Since our simulation is not calibrated to reproduce the observed metal statistics, we use the gas that is traced by \civ~(red box) and \mgii~(blue box) in our analysis.
    The dashed vertical line is the density above which the gas follows an artificial EOS, and which is excluded from the analysis.
    }
        \label{Fig:PhaseSpaceCIVandMgII}
\end{figure}

\subsection{Selection of Gas Phases}
\label{Sec:SelectionGasPhases}
In observations, different ions are used to trace different gas phases of the CGM depending on the temperature and density at which they are ionized.
For example, \mgii~traces the cool phase of $T\approx 10^4$ K, while \civ~traces the cool/warm phase of $T\approx 10^{4-5}$ K in lower density gas \citep[for a more detailed description see][]{Tumlinson2017CGMreview}.
Since our simulation is not calibrated on the metals, we were not able to reproduce observables, such as the number of absorbers per unit redshift, when calculating the ionization states of \mgii~and \civ.
However, we can still do a comparative analysis to investigate whether there are systematic differences between the CGM of group and isolated galaxies by directly analyzing the gas phases traced by the ions.

In Fig. \ref{Fig:PhaseSpaceCIVandMgII}, we show the phase space diagram of the gas around a galaxy with a halo mass of $M_{200}\approx 2 \times 10^{11}$ M$_\odot$.
The dashed vertical line is the density above which gas follows an artificial equation of state (EoS) as described in Sec. \ref{sec:simulation}.
We exclude this part of the phase space diagram from our analysis since it would give unphysical results.
For our analysis, we perform a cut in density and temperature to roughly select gas traced by \civ~and \mgii~in our simulation.
For \civ, we take gas with temperatures $3.5<\text{log}_{10}(T)\text{[K]}<5.3$ and densities $-5<\text{log}_{10}(\rho) [\text{cm}^{-3}]<-3$, while for \mgii, we take gas with temperatures $3.5<\text{log}_{10}(T)\text{[K]}<4.5$ and densities $-3<\text{log}_{10}(\rho) [\text{cm}^{-3}]<1$.
The cuts in temperature and density we use for our analysis are marked as red box for \civ~and blue box for \mgii~in Fig. \ref{Fig:PhaseSpaceCIVandMgII}.

\begin{figure*}
        \centering
        \includegraphics[width=\textwidth]{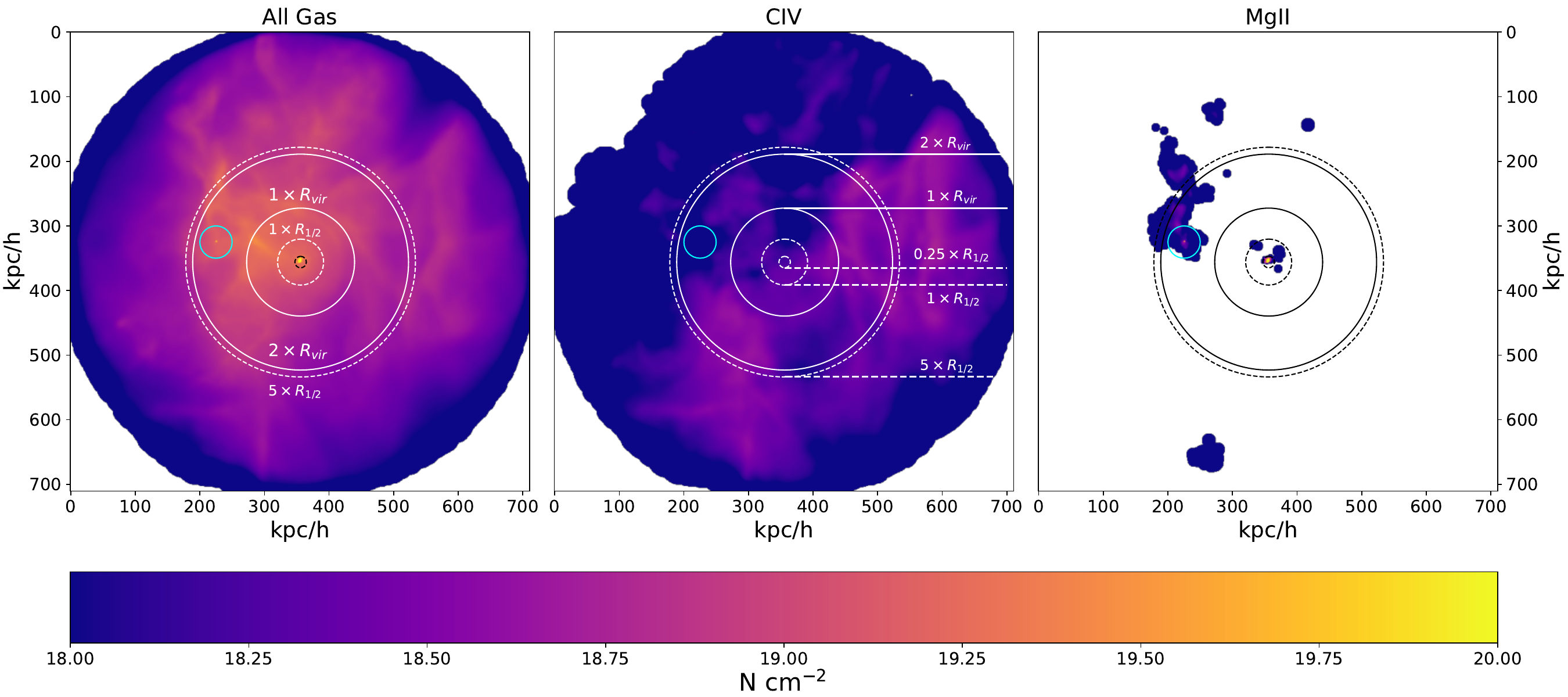}
        \caption{Gas within a sphere of radius $10\times$R$_{1/2}$ around a galaxy of mass M$_{200}\approx 2\times 10^{11}$ M$_\odot$.
        The left panel shows all the gas, while the middle panel shows gas traced by \civ\ (red box in Fig. \ref{Fig:PhaseSpaceCIVandMgII}), and the right panel gas traced by \mgii\ (blue box in Fig. \ref{Fig:PhaseSpaceCIVandMgII}).
        We mark the radii corresponding to R$_{vir}$ and $2\times$R$_{vir}$ with solid circles, while the radii corresponding to R$_{1/2}$ and $5\times$R$_{1/2}$ are marked with dashed circles.
        The innermost dashed circle of $0.25\times$R$_{1/2}$ marks the region excluded from our analysis to avoid including gas of the galaxy itself.
        In the left panel, we see the galaxy as a bright dot in the center and another smaller bright dot on the left close to $2\times$R$_{vir}$, which is an infalling satellite galaxy (cyan circle).
        In the middle panel, which represents the warm phase traced by \civ, both these structures are missing and instead we see a diffuse structure towards the lower right.
        In the right panel instead, we only see scattered blobs of cold gas which is traced by \mgii, mostly at the center and to the upper left where the satellite is falling in.
        }
        \label{Fig:ExampleGalaxy_AllGasCIVandMgII}
\end{figure*}

The results of selecting gas in this way are illustrated in Fig. \ref{Fig:ExampleGalaxy_AllGasCIVandMgII}, where we show the gas around the same galaxy of halo mass $M_{200}\approx 2 \times 10^{11}$ M$_\odot$ used to construct the phase space diagram in Fig. \ref{Fig:PhaseSpaceCIVandMgII}.
In the left panel, we show all the gas without any cuts in temperature and density, in the middle panel we show the gas traced by \civ, and in the right panel we show the gas traced by \mgii~according to our cuts.
The gas is selected in a sphere with radius $r=10\times R_{1/2}$ around the galaxy and then projected onto the 2D plane.
We mark $R_{1/2}$ and $5\times R_{1/2}$ with dashed circles, while $R_{vir}$ and $2\times R_{vir}$ are marked with solid circles.
In the left panel, showing all the gas without any cuts in temperature and density, the galaxy in the middle can be identified by its high column density.
Towards the left there is an infalling satellite galaxy which we mark with a cyan circle.
Additionally, we can also see some diffuse filamentary structure.
If we look at the middle panel, where we show the gas traced by \civ, both the central galaxy and the satellite are absent and only a diffuse structure in the lower right half is visible.
In the right panel, where we show the gas traced by \mgii, gas is only visible in the middle where there is the central galaxy, and towards the top left where there is the satellite.
Based on visual inspection, this satellite fell into the central galaxy from the right and then made its way towards the left above the central galaxy with the tail of cold gas behind the satellite coming from ram pressure stripping of cold gas.
We conclude that our selection of gas phases with the cuts in temperature and density is likely to trace well the cool and warm gas phases.

\subsection{Calculation of Covering Fractions}
To analyze the influence of the environment on the CGM of galaxies, we calculate the covering fractions of gas traced by \mgii~and gas traced by \civ~for group and isolated galaxies.
This tells us about the abundance and distribution of different gas phases around the galaxies and whether these change according to the environment.
We select the gas in a sphere with radius $10\times R_{1/2}$ around the galaxy as shown in Fig. \ref{Fig:ExampleGalaxy_AllGasCIVandMgII}.
In appendix \ref{Sec:GasvLOS}, we show that our decision to select gas in 3D instead of velocity space, as done in observations, has no influence on the environmental dependence of the CGM.
We then project this gas onto the 2D plane and calculate the column density in pixels with a side length of 1 kpc/h~$\times$~1 kpc/h.
The covering fraction $f_c$ is the number of sightlines that detect a certain ion, or gas phase, in our case, divided by the total number of sightlines, i.e.,
\begin{equation}
    f_c = \frac{\text{\# detections}}{\text{\# sightlines}}.
\end{equation}
In our analysis, we avoid all sightlines that directly pass through the galaxy by discarding all sighlines within $r<0.25\times R_{1/2}$.
We marked this area by the innermost dashed circle in Fig. \ref{Fig:ExampleGalaxy_AllGasCIVandMgII}.
To calculate the covering fraction, we have to define a threshold in column density for the gas phases above which we count it as a detection for a certain sightline.
In our analysis, we will show both the results for a fixed column density limit, and how they change with the detection limit.
Finally, to calculate the covering fraction as a function of radius, we calculate the covering fraction in annuli with inner radius $r_{in}=0.25\times R_{1/2}$ and out radii $r_{out}=\text{2.5, 5, 7.5, and 10}\times R_{1/2}$.

\section{Results}
\label{Sec:Results}

\subsection{Covering fractions in the CGM}
\label{Sec:CoveringFractionsInTheCGM}
To eliminate potential biases connected to the mass of galaxies, we created matched samples in stellar mass between the samples of group and isolated galaxies.
To build our matched samples, we check whether there is an isolated galaxy within 0.3 dex in stellar mass for each group galaxy.
If we find a matching galaxy, the galaxy pair is put in the sample of matched galaxies.
If there is more than one matching galaxy, we randomly choose one of them.
If no matching galaxy is found, the group galaxy is discarded.
At redshift $z=0$, the matched sample consists of 131 group and 131 isolated galaxies, if satellites are included in the matched sample.
However, if we exclude satellites and only use central galaxies, the matched sample consists of 27 group and 27 isolated galaxies.
We then performed our analysis with the samples of matched group and isolated galaxies found in this way.

\begin{figure*}
        \centering
        \includegraphics[width=\textwidth]{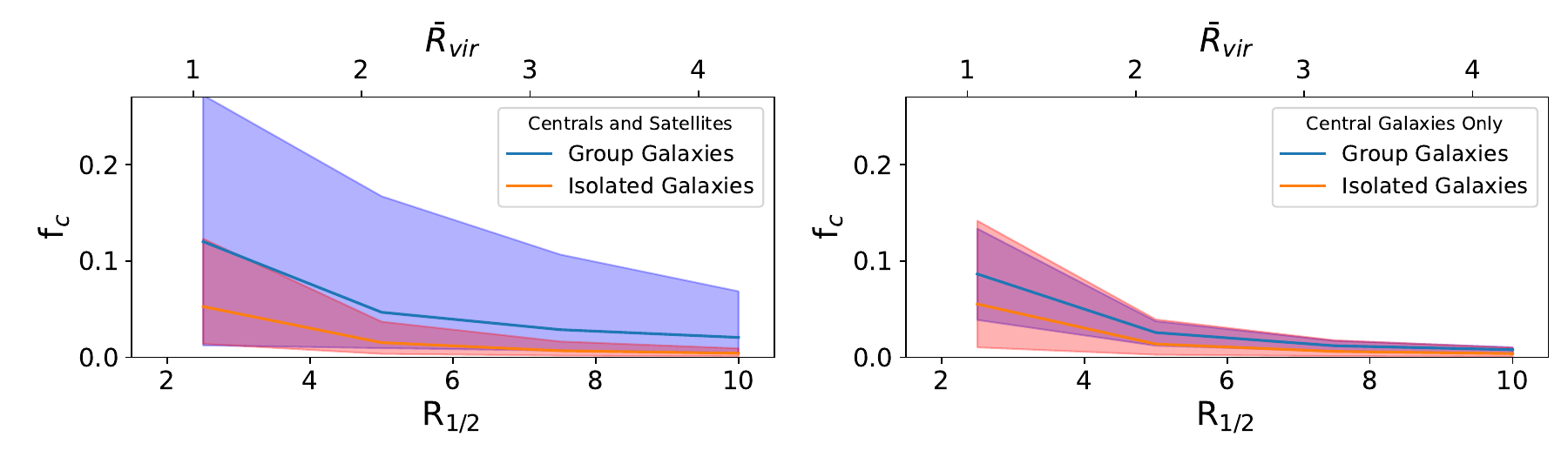}
        \caption{Median covering fractions for group (blue lines) and isolated galaxies (red lines) with the 16/84 percentiles (shaded areas) showing the 1$\sigma$ halo-to-halo variation.
        The left panel shows the covering fraction for gas traced by MgII if satellite galaxies are included in the sample.
        In this case, we can reproduce the observational trend of a higher covering fraction in group galaxies although the scatter is large.
        The right panel shows the covering fractions of gas traced by MgII when only central galaxies are considered.
        In this case, while the median covering fraction of group galaxies is still higher than the one of isolated galaxies, the two distribution are similar within the scatter.
        }
        \label{Fig:ReproduceTrend}
\end{figure*}

\begin{figure*}
        \centering
        \includegraphics[width=\textwidth]{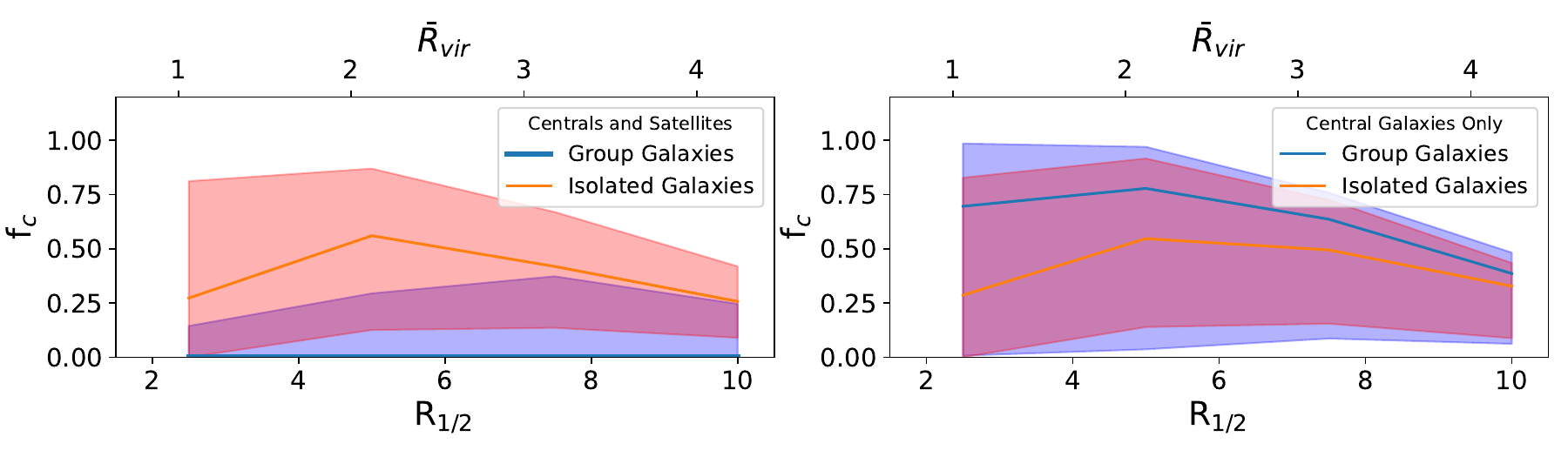}
        \caption{Same as Fig. \ref{Fig:ReproduceTrend} but for gas traced by \civ.
        If satellites are included in the sample, the median covering fraction for group galaxies is zero (left panel).
        When only central galaxies are considered, we recover the higher median covering fraction of group galaxies (right panel).
        However, the scatter is large and the two distributions overlap.
        }
        \label{Fig:ReproduceTrendCIV}
\end{figure*}

Using this method, we were able to reproduce the result of a higher covering fraction in group galaxies compared to isolated galaxies as found in observations \citep[c.f. for example][]{Galbiati+2024, Galbiati2023, Dutta2020, Dutta2021, Lofthouse2023} under certain conditions in our simulation.
The clearest case we present in the left panel of Fig. \ref{Fig:ReproduceTrend} where we show the covering fractions of gas traced by \mgii~around group galaxies (blue) and isolated galaxies (red) when adopting a detection limit of $10^{19}$ cm$^{-2}$ for the total gas column density within the temperature and density cut for \mgii~as defined by the blue box in Fig. \ref{Fig:PhaseSpaceCIVandMgII}.
If we assume solar metalicities following \cite{Asplund2009}, the detection limit of $10^{19}$ cm$^{-2}$ for all the gas translates to $10^{14.6}$ cm$^{-2}$ for Magnesium.
Using \texttt{Trident} \citep{TridentMethodsPaper}, which interpolates over ion tables constructed with \texttt{Cloudy} \citep{Ferland2013Cloudy}, we calculate the ionization fraction $f_i$ of \mgii~in the area of the phase space diagram we are probing and find that it varies between $1\%<f_i<93\%$.
Assuming an ionization fraction of $1\%\sim 10\%$, we arrive at detection limits of N(\mgii) $\sim10^{12.6-13.6}$ cm$^{-2}$.
This is close to the typical detection limits for \mgii~used in the literature of rest-frame equivalent widths $\sim0.03-0.3$\,\AA\ \citep[e.g.,][]{Dutta2020}, which is approximately N(\mgii) $\sim10^{12-13}$ cm$^{-2}$ assuming the linear part of the curve-of-growth.
Thus, the detection limit of 19 cm$^{-2}$ for all gas is reasonable, especially given all the uncertainties in our modeling.
Later we also show how the results change when adopting different detection limits.

Returning to the left panel of Fig. \ref{Fig:ReproduceTrend}, where we include satellites when constructing the matched sample, we can clearly see that the covering fraction of \mgii~gas is higher in group galaxies compared to isolated galaxies.
For example, at $r\approx 2\times R_{vir}$, the median of group galaxies is a factor of 3 higher than the median of isolated galaxies.
This is consistent with a factor of $2\sim 3$ enhancement in \mgii~covering fraction at any given distance as found in \cite{Dutta2021}.
However, the $1\sigma$ halo-to-halo variation, especially for group galaxies, is large.
In the right panel of Fig. \ref{Fig:ReproduceTrend}, we show again the covering fractions of \mgii~gas with detection limit of $10^{19}$ cm$^{-2}$ for the total gas column density.
However, here we used the matched sample consisting of only central galaxies.
Although the median covering fraction of group galaxies is still higher than the median covering fraction of isolated galaxies, the two distributions are similar within the scatter.

In Fig. \ref{Fig:ReproduceTrendCIV} we repeat the analysis for gas traced by \civ.
In this case we chose a detection limit of $10^{18.5}$ cm$^{-2}$ for all gas to calculate the covering fractions.
If we apply the same simplistic modeling as before, the detection limit of $10^{18.5}$ cm$^{-2}$ for all gas that has the right temperature and density to be ionized into \civ~as defined by the red box in Fig. \ref{Fig:PhaseSpaceCIVandMgII} translates into a detection limit of $10^{14.93}$ cm$^{-2}$ for Carbon \citep{Asplund2009}.
We calculate again the ionization fraction $f_i$ of \civ~with \texttt{Trident} \citep{TridentMethodsPaper} in the area of the phase space diagram shown in Fig. \ref{Fig:PhaseSpaceCIVandMgII} and find that it varies between $1\%<f_i<29\%$.
Assuming ionization fractions of 1\% to 10\%, we arrive at a detection limit of N(\civ) $\sim10^{13-14}$ cm$^{-2}$, which is in the range of values $\sim10^{13-14}$ cm$^{-2}$ (rest-frame equivalent width $\sim0.1-0.3$\,\AA\ assuming linear part of the curve-of-growth) used in observations \citep[e.g.,][]{Dutta2021,Galbiati+2024}.
Therefore, our detection limit of $10^{18.5}$ cm$^{-2}$ for the total gas content is again a reasonable value given the simplicity of our modeling.

Looking at the left panel of Fig. \ref{Fig:ReproduceTrendCIV} where we include satellites in our sample, we see a different behavior compared to the case of gas traced by \mgii.
The median covering fraction of gas traced by \civ~for group galaxies is zero, while the median for isolated galaxies is well above zero.
While this is in contradiction to results from observations at higher redshifts where the \civ~covering fraction is higher in groups compared to isolated galaxies \citep[c.f. for example][]{Dutta2021, Galbiati2023},
it actually reproduces the trend seen by the study of \cite{Burchett+2016}, which reports an absence of \civ~detections in group galaxies at redshift $z\approx 0$.
However, if we only use central galaxies when constructing the matched sample, we obtain a higher median covering fraction for group galaxies although with a large $1\sigma$ halo-to-halo variation.
We show this case in the right panel of Fig. \ref{Fig:ReproduceTrendCIV}.

\begin{figure*}
    \centering
    \begin{subfigure}
    \centering
    \includegraphics[width=\textwidth]{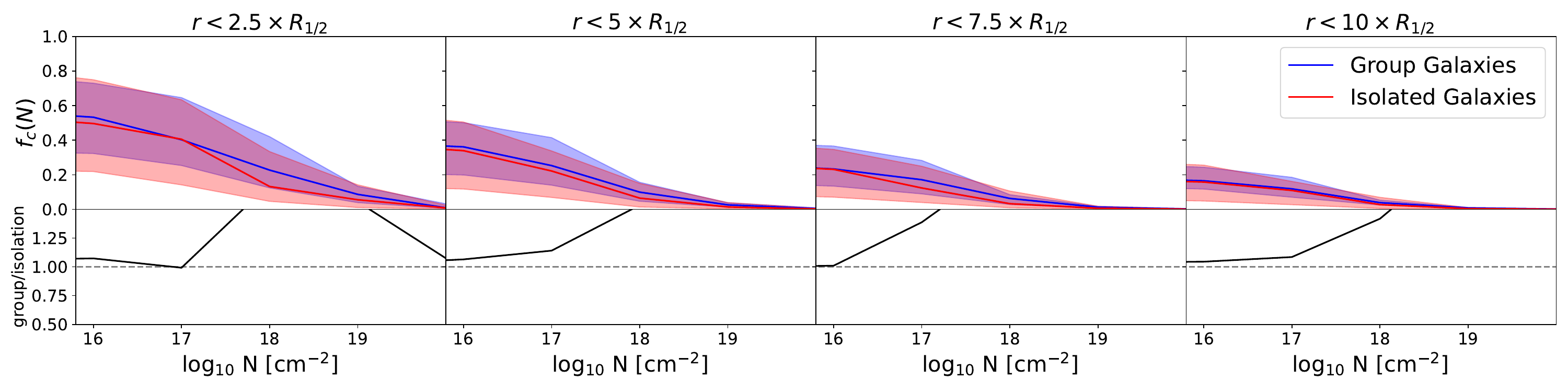}
    \caption{Covering fraction of gas traced by \mgii~as a function of detection limit (upper panels) and the ratio between the medians (lower panels) for groups and isolated galaxies.
    We show the results at a radius of 2.5, 5, 7.5 and $10\times R_{1/2}$ corresponding to approximately 1, 2, 3 and $4\times R_{vir}$.
    The solid lines are the median covering fractions while the shaded areas are the 16/84 percentiles.
    While the strength of the trend of a higher median covering fraction in groups depends on the chosen detection limit, the two distributions are similar within the scatter for all detection limits.
    }
    \label{Fig:6a}
    \end{subfigure}
    \hfill
    \begin{subfigure}
    \centering
    \includegraphics[width=\textwidth]{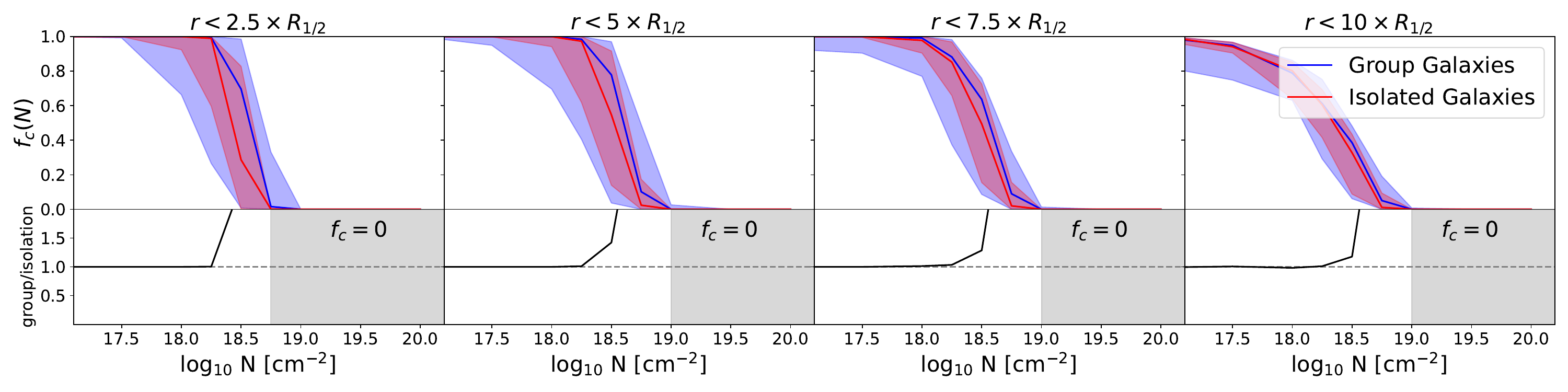}
    \caption{Same as in Fig. \ref{Fig:6a} but for gas traced by \civ.
    The transition between a covering fraction of $f_c=1$ and $f_c=0$ happens within one order of magnitude in column density. Group galaxies have a higher median covering fraction than isolated galaxies.
    The grey shaded area in the lower panels are the detection limits where the median covering fraction for both distributions are zero.
    }
    \label{Fig:6b}
    \end{subfigure}

\label{Fig:CumulativeCoveringFractions}
\end{figure*}

Next, we show in Figs. \ref{Fig:6a} and \ref{Fig:6b} the covering fractions of group and isolated galaxies as a function of detection limit.
We restrict ourselves to a matched sample consisting of only central galaxies since including satellites in the sample is introducing a bias as we will show in Sec. \ref{Sec:3DStructure}.
In Fig. \ref{Fig:6a} where we show the results for \mgii~gas, we see that group galaxies have a higher median covering fraction than isolated galaxies with the trend diminishing when going to lower detection limits.
However, for all detection limits, the 16/84 percentiles for group galaxies overlap with the 16/84 percentile of isolated galaxies.
We see a similar picture for \civ~gas in Fig. \ref{Fig:6b} with a higher median covering fraction for group galaxies, a diminishing trend towards lower detection limits, and a scatter of the two distributions that overlaps.
The diminishing of the trend with decreasing detection limit is not surprising as with lower detection limits we also start to pick out more diffuse components in the CGM, which is likely not as prone to environmental influence as the denser part that comes from the perturbed ISM.
Having found the above trends of covering fraction, we next investigate whether there is any difference in the physical conditions of the CGM gas in groups and isolated galaxies.

\begin{figure}
        \centering
        \includegraphics[width=\columnwidth]{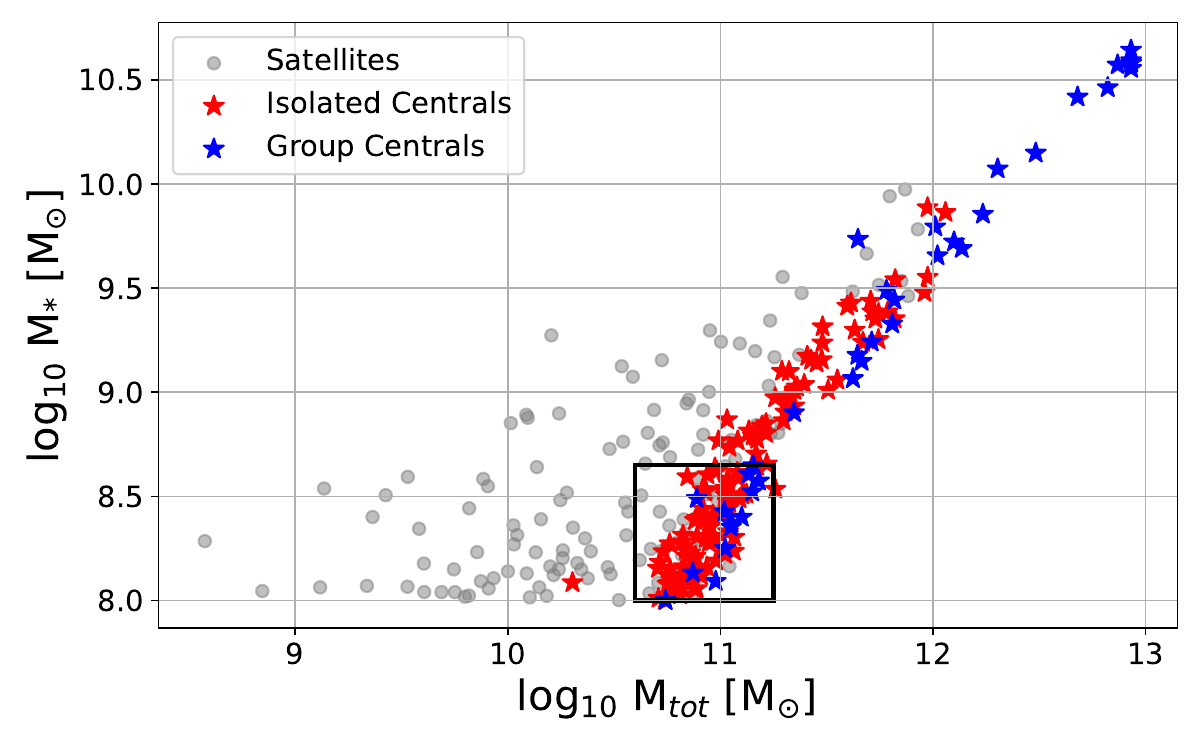}
        \caption{Stellar mass plotted against the total bound mass of our galaxy sample with $M_*>10^8$ M$_\odot$.
        Isolated central galaxies are shown as red stars, central galaxies in groups as blue stars, and satellites as grey dots.
        At high masses, all central galaxies are part of a group, while at stellar masses $M_*<10^{10}$ M$_\odot$, we find both, centrals in groups and isolation.
        The black square marks the subsample selected within 0.65 dex in stellar mass and total bound mass for further analysis.
        }
        \label{Fig:SubsampleSelection}
\end{figure}

\begin{figure*}
        \centering
        \includegraphics[width=\textwidth]{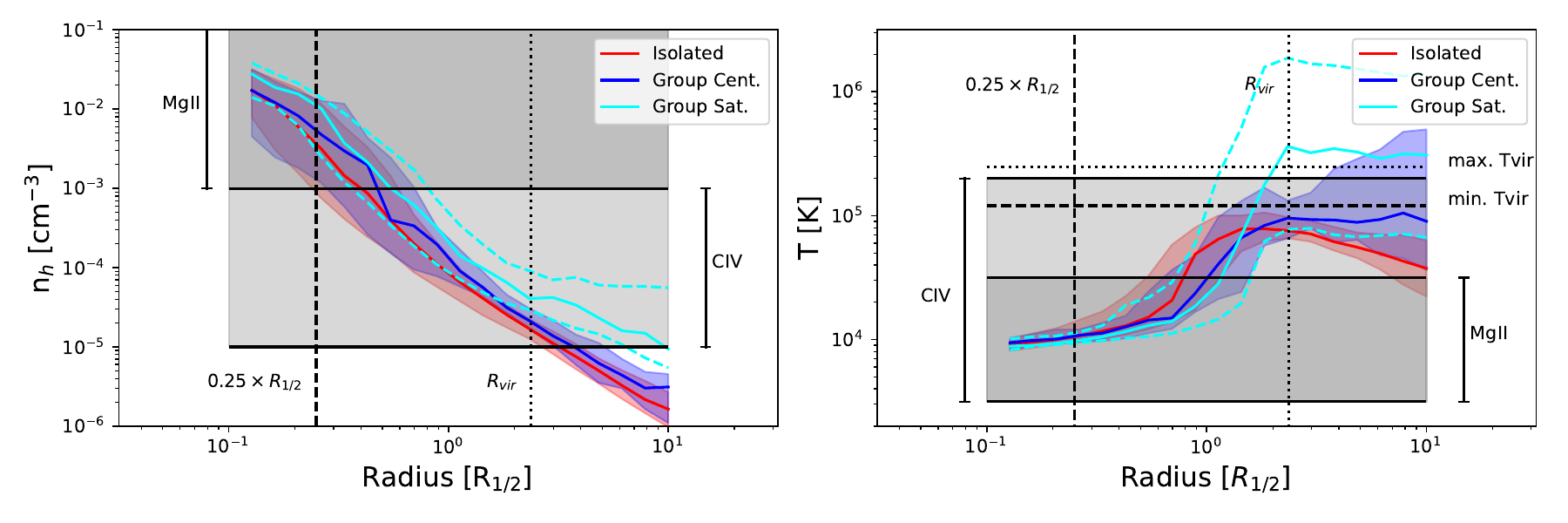}
        \caption{Radial density and temperature profiles of central and satellite galaxies within the square shown in Fig. \ref{Fig:SubsampleSelection}.
        The solid lines are the medians and the shaded areas the 16/84 percentiles.
        We show the density and temperature cuts used for \mgii~and \civ~as grey shaded areas.
        The vertical dashed line marks $0.25\times R_{1/2}$.
        Gas within this radius was not used in our analysis of covering fractions.
        The vertical dotted line marks approximately $R_{vir}$.
        In the right panel, we mark the minimum and maximum virial temperature of our sample of central galaxies within the mass bin by a horizontal dashed line (minimum) and dotted line (maximum).
        In the left panel we see that central galaxies matched in stellar mass and total bound mass show similar density profiles up to $10\times R_{1/2}$ no matter whether they are classified as group galaxies or isolated galaxies, while satellite galaxies show higher densities.
        In the right panel, the temperature profiles of group and isolated central galaxies are similar up until $2.5\times R_{1/2}$ or about $R_{vir}$.
        Only at larger radii, group galaxies show a slightly higher median temperature than isolated galaxies.
        However, satellite galaxies reach temperatures much higher than central galaxies, exceeding even the virial temperature that can be reached by central galaxies in our chosen mass bin.
        }
        \label{Fig:RadialProfilesTempDens}
\end{figure*}

\subsection{The physical properties of the CGM}
\label{Sec:3DStructure}
The quantities we want to analyze are the temperature and density of the gas in the CGM, since a difference in these would have a direct influence on the measured covering fractions of the gas phases we chose.
However, the temperature that can be reached inside a galaxy depends on its halo mass.
Using our matched sample that extends over several orders of magnitude in halo and stellar masses would introduce an artificial spread in temperature. 
To eliminate this variation in temperature from our analysis, we restrict ourselves to galaxies in a relatively small bin of stellar mass and halo mass.
In Fig. \ref{Fig:SubsampleSelection}, we show the stellar mass versus halo mass relation for all galaxies with a stellar mass $M_*>10^8$ M$_\odot$ in our simulation.\footnote{As the virial radius and therefore also the virial mass is not well defined for satellites, we decided to use the total bound mass instead of the virial mass as a proxy for the halo mass.}
While central galaxies in groups occupy the high mass end, central galaxies in isolation and all satellites have a stellar mass below $10^{10}$ M$_\odot$.
Only at the very low mass end, we find several group central galaxies with a similar distribution of stellar mass and halo mass as isolated central galaxies.
Therefore, we select only galaxies with stellar mass, $8<\text{log}_{10}M_*/M_\odot<8.65$, and halo mass, $10.6<\text{log}_{10}M_{tot}/M_\odot<11.25$.
Our selection window is marked with a black rectangle in Fig. \ref{Fig:SubsampleSelection}.
This selection gives us 76 isolated galaxies, which are all central galaxies, and 37 group galaxies, of which 13 are central galaxies and 24 satellites.
In the following analysis, we treat central galaxies and satellite galaxies in groups as two different samples to investigate the effect of satellite galaxies on the observed properties of the CGM.

We calculate  the radial density and temperature profiles for these three subsamples by first calculating the median temperature and median density in shells around each galaxy.
We then take the median temperature and median density of each subsample in each shell and plot it against the radius.
We show the results in Fig. \ref{Fig:RadialProfilesTempDens}, where central galaxies in groups are shown in blue, satellite galaxies in groups are shown in cyan, and isolated galaxies are shown in red.
The solid line is the median and the shaded areas are the 16/84 percentiles.
For better visibility, we show the 16/84 percentiles of satellite galaxies in groups as dashed lines.
We mark $0.25\times R_{1/2}$, i.e., the inner part which we do not consider for our analysis, with a dashed line and $R_{vir}$ with a dotted line.
The density and temperature cuts used to select \mgii~and \civ~gas are marked as gray shaded areas.

In the left panel of Fig. \ref{Fig:RadialProfilesTempDens} we show the density profiles.
Central galaxies in groups and central galaxies in isolation have similar density profiles out to $10\times R_{1/2}$ or approximately $4\times R_{vir}$.
However, satellite galaxies show a profile with a higher density compared to both group and isolated central galaxies.
In the inner part, up to about $0.5\times R_{1/2}$, this difference is still small but then it steadily grows, and from about $R_{1/2}$ the profile clearly diverges from the profiles of central galaxies.
At the approximate $R_{vir}$, the median gas density of satellites is about a factor of 2 larger than the one of central galaxies.

In the right panel, we show the radial temperature profiles and see a similar picture.
Comparing only central galaxies in groups and isolation, we see again that their temperature profiles are similar, especially in the inner part with $r<R_{vir}$.
Only at radii larger than the virial radius, the two profiles start to diverge, and group galaxies reach a temperature about a factor of 2 higher than isolated galaxies at $4\times R_{vir}$.
The higher temperature in group galaxies at large radii is mostly due to the admixture of gas bound to other galaxies in the FoF group, i.e., it is not the CGM of the galaxy itself, but the CGM of another galaxy in the FoF group.
The picture is completely different if we look at the profile for satellite galaxies.
In the inner part it is similar to central galaxies, but starts to diverge from the profiles of central galaxies at $R_{1/2}$ and reaches on average a temperature which is a factor of 4 higher compared to central galaxies at $R_{vir}$.
What is especially interesting is that satellite galaxies reach temperatures much higher than what is expected from their virial mass within approximately their virial radius.\footnote{Since virial properties are ill-defined for satellites as they depend on their location within the host halo, here we use $2.4\times R_{1/2}\approx R_{vir}$ as approximate virial radius.}
This behavior is due to satellite galaxies being embedded in the halo of more massive galaxies, i.e., what is probed here is not the CGM of the satellite but the CGM of its more massive host.

With the temperature and density profiles, we can now explain why the covering fractions for \civ~gas is much lower for group galaxies if satellite galaxies are included in the sample (see left panel of Fig. \ref{Fig:ReproduceTrendCIV}).
As becomes apparent from the right panel of Fig. \ref{Fig:RadialProfilesTempDens}, most of the gas around satellites is at a temperature above the temperature cut we employed to select \civ~gas.
Therefore, when calculating the covering fractions with satellites in the sample, the absence of gas in the right temperature phase around satellites leads to a lower covering fraction for the sample of group galaxies.
In other words, as satellites are embedded in the halos of larger host galaxies, the CGM of satellites reaches higher temperatures than the CGM of central galaxies of a similar mass.
This difference in temperature then leads to less gas in the right phase to be ionized to \civ.

The interpretation for \mgii~gas is not as straight forward.
While the density profile for satellites starts to diverge from the profiles of central galaxies at about $R_{1/2}$ and consequently lies above the profiles for isolated galaxies and group central galaxies, this happens at densities that are beyond the cut we used to select \mgii.
However, from the right panel of Fig. \ref{Fig:ExampleGalaxy_AllGasCIVandMgII}, we see that there is \mgii~gas at about $2\times R_{vir}$, where the median density profile is below our density cut for \mgii.
However, as we calculate the density profiles using median statistics, this gas does not show up in the profile. 
As mentioned earlier, the gas we see on the top left of the right panel of Fig. \ref{Fig:ExampleGalaxy_AllGasCIVandMgII} belongs to an infalling satellite.
Comparing this with the central galaxy in the middle, we see that the area of \mgii~gas around the satellite is larger than that around the central.
This implies that there are potentially more sightlines to detect \mgii~around satellites (hence also a higher covering fraction) than that around central galaxies due to the gas that is stripped from satellites as they move through the CGM of their host galaxies.
This then translates into a higher covering fraction around satellites compared to centrals in both isolation and groups.

\begin{figure}
        \centering
        \includegraphics[width=\columnwidth]{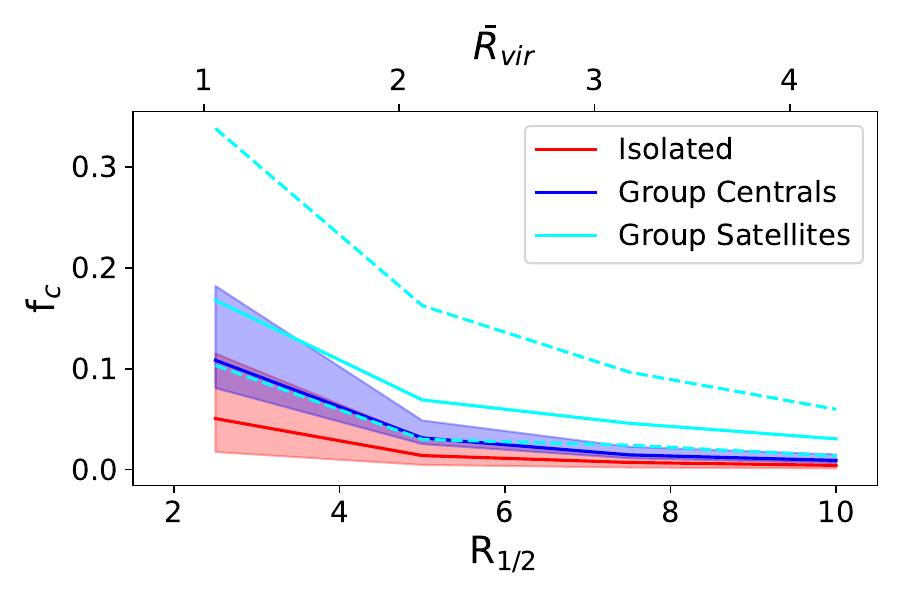}
        \caption{Covering fraction of \mgii~gas for isolated galaxies (red), central galaxies in groups (blue) and satellite galaxies in groups (cyan) in the stellar mass-halo mass bin shown in Fig. \ref{Fig:SubsampleSelection}. Shaded areas and dashed lines represent the $1\sigma$ halo-to-halo variation.
        We use a detection limit of 10$^{19}$ cm$^{-2}$ total gas column density.
        While group centrals have a slightly higher covering fraction than isolated centrals, the bigger difference is between satellites galaxies and central galaxies, both in isolation and groups.
        }
        \label{Fig:IsolationGroupsAndSatellitesInBin}
\end{figure}

We show this in Fig. \ref{Fig:IsolationGroupsAndSatellitesInBin}, where the median covering fraction of satellite galaxies (cyan) is higher than that of isolated central galaxies (red) by a factor of 3 at $R_{vir}$ and a factor of 7 at $4\times R_{vir}$, and is higher than that of central galaxies in groups (blue) by a factor of 1.5 at $R_{vir}$ and a factor of 3 at $4\times R_{vir}$.
Furthermore, central galaxies in groups have about a factor of 2 higher median covering fraction than isolated galaxies.
However, we note that by lowering the detection limit of total gas from 10$^{19}$ cm$^{-2}$ as used in Fig. \ref{Fig:IsolationGroupsAndSatellitesInBin} by one order of magnitude to 10$^{18}$ cm$^{-2}$, the difference between group central galaxies and isolated galaxies goes away, while satellite galaxies still have a higher covering fraction than both group central galaxies and isolated galaxies.

Therefore, we can contrast two scenarios in our model.
First, if the two samples are matched in all relevant aspects, that is stellar mass, halo mass and being the central galaxy of a DM halo, there is no significant difference in the physical properties of the CGM between galaxies in groups and galaxies in isolation.
While the median covering fraction of central galaxies in groups is about a factor of 2 higher than the median of isolated galaxies, the $1\sigma$ halo-to-halo variance is large.
Furthermore, the difference in median covering fractions goes away when going to lower detection limits.
In other words, in our model the large scale structure a galaxy is embedded in has no significant influence on its CGM when only central galaxies are considered.
This is in contrast to the mounting evidence from observations that finds a dependence on the large-scale structure of the incidence and distribution of cool gas in the CGM of galaxies \citep[see e.g.][]{Galbiati+2024, Galbiati2023, Lofthouse2023, Dutta2020, Dutta2021}.
However, the picture changes when satellite galaxies are considered.
In this case, we see a difference in covering fractions for both \mgii- and \civ~like gas.
In case of \civ-like gas, inclusion of satellites leads to an absence of detections in groups due to the different temperature structure in the CGM of satellite galaxies.
Contrary, for \mgii-like gas, the inclusion of satellites leads to an increase in covering fraction due to the cold gas being stripped from the satellites.
Therefore, there is a systematic difference between the CGM of central galaxies and satellite galaxies.
This systematic difference in the CGM between these two populations is also visible in the temperature and density profiles, where satellites reach higher densities and temperatures than their central galaxy counterparts.
The deviation from the temperature profile of central galaxies happens within approximately the virial radius of satellite galaxies.

\section{Discussion}
\label{Sec:Discussion}

Our results are not able to fully capture the trend of higher covering fraction of cool and warm gas found by a growing body of observations up to $z\sim 4$.
The only exception is the admixture of satellites to our sample, which allows to explain the observed difference in covering fractions with the current model.
In this section, we discuss possible explanations for this discrepancy and discuss them in light of the existing literature.

\subsection{The Role of Satellite Galaxies}
\label{Sec:MissclassificationOfSatellites}
The only physical scenario in our model in which the covering fractions of group and isolated galaxies are clearly different is when we include satellite galaxies in our sample of group galaxies.
Furthermore, we showed that the physical properties of the CGM of central galaxies and satellite galaxies are intrinsically different and hence also the covering fraction of different gas phases.
We want to stress that this result is independent of the included physical processes and the employed subgrid physics prescriptions in the simulation since it relies only on the different positions of central and satellite galaxies within the parent DM haloes and the densities and virial temperatures that can be reached there.

This has significant implications for all observational studies on the environmental dependence of the CGM.
Since it is challenging to accurately determine whether a galaxy is a central galaxy or a satellite in observations, it is possible that at least some group central galaxies could actually be satellite galaxies when constructing matched samples for comparing covering fractions in observations.
Thus, the samples of group and isolated galaxies in observations are likely not truly matched in all relevant aspects, and one could be comparing two intrinsically different samples apart from their large scale environment.
To ameliorate this problem in observations, one has to rely on proxies such as picking the most massive galaxy or the galaxy closest in projected distance from the quasar sightline.
\cite{Dutta2020, Dutta2021} experimented with such different ways of defining groups and central galaxies, but found that their results on environmental effects hold within the uncertainties irrespective of the different definitions.
However, if a more massive central galaxy is not part of the FoF group because it is outside the field of view, or it is not detected due to being located in front of the bright quasar, or being heavily dust obscured, none of the above ways to define groups and central galaxies will solve the problem of misclassified satellite galaxies.

\subsection{Missing Physics}
The simulation we used does not include several key physical mechanisms that can influence the distribution of ions in the CGM.
While cosmic rays and magnetic fields are not included at all in the \texttt{EAGLE} model, AGN feedback is not included in the simulation we use \citep{BLF2020}.
However, each of these processes can influence the CGM.

\subsubsection{AGNs}
\label{Sec:MissingPhysicsAGNs}
In the standard \texttt{EAGLE} model, the AGN feedback is implemented thermally and released stochastically whenever the reservoir of feedback energy exceeds a certain threshold \citep[][]{Schaye2015}.
\cite{Rahmati2016EagleAGNsAndIons} showed that turning off AGN feedback in the \texttt{EAGLE} model does not influence the cosmic distribution of metal ions, while the strength of the star formation feedback does.
However, \cite{Segers2017FluctuatingAGNs} showed that fluctuating AGN activity can affect the covering fraction of ions in the CGM due to the changing ionizing radiation.
This picture is supported by \cite{Hani2019DiversityOfCGMauriga}.
They use the \texttt{Auriga} simulations and show that if the AGN feedback is turned off, the metal column densities do not change, but the column densities of ions do.
Thus, if galaxies in overdensities have different AGN luminosities than galaxies in isolation, this could produce a difference in ion covering fractions as seen in observations.
Observationally, \cite{Man2019AGNandEnvironmentSDSS} find an increase of AGN activity in satellite galaxies in denser environments.
However, studies with MUSE that found an environmental dependence of the CGM, such as \cite{Dutta2020} and \cite{Galbiati+2024}, did not find any signatures of current AGN activity in the optical spectra of the galaxies in their sample.

\subsubsection{Cosmic Rays}
Cosmic rays can have a significant effect on the CGM and the observables we are interested in.
For example, \cite{DeFilippis2024CGMandCosmicRays} showed that the inclusion of cosmic rays increases the covering fractions of \civ~and \ovi~in the CGM, while that of \mgii~was not affected.
Their results are in agreement with \cite{Salem2016CosmicRaysCGM}, who find that the inclusion of cosmic rays increases the metalicity in the CGM due to cosmic ray driven outflows, and that the covering fraction of \ciii, \siiv, and \ovi~is higher if cosmic rays are included.
The dominant source of cosmic rays for galaxies up to a halo mass of $10^{12}$ M$_\odot$ are supernovae (SNe), while above that, it is the AGN at the center of the galaxy \citep{RuszkowskiPfrommer2023CosmicRayReview}, i.e., both of them are internal to the galaxy.
While we discussed the environmental dependence of AGN activity in Sec. \ref{Sec:MissingPhysicsAGNs}, the rate of SNe depends on the mass of the galaxy, the color (or specific star formation rate, sSFR), and the Hubble type of the galaxy \citep{Li2011SNeRates, Ma2025SNeRates}.
\cite{Dutta2021} find a dependence of the \mgii~covering fraction on the stellar mass and SFR and a weaker dependence of the \civ\ covering fraction on the same. However, they control for the stellar mass and SFR when checking for environmental dependence of the CGM.

\subsubsection{Magnetic Fields}
Similar to cosmic rays, magnetic fields have been shown to influence the properties of the CGM.
For example, \cite{VanDeVoort2021MagneticFieldsAndCGM} showed that the inclusion of magnetic fields increases the pressure of the CGM and temperature in the outer halo, while the mixing of the metals is reduced.
An increase in temperature in the CGM when magnetic fields are included is also found by \cite{Buie2022MagneticFieldsAndCGM}.
Furthermore, magnetic fields decrease the effectiveness of ram pressure stripping and reduce the volume of the high-metallicity gas \citep{Rintoul2025MagneticFieldsAndRamPressure}.
However, for magnetic fields to play a role in the dependence of ion covering fractions on the environment, there needs to be a dependence of the magnetic field strength on the large scale environment in which the galaxies are embedded.
Using simulations, \cite{Marinacci2015LargeScalePropertiesMagneticFields, Marinacci2018FirstResultsTNGmagneticfields} showed that magnetic fields are stronger in baryon overdensities, but that the enhanced magnetic field is mostly contained within the virial radius of the halo.
\cite{McDonough2025MagneticFieldsInSatellites} showed that the median magnetic field strength in isolated central galaxies and non-isolated central galaxies is the same in the TNG100 simulations.
However, satellite galaxies have been found to show a stronger magnetic field than central galaxies of the same mass in simulations \citep{Werhahn2025MagneticFieldsInSatellites, McDonough2025MagneticFieldsInSatellites}.

\subsection{Comparison with the literature}
\cite{Nelson2018DistributionOfHighlyIonizedOxygenInTNG} test for the environmental dependence of \ovi~covering fraction in TNG100 (see their Figure 13 and discussion thereof).
Using a similar criterion as in this work to split their sample into isolated galaxies and non-isolated galaxies, they do not find a significant difference in the \ovi~covering fractions with the environment.
However, they do not investigate whether adding satellites to the sample of non-isolated galaxies changes their result.
We note that \ovi~traces a different gas phase compared to \mgii~and \civ.
Therefore, the environmental effects are likely to be different and a direct comparison with our results is not possible.

\cite{Nelson2021ColdCircumgalacticMedium} use TNG50 to investigate the cold gas in emission.
They find a dependence of the size of the \mgii~emission haloes on the environment, where the half light radii of the haloes are larger in denser environments.
This is due to additional cold gas within nearby satellites and gas stripped from these satellites.
\cite{Dutta2023MgIIemission} studied the average \mgii~emission in observations and also find a slightly more extended \mgii~emission on average around group galaxies compared to isolated galaxies, after matching in stellar mass and redshift.
However, \mgii~emission typically traces denser gas closer to the galaxy in the disk-halo interface and inner CGM, in contrast to \mgii~absorption, which traces the more diffuse gas in the CGM.
Therefore, here as well, no direct comparison with our results is possible.

\cite{Sims2026CAMELSEnvironment} use the \texttt{CAMELS} simulations to investigate how the large-scale environment influences the baryon fraction, CGM fraction, gas mass fraction, and star formation rate in the SIMBA, IllustrisTNG, ASTRID, and Swift-EAGLE feedback models.
They find an environmental dependence of the baryon fraction and CGM fractions, but this environmental dependence is itself dependent on the feedback model.
SIMBA and IllustrisTNG have higher baryon and CGM fractions in underdense regions for haloes with a mass below $10^{12}$ M$_\odot$, while the trend is reversed in EAGLE-Swift, i.e., the baryon and CGM fractions are higher in overdense regions.
Therefore, the higher median covering fractions we find in group galaxies might be due to the \texttt{EAGLE} feedback model and could be absent when using simulation boxes based on other feedback prescriptions.

Finally, the misclassification of satellites as central galaxies was tested by \cite{Shreeram2025CGMprojectionXray}.
They showed that including satellites in the sample increases the stacked X-ray surface brightness profile compared to a sample of only central galaxies.
However, since they probe the hot gas phase, a direct comparison with our work is not possible here as well.

\subsection{Caveats}
We chose one of the highest-resolution cosmological boxes currently available for our study to get both, a statistical sample of galaxies in different environments and a high resolution.
Nevertheless, our choice of simulation also leads to some caveats.

\subsubsection{Sample Size}
\label{Sec:SampleSize}
In our simulation, we have 289 galaxies at $z=0$ above our stellar mass cut of $10^8$ M$_\odot$.
Splitting this sample into group and isolated galaxies gives us a mass-matched sample of 131 galaxy pairs if satellites are included.
This sample size is big enough for a statistical analysis.
However, to show the systematic difference between central and satellite galaxies, we had to further split our sample of group galaxies into centrals and satellites, which left us with a matched sample of 27 galaxy pairs.
To show the effect on the temperature and density profiles, we had to further restrict ourselves to galaxies in a small bin of stellar mass and halo mass, leading to only 13 group central galaxies.
A sample of such a small size is prone to small number statistics and these results have therefore to be taken with caution.

\subsubsection{Resolution}
\label{Sec:Resolution}
A high enough spatial resolution is important to get the right amount of cold gas in the CGM \citep[c.f.][]{Hummels2019CGMandResolution, VanDeVoort2019CGMHighResolution, Peeples2019Foggie, Ramesh2024GIBLEMagneticFields, Ramesh2024GIBLESmallScaleGas}.
Some of the additional cold gas in high-resolution runs comes from precipitation in filaments, which can be resolved in the higher resolution runs but not in the low resolution runs \citep{Hummels2019CGMandResolution}.
A similar effect would also happen for gas stripped from infalling satellites.
Thus, our results might be biased due to the low resolution, and the small difference we see between group central galaxies and isolated galaxies might in reality be much bigger if all the processes involved could be resolved.

However, all the work on the CGM and resolution has been done with grid based and moving mesh codes, while we are using a simulation based on SPH which is different in two important ways.
First, the spatial resolution is tied to the particle density and therefore is adaptive in the sense that higher density gas has automatically higher spatial resolution.
Second, particle based codes suppress fluid mixing and better preserve small structures \citep{Agertz2007, Hummels2019CGMandResolution}, which can also influence the amount of cold gas.
Currently, no work exists that examines the influence of increased resolution on the CGM using SPH similar to what has been done for grid based and moving mesh codes.
Therefore, while we expect that SPH codes will show a similar behavior concerning the amount of cold gas in the CGM when increasing the resolution, it is not clear whether this is indeed the case.

\subsubsection{The mass of ram pressure stripped galaxies}
Environmental interactions in groups can remove gas from galaxies and increase the projected area in which an ion can be found compared to isolated galaxies \citep[see e.g.][]{Fossati2019MUDF2,Dutta2021}.
In the right panel of Fig. \ref{Fig:ExampleGalaxy_AllGasCIVandMgII}, we see an example of ram pressure stripping causing such an effect in our simulation.
In \cite{Herzog2023}, we showed that about 25\% of all galaxies in our simulation box loose gas due to ram pressure stripping in galaxy haloes of more massive hosts and the cosmic web filaments connecting them.
Thus, we expected to see the effect of ram pressure stripping causing an increased covering fraction of cold gas in group galaxies.
However, when matched in all relevant aspects, we cannot fully reproduce the observed higher covering fractions in the CGM of group galaxies.

This might be due to either the limited spatial resolution of our simulation (see Sec. \ref{Sec:Resolution}) or the low mass of the infalling satellites. 
In their work, \cite{ManamiRoy2024} checked the effect of infalling satellites with different masses and showed that an infalling galaxy needs a stellar mass of about $10^{10}$ M$_{\odot}$ to add a significant amount of cold gas to the host CGM.
In our simulation box, the most massive structure that formed has a halo mass of about $8\times 10^{12}$ M$_{\odot}$ and the galaxies that were stripped by ram pressure have a halo mass of less than $10^{10}$ M$_{\odot}$, i.e., a stellar mass of about $10^{8}$ M$_{\odot}$.
Thus, we might not be seeing a clearer difference in the covering fractions around isolated and group galaxies due to the low mass of infalling galaxies in our simulation.

\subsubsection{The metal content of our simulation} 
Since our simulation does not reproduce the right metal content on cosmological scales, we used the gas phases that are traced by \mgii~and \civ~instead of the ions themselves.
However, it is supported by both observations and simulations that environmental interactions in overdense regions can enhance the metallicity of the surrounding gas.
For example, the M86 group falling into the Virgo Cluster shows a ram pressure stripped tail, with a metal abundance almost twice the abundance of the surrounding gas \citep{Ehlert2013MetalsM86, Gastaldello2021}.  
Furthermore, major mergers can lead to a metal enrichment of the CGM surrounding the galaxies because of gas that is removed by dynamical interactions between the merging galaxies \citep{DiCesare2024MetalsAndMergers} and by outflows due to AGN activity triggered by the merger \citep{Hani2018MergersAndCGM}.
Thus, if the difference in the ion covering fractions with environment seen in observations is caused by a difference in the metal content and distribution, instead of differences in temperature and density, we cannot investigate it with our method.

\section{Conclusion}
\label{Sec:Conclusion}

We investigated the environmental dependence of the CGM using a high-resolution ($m_{gas}\approx 4.5\times 10^4$ M$_{\odot}$, $m_{dm}\approx 2.4\times 10^5$ M$_{\odot}$) cosmological simulation based on the \texttt{EAGLE} model of galaxy formation.
Selecting the gas phases traced by \mgii~and \civ, and dividing our sample of galaxies at $z=0$ into group and isolated galaxies using a FoF algorithm, we found the following:
\begin{itemize}
    \item if the two samples of galaxies are matched in all relevant aspects, that is stellar mass, halo mass, and being the central galaxy of a DM halo, there is no significant difference in the covering fractions of gas traced by \mgii~and \civ~between group and isolated galaxies (see right panels of Figs. \ref{Fig:ReproduceTrend} and \ref{Fig:ReproduceTrendCIV}).
    \item this result is independent of the chosen detection limit (see Figs. \ref{Fig:6a} and \ref{Fig:6b}).
    \item when selecting group and isolated central galaxies within a small bin of stellar mass and halo mass (see Fig. \ref{Fig:SubsampleSelection}), their density profiles are the same out to $10\times R_{1/2}$, while the temperature profiles agree out to $2.5\times R_{1/2}\approx R_{vir}$.
    Beyond that, group galaxies have a slightly higher temperature, which is due to the presence of other galaxies (see Fig. \ref{Fig:RadialProfilesTempDens}).
    \item if satellite galaxies are included in the sample of group galaxies, we can reproduce the observed trend of higher covering fractions in groups for gas traced by \mgii~(see left panel of Fig. \ref{Fig:ReproduceTrend} and Fig. \ref{Fig:IsolationGroupsAndSatellitesInBin}). For \civ, we can reproduce the observed lack of detections in groups at low redshift (see left panel of Fig. \ref{Fig:ReproduceTrendCIV}).
    Furthermore, we showed that satellite galaxies have different temperature and density profiles than central galaxies at the same stellar and halo mass (see Fig. \ref{Fig:RadialProfilesTempDens}).
\end{itemize}

Our difficulty to fully reproduce the observed differences in covering fractions between group and isolated galaxies if only central galaxies are considered is a direct consequence of the difficulty in modelling the multiphase CGM.
However, it gives a clear path for future research, which needs to overcome the conflict between sample size and resolution with a set of zoom-in simulations of central and satellite galaxies in different environments.
While the codes and technology to do this already exists, it requires an amount of computing power that is currently not available.
Our result of a different density and temperature structure of satellites is independent of the employed subgrid physics since it only depends on their position within the DM halo of a more massive host.
Thus, any observational study on the CGM that cannot accurately distinguish between central galaxies and satellites might suffer from a potential bias of comparing two intrinsically different samples when comparing the CGM of group galaxies (potentially containing satellites) and isolated galaxies (without satellites).
To discern the role of the environment on the CGM, we, therefore, need to clearly disentangle central galaxies and satellites in observational studies.

\clearpage

%%%%%%%%%%%%%%%%%%%%%%%%%%%%%%%%%%%%%%%%%%%%%%%%%%%%%%%%%%%%%%
\begin{acknowledgements}
      We want to thank Alejandro Ben\'itez-Llambay for his help and advice in handling the simulation as well as the fruitful discussions and comments.
      We also want to thank Adriana Gargiulo for her comments on the paper draft which helped to clarify the final version.
      This work used the DiRAC@Durham facility managed by the Institute for Computational Cosmology on behalf of the STFC DiRAC HPC Facility (www.dirac.ac.uk).
      The equipment was funded by BEIS capital funding via STFC capital grants ST/K00042X/1, ST/P002293/1, ST/R002371/1 and ST/S002502/1, Durham University and STFC operations grant ST/R000832/1.
      DiRAC is part of the National e-Infrastructure.
\end{acknowledgements}

%%%%%%%%%%%%%%%%%%%%%%%%%%%%%%%%%%%%%%%%%%%%%%%%%%%%%%%%%%%%%%
% WARNING
% Please note that we have included the references below in
% order to compile the document, but we ask you to:
%
% - use BibTeX with the regular commands:
%   \bibliographystyle{aa} % style aa.bst
%   \bibliography{Yourfile} % your references Yourfile.bib
% - join the .bib files when you upload your source files
%%%%%%%%%%%%%%%%%%%%%%%%%%%%%%%%%%%%%%%%%%%%%%%%%%%%%%%%%%%%%%
\bibliographystyle{bibtex/aa}
\bibliography{CGMenvironment}

@ARTICLE{Dutta2021,
       author = {{Dutta}, Rajeshwari and {Fumagalli}, Michele and {Fossati}, Matteo and {Bielby}, Richard M. and {Stott}, John P. and {Lofthouse}, Emma K. and {Cantalupo}, Sebastiano and {Cullen}, Fergus and {Crain}, Robert A. and {Tripp}, Todd M. and {Prochaska}, J. Xavier and {Arrigoni Battaia}, Fabrizio and {Burchett}, Joseph N. and {Fynbo}, Johan P.~U. and {Murphy}, Michael T. and {Schaye}, Joop and {Tejos}, Nicolas and {Theuns}, Tom},
        title = "{Metal-enriched halo gas across galaxy overdensities over the last 10 billion years}",
      journal = {\mnras},
         year = 2021,
        month = dec,
       volume = {508},
       number = {3},
        pages = {4573-4599},
          doi = {10.1093/mnras/stab2752},
archivePrefix = {arXiv},
       eprint = {2109.10927},
 primaryClass = {astro-ph.GA},
       adsurl = {https://ui.adsabs.harvard.edu/abs/2021MNRAS.508.4573D}
}

@ARTICLE{Galbiati2023,
       author = {{Galbiati}, Marta and {Fumagalli}, Michele and {Fossati}, Matteo and {Lofthouse}, Emma K. and {Dutta}, Rajeshwari and {Prochaska}, J. Xavier and {Murphy}, Michael T. and {Cantalupo}, Sebastiano},
        title = "{MUSE Analysis of Gas around Galaxies (MAGG) -- V: Linking ionized gas traced by CIV and SiIV absorbers to Ly${\alpha}$ emitting galaxies at $z\approx 3.0-4.5$}",
      journal = {arXiv e-prints},
         year = 2023,
        month = jan,
          eid = {arXiv:2302.00021},
        pages = {arXiv:2302.00021},
          doi = {10.48550/arXiv.2302.00021},
archivePrefix = {arXiv},
       eprint = {2302.00021},
 primaryClass = {astro-ph.GA},
       adsurl = {https://ui.adsabs.harvard.edu/abs/2023arXiv230200021G}
}

@ARTICLE{Galbiati+2024,
       author = {{Galbiati}, Marta and {Dutta}, Rajeshwari and {Fumagalli}, Michele and {Fossati}, Matteo and {Cantalupo}, Sebastiano},
        title = "{MUSE Analysis of Gas around Galaxies (MAGG): VI. The cool and enriched gas environment of z {\ensuremath{\gtrsim}} 3 Ly{\ensuremath{\alpha}} emitters}",
      journal = {\aap},
         year = 2024,
        month = oct,
       volume = {690},
          eid = {A7},
        pages = {A7},
          doi = {10.1051/0004-6361/202450741},
archivePrefix = {arXiv},
       eprint = {2406.10350},
 primaryClass = {astro-ph.GA},
       adsurl = {https://ui.adsabs.harvard.edu/abs/2024A&A...690A...7G}
}

@ARTICLE{Tumlinson2017CGMreview,
       author = {{Tumlinson}, Jason and {Peeples}, Molly S. and {Werk}, Jessica K.},
        title = "{The Circumgalactic Medium}",
      journal = {\araa},
         year = 2017,
        month = aug,
       volume = {55},
       number = {1},
        pages = {389-432},
          doi = {10.1146/annurev-astro-091916-055240},
archivePrefix = {arXiv},
       eprint = {1709.09180},
 primaryClass = {astro-ph.GA},
       adsurl = {https://ui.adsabs.harvard.edu/abs/2017ARA&A..55..389T}
}

@ARTICLE{Peroux2020CosmicBaryonCycle,
       author = {{P{\'e}roux}, C{\'e}line and {Howk}, J. Christopher},
        title = "{The Cosmic Baryon and Metal Cycles}",
      journal = {\araa},
         year = 2020,
        month = aug,
       volume = {58},
        pages = {363-406},
          doi = {10.1146/annurev-astro-021820-120014},
archivePrefix = {arXiv},
       eprint = {2011.01935},
 primaryClass = {astro-ph.GA},
       adsurl = {https://ui.adsabs.harvard.edu/abs/2020ARA&A..58..363P}
}

@ARTICLE{Boselli2022RamPressureReview,
       author = {{Boselli}, Alessandro and {Fossati}, Matteo and {Sun}, Ming},
        title = "{Ram pressure stripping in high-density environments}",
      journal = {\aapr},
         year = 2022,
        month = dec,
       volume = {30},
       number = {1},
          eid = {3},
        pages = {3},
          doi = {10.1007/s00159-022-00140-3},
archivePrefix = {arXiv},
       eprint = {2109.13614},
 primaryClass = {astro-ph.GA},
       adsurl = {https://ui.adsabs.harvard.edu/abs/2022A&ARv..30....3B}
}

@ARTICLE{BlantonAndMoustakas2009,
       author = {{Blanton}, Michael R. and {Moustakas}, John},
        title = "{Physical Properties and Environments of Nearby Galaxies}",
      journal = {\araa},
         year = 2009,
        month = sep,
       volume = {47},
       number = {1},
        pages = {159-210},
          doi = {10.1146/annurev-astro-082708-101734},
archivePrefix = {arXiv},
       eprint = {0908.3017},
 primaryClass = {astro-ph.GA},
       adsurl = {https://ui.adsabs.harvard.edu/abs/2009ARA&A..47..159B}
}

@ARTICLE{NaabAndOstriker2017,
       author = {{Naab}, Thorsten and {Ostriker}, Jeremiah P.},
        title = "{Theoretical Challenges in Galaxy Formation}",
      journal = {\araa},
         year = 2017,
        month = aug,
       volume = {55},
       number = {1},
        pages = {59-109},
          doi = {10.1146/annurev-astro-081913-040019},
archivePrefix = {arXiv},
       eprint = {1612.06891},
 primaryClass = {astro-ph.GA},
       adsurl = {https://ui.adsabs.harvard.edu/abs/2017ARA&A..55...59N}
}

@ARTICLE{Benavides2021,
       author = {{Benavides}, Jos{\'e} A. and {Sales}, Laura V. and {Abadi}, Mario. G. and {Pillepich}, Annalisa and {Nelson}, Dylan and {Marinacci}, Federico and {Cooper}, Michael and {Pakmor}, Ruediger and {Torrey}, Paul and {Vogelsberger}, Mark and {Hernquist}, Lars},
        title = "{Quiescent ultra-diffuse galaxies in the field originating from backsplash orbits}",
      journal = {Nature Astronomy},
         year = 2021,
        month = sep,
       volume = {5},
        pages = {1255-1260},
          doi = {10.1038/s41550-021-01458-1},
archivePrefix = {arXiv},
       eprint = {2109.01677},
 primaryClass = {astro-ph.GA},
       adsurl = {https://ui.adsabs.harvard.edu/abs/2021NatAs...5.1255B}
}

@ARTICLE{Herzog2023,
       author = {{Herzog}, Georg and {Ben{\'\i}tez-Llambay}, Alejandro and {Fumagalli}, Michele},
        title = "{The present-day gas content of simulated field dwarf galaxies}",
      journal = {\mnras},
         year = 2023,
        month = feb,
       volume = {518},
       number = {4},
        pages = {6305-6317},
          doi = {10.1093/mnras/stac3282},
archivePrefix = {arXiv},
       eprint = {2209.11782},
 primaryClass = {astro-ph.GA},
       adsurl = {https://ui.adsabs.harvard.edu/abs/2023MNRAS.518.6305H}
}

@ARTICLE{Applebaum2021,
       author = {{Applebaum}, Elaad and {Brooks}, Alyson M. and {Christensen}, Charlotte R. and {Munshi}, Ferah and {Quinn}, Thomas R. and {Shen}, Sijing and {Tremmel}, Michael},
        title = "{Ultrafaint Dwarfs in a Milky Way Context: Introducing the Mint Condition DC Justice League Simulations}",
      journal = {\apj},
         year = 2021,
        month = jan,
       volume = {906},
       number = {2},
          eid = {96},
        pages = {96},
          doi = {10.3847/1538-4357/abcafa},
archivePrefix = {arXiv},
       eprint = {2008.11207},
 primaryClass = {astro-ph.GA},
       adsurl = {https://ui.adsabs.harvard.edu/abs/2021ApJ...906...96A}
}

@ARTICLE{Bordoloi2011,
       author = {{Bordoloi}, R. and {Lilly}, S.~J. and {Knobel}, C. and {Bolzonella}, M. and {Kampczyk}, P. and {Carollo}, C.~M. and {Iovino}, A. and {Zucca}, E. and {Contini}, T. and {Kneib}, J. -P. and {Le Fevre}, O. and {Mainieri}, V. and {Renzini}, A. and {Scodeggio}, M. and {Zamorani}, G. and {Balestra}, I. and {Bardelli}, S. and {Bongiorno}, A. and {Caputi}, K. and {Cucciati}, O. and {de la Torre}, S. and {de Ravel}, L. and {Garilli}, B. and {Kova{\v{c}}}, K. and {Lamareille}, F. and {Le Borgne}, J. -F. and {Le Brun}, V. and {Maier}, C. and {Mignoli}, M. and {Pello}, R. and {Peng}, Y. and {Perez Montero}, E. and {Presotto}, V. and {Scarlata}, C. and {Silverman}, J. and {Tanaka}, M. and {Tasca}, L. and {Tresse}, L. and {Vergani}, D. and {Barnes}, L. and {Cappi}, A. and {Cimatti}, A. and {Coppa}, G. and {Diener}, C. and {Franzetti}, P. and {Koekemoer}, A. and {L{\'o}pez-Sanjuan}, C. and {McCracken}, H.~J. and {Moresco}, M. and {Nair}, P. and {Oesch}, P. and {Pozzetti}, L. and {Welikala}, N.},
        title = "{The Radial and Azimuthal Profiles of Mg II Absorption around 0.5 < z < 0.9 zCOSMOS Galaxies of Different Colors, Masses, and Environments}",
      journal = {\apj},
         year = 2011,
        month = dec,
       volume = {743},
       number = {1},
          eid = {10},
        pages = {10},
          doi = {10.1088/0004-637X/743/1/10},
archivePrefix = {arXiv},
       eprint = {1106.0616},
 primaryClass = {astro-ph.CO},
       adsurl = {https://ui.adsabs.harvard.edu/abs/2011ApJ...743...10B}
}

@ARTICLE{Burchett+2016,
       author = {{Burchett}, Joseph N. and {Tripp}, Todd M. and {Bordoloi}, Rongmon and {Werk}, Jessica K. and {Prochaska}, J. Xavier and {Tumlinson}, Jason and {Willmer}, C.~N.~A. and {O'Meara}, John and {Katz}, Neal},
        title = "{A Deep Search for Faint Galaxies Associated with Very Low Redshift C IV Absorbers. III. The Mass- and Environment-dependent Circumgalactic Medium}",
      journal = {\apj},
         year = 2016,
        month = dec,
       volume = {832},
       number = {2},
          eid = {124},
        pages = {124},
          doi = {10.3847/0004-637X/832/2/124},
archivePrefix = {arXiv},
       eprint = {1512.00853},
 primaryClass = {astro-ph.GA},
       adsurl = {https://ui.adsabs.harvard.edu/abs/2016ApJ...832..124B}
}

@INPROCEEDINGS{MUSEinstrument2006,
       author = {{Bacon}, R. and {Bauer}, S. and {Boehm}, P. and {Boudon}, D. and {Brau-Nogu{\'e}}, S. and {Caillier}, P. and {Capoani}, L. and {Carollo}, C.~M. and {Champavert}, N. and {Contini}, T. and {Daguis{\'e}}, E. and {Dall{\'e}}, D. and {Delabre}, B. and {Devriendt}, J. and {Dreizler}, S. and {Dubois}, J. and {Dupieux}, M. and {Dupin}, J.~P. and {Emsellem}, E. and {Ferruit}, P. and {Franx}, M. and {Gallou}, G. and {Gerssen}, J. and {Guiderdoni}, B. and {Hahn}, T. and {Hofmann}, D. and {Jarno}, A. and {Kelz}, A. and {Koehler}, C. and {Kollatschny}, W. and {Kosmalski}, J. and {Laurent}, F. and {Lilly}, S.~J. and {Lizon}, J. and {Loupias}, M. and {Lynn}, S. and {Manescau}, A. and {McDermid}, R.~M. and {Monstein}, C. and {Nicklas}, H. and {Par{\`e}s}, L. and {Pasquini}, L. and {P{\'e}contal-Rousset}, A. and {P{\'e}contal}, E. and {Pello}, R. and {Petit}, C. and {Picat}, J. -P. and {Popow}, E. and {Quirrenbach}, A. and {Reiss}, R. and {Renault}, E. and {Roth}, M. and {Schaye}, J. and {Soucail}, G. and {Steinmetz}, M. and {Stroebele}, S. and {Stuik}, R. and {Weilbacher}, P. and {Wozniak}, H. and {de Zeeuw}, P.~T.},
        title = "{Probing unexplored territories with MUSE: a second generation instrument for the VLT}",
    booktitle = {Ground-based and Airborne Instrumentation for Astronomy},
         year = 2006,
       editor = {{McLean}, Ian S. and {Iye}, Masanori},
       series = {Society of Photo-Optical Instrumentation Engineers (SPIE) Conference Series},
       volume = {6269},
        month = jun,
          eid = {62690J},
        pages = {62690J},
          doi = {10.1117/12.669772},
archivePrefix = {arXiv},
       eprint = {astro-ph/0606329},
 primaryClass = {astro-ph},
       adsurl = {https://ui.adsabs.harvard.edu/abs/2006SPIE.6269E..0JB}
}

@INPROCEEDINGS{MUSEinstrument2010,
       author = {{Bacon}, R. and {Accardo}, M. and {Adjali}, L. and {Anwand}, H. and {Bauer}, S. and {Biswas}, I. and {Blaizot}, J. and {Boudon}, D. and {Brau-Nogue}, S. and {Brinchmann}, J. and {Caillier}, P. and {Capoani}, L. and {Carollo}, C.~M. and {Contini}, T. and {Couderc}, P. and {Daguis{\'e}}, E. and {Deiries}, S. and {Delabre}, B. and {Dreizler}, S. and {Dubois}, J. and {Dupieux}, M. and {Dupuy}, C. and {Emsellem}, E. and {Fechner}, T. and {Fleischmann}, A. and {Fran{\c{c}}ois}, M. and {Gallou}, G. and {Gharsa}, T. and {Glindemann}, A. and {Gojak}, D. and {Guiderdoni}, B. and {Hansali}, G. and {Hahn}, T. and {Jarno}, A. and {Kelz}, A. and {Koehler}, C. and {Kosmalski}, J. and {Laurent}, F. and {Le Floch}, M. and {Lilly}, S.~J. and {Lizon}, J. -L. and {Loupias}, M. and {Manescau}, A. and {Monstein}, C. and {Nicklas}, H. and {Olaya}, J. -C. and {Pares}, L. and {Pasquini}, L. and {P{\'e}contal-Rousset}, A. and {Pell{\'o}}, R. and {Petit}, C. and {Popow}, E. and {Reiss}, R. and {Remillieux}, A. and {Renault}, E. and {Roth}, M. and {Rupprecht}, G. and {Serre}, D. and {Schaye}, J. and {Soucail}, G. and {Steinmetz}, M. and {Streicher}, O. and {Stuik}, R. and {Valentin}, H. and {Vernet}, J. and {Weilbacher}, P. and {Wisotzki}, L. and {Yerle}, N.},
        title = "{The MUSE second-generation VLT instrument}",
    booktitle = {Ground-based and Airborne Instrumentation for Astronomy III},
         year = 2010,
       editor = {{McLean}, Ian S. and {Ramsay}, Suzanne K. and {Takami}, Hideki},
       series = {Society of Photo-Optical Instrumentation Engineers (SPIE) Conference Series},
       volume = {7735},
        month = jul,
          eid = {773508},
        pages = {773508},
          doi = {10.1117/12.856027},
archivePrefix = {arXiv},
       eprint = {2211.16795},
 primaryClass = {astro-ph.IM},
       adsurl = {https://ui.adsabs.harvard.edu/abs/2010SPIE.7735E..08B}
}

@ARTICLE{Fossati2019MUDF2,
       author = {{Fossati}, M. and {Fumagalli}, M. and {Lofthouse}, E.~K. and {D'Odorico}, V. and {Lusso}, E. and {Cantalupo}, S. and {Cooke}, R.~J. and {Cristiani}, S. and {Haardt}, F. and {Morris}, S.~L. and {Peroux}, C. and {Prichard}, L.~J. and {Rafelski}, M. and {Smail}, I. and {Theuns}, T.},
        title = "{The MUSE Ultra Deep Field (MUDF). II. Survey design and the gaseous properties of galaxy groups at 0.5 < z < 1.5}",
      journal = {\mnras},
         year = 2019,
        month = nov,
       volume = {490},
       number = {1},
        pages = {1451-1469},
          doi = {10.1093/mnras/stz2693},
archivePrefix = {arXiv},
       eprint = {1909.04672},
 primaryClass = {astro-ph.GA},
       adsurl = {https://ui.adsabs.harvard.edu/abs/2019MNRAS.490.1451F}
}

@ARTICLE{Dutta2020,
       author = {{Dutta}, Rajeshwari and {Fumagalli}, Michele and {Fossati}, Matteo and {Lofthouse}, Emma K. and {Prochaska}, J. Xavier and {Arrigoni Battaia}, Fabrizio and {Bielby}, Richard M. and {Cantalupo}, Sebastiano and {Cooke}, Ryan J. and {Murphy}, Michael T. and {O'Meara}, John M.},
        title = "{MUSE Analysis of Gas around Galaxies (MAGG) - II: metal-enriched halo gas around z {\ensuremath{\sim}} 1 galaxies}",
      journal = {\mnras},
         year = 2020,
        month = dec,
       volume = {499},
       number = {4},
        pages = {5022-5046},
          doi = {10.1093/mnras/staa3147},
archivePrefix = {arXiv},
       eprint = {2009.14219},
 primaryClass = {astro-ph.GA},
       adsurl = {https://ui.adsabs.harvard.edu/abs/2020MNRAS.499.5022D}
}

@ARTICLE{Lofthouse2023,
       author = {{Lofthouse}, Emma K. and {Fumagalli}, Michele and {Fossati}, Matteo and {Dutta}, Rajeshwari and {Galbiati}, Marta and {Arrigoni Battaia}, Fabrizio and {Cantalupo}, Sebastiano and {Christensen}, Lise and {Cooke}, Ryan J. and {Longobardi}, Alessia and {Murphy}, Michael T. and {Prochaska}, J. Xavier},
        title = "{MUSE Analysis of Gas around Galaxies (MAGG) - IV. The gaseous environment of z   3-4 Ly {\ensuremath{\alpha}} emitting galaxies}",
      journal = {\mnras},
         year = 2023,
        month = jan,
       volume = {518},
       number = {1},
        pages = {305-331},
          doi = {10.1093/mnras/stac3089},
archivePrefix = {arXiv},
       eprint = {2209.15021},
 primaryClass = {astro-ph.GA},
       adsurl = {https://ui.adsabs.harvard.edu/abs/2023MNRAS.518..305L}
}

@ARTICLE{Schaye2015,
       author = {{Schaye}, Joop and {Crain}, Robert A. and {Bower}, Richard G. and {Furlong}, Michelle and {Schaller}, Matthieu and {Theuns}, Tom and {Dalla Vecchia}, Claudio and {Frenk}, Carlos S. and {McCarthy}, I.~G. and {Helly}, John C. and {Jenkins}, Adrian and {Rosas-Guevara}, Y.~M. and {White}, Simon D.~M. and {Baes}, Maarten and {Booth}, C.~M. and {Camps}, Peter and {Navarro}, Julio F. and {Qu}, Yan and {Rahmati}, Alireza and {Sawala}, Till and {Thomas}, Peter A. and {Trayford}, James},
        title = "{The EAGLE project: simulating the evolution and assembly of galaxies and their environments}",
      journal = {\mnras},
         year = 2015,
        month = jan,
       volume = {446},
       number = {1},
        pages = {521-554},
          doi = {10.1093/mnras/stu2058},
archivePrefix = {arXiv},
       eprint = {1407.7040},
 primaryClass = {astro-ph.GA},
       adsurl = {https://ui.adsabs.harvard.edu/abs/2015MNRAS.446..521S}
}

@ARTICLE{Crain2015,
       author = {{Crain}, Robert A. and {Schaye}, Joop and {Bower}, Richard G. and {Furlong}, Michelle and {Schaller}, Matthieu and {Theuns}, Tom and {Dalla Vecchia}, Claudio and {Frenk}, Carlos S. and {McCarthy}, Ian G. and {Helly}, John C. and {Jenkins}, Adrian and {Rosas-Guevara}, Yetli M. and {White}, Simon D.~M. and {Trayford}, James W.},
        title = "{The EAGLE simulations of galaxy formation: calibration of subgrid physics and model variations}",
      journal = {\mnras},
         year = 2015,
        month = jun,
       volume = {450},
       number = {2},
        pages = {1937-1961},
          doi = {10.1093/mnras/stv725},
archivePrefix = {arXiv},
       eprint = {1501.01311},
 primaryClass = {astro-ph.GA},
       adsurl = {https://ui.adsabs.harvard.edu/abs/2015MNRAS.450.1937C}
}

@ARTICLE{Fumagalli+2011,
       author = {{Fumagalli}, Michele and {Prochaska}, J. Xavier and {Kasen}, Daniel and {Dekel}, Avishai and {Ceverino}, Daniel and {Primack}, Joel R.},
        title = "{Absorption-line systems in simulated galaxies fed by cold streams}",
      journal = {\mnras},
         year = 2011,
        month = dec,
       volume = {418},
       number = {3},
        pages = {1796-1821},
          doi = {10.1111/j.1365-2966.2011.19599.x},
archivePrefix = {arXiv},
       eprint = {1103.2130},
 primaryClass = {astro-ph.CO},
       adsurl = {https://ui.adsabs.harvard.edu/abs/2011MNRAS.418.1796F}
}

@ARTICLE{Suresh+2015,
       author = {{Suresh}, Joshua and {Bird}, Simeon and {Vogelsberger}, Mark and {Genel}, Shy and {Torrey}, Paul and {Sijacki}, Debora and {Springel}, Volker and {Hernquist}, Lars},
        title = "{The impact of galactic feedback on the circumgalactic medium}",
      journal = {\mnras},
         year = 2015,
        month = mar,
       volume = {448},
       number = {1},
        pages = {895-909},
          doi = {10.1093/mnras/stu2762},
archivePrefix = {arXiv},
       eprint = {1501.02267},
 primaryClass = {astro-ph.GA},
       adsurl = {https://ui.adsabs.harvard.edu/abs/2015MNRAS.448..895S}
}

@ARTICLE{Turner+2017,
       author = {{Turner}, Monica L. and {Schaye}, Joop and {Crain}, Robert A. and {Rudie}, Gwen and {Steidel}, Charles C. and {Strom}, Allison and {Theuns}, Tom},
        title = "{A comparison of observed and simulated absorption from H I, C IV, and Si IV around z {\ensuremath{\approx}} 2 star-forming galaxies suggests redshift-space distortions are due to inflows}",
      journal = {\mnras},
         year = 2017,
        month = oct,
       volume = {471},
       number = {1},
        pages = {690-705},
          doi = {10.1093/mnras/stx1616},
archivePrefix = {arXiv},
       eprint = {1703.00086},
 primaryClass = {astro-ph.GA},
       adsurl = {https://ui.adsabs.harvard.edu/abs/2017MNRAS.471..690T}
}

@ARTICLE{Hafen2017,
       author = {{Hafen}, Zachary and {Faucher-Gigu{\`e}re}, Claude-Andr{\'e} and {Angl{\'e}s-Alc{\'a}zar}, Daniel and {Kere{\v{s}}}, Du{\v{s}}an and {Feldmann}, Robert and {Chan}, T.~K. and {Quataert}, Eliot and {Murray}, Norman and {Hopkins}, Philip F.},
        title = "{Low-redshift Lyman limit systems as diagnostics of cosmological inflows and outflows}",
      journal = {\mnras},
         year = 2017,
        month = aug,
       volume = {469},
       number = {2},
        pages = {2292-2304},
          doi = {10.1093/mnras/stx952},
archivePrefix = {arXiv},
       eprint = {1608.05712},
 primaryClass = {astro-ph.GA},
       adsurl = {https://ui.adsabs.harvard.edu/abs/2017MNRAS.469.2292H}
}

@ARTICLE{Oppenheimer+2018,
       author = {{Oppenheimer}, Benjamin D. and {Schaye}, Joop and {Crain}, Robert A. and {Werk}, Jessica K. and {Richings}, Alexander J.},
        title = "{The multiphase circumgalactic medium traced by low metal ions in EAGLE zoom simulations}",
      journal = {\mnras},
         year = 2018,
        month = nov,
       volume = {481},
       number = {1},
        pages = {835-859},
          doi = {10.1093/mnras/sty2281},
archivePrefix = {arXiv},
       eprint = {1709.07577},
 primaryClass = {astro-ph.GA},
       adsurl = {https://ui.adsabs.harvard.edu/abs/2018MNRAS.481..835O}
}

@ARTICLE{Fielding2020,
       author = {{Fielding}, Drummond and {Tonnesen}, Stephanie and {DeFelippis}, Daniel and {Li}, Miao and {Su}, Kung-Yi and {Bryan}, Greg L. and {Kim}, Chang-Goo and {Forbes}, John C. and {Somerville}, Rachel S. and {Battaglia}, Nicholas and {Schneider}, Evan E. and {Li}, Yuan and {Choi}, Ena and {Hayward}, Christopher C. and {Hernquist}, Lars},
        title = "{First Results from SMAUG: Uncovering the Origin of the Multiphase Circumgalactic Medium with a Comparative Analysis of Idealized and Cosmological Simulations}",
      journal = {\apj},
         year = 2020,
        month = nov,
       volume = {903},
       number = {1},
          eid = {32},
        pages = {32},
          doi = {10.3847/1538-4357/abbc6d},
archivePrefix = {arXiv},
       eprint = {2006.16316},
 primaryClass = {astro-ph.GA},
       adsurl = {https://ui.adsabs.harvard.edu/abs/2020ApJ...903...32F}
}

@ARTICLE{Ho2020,
       author = {{Ho}, Stephanie H. and {Martin}, Crystal L. and {Schaye}, Joop},
        title = "{Morphological and Rotation Structures of Circumgalactic Mg II Gas in the EAGLE Simulation and the Dependence on Galaxy Properties}",
      journal = {\apj},
         year = 2020,
        month = nov,
       volume = {904},
       number = {1},
          eid = {76},
        pages = {76},
          doi = {10.3847/1538-4357/abbe88},
archivePrefix = {arXiv},
       eprint = {2010.02944},
 primaryClass = {astro-ph.GA},
       adsurl = {https://ui.adsabs.harvard.edu/abs/2020ApJ...904...76H}
}

@ARTICLE{WengPeroux2024,
       author = {{Weng}, Simon and {P{\'e}roux}, C{\'e}line and {Ramesh}, Rahul and {Nelson}, Dylan and {Sadler}, Elaine M. and {Zwaan}, Martin and {Bollo}, Victoria and {Casavecchia}, Benedetta},
        title = "{The physical origins of gas in the circumgalactic medium using observationally motivated TNG50 mocks}",
      journal = {\mnras},
         year = 2024,
        month = jan,
       volume = {527},
       number = {2},
        pages = {3494-3516},
          doi = {10.1093/mnras/stad3426},
archivePrefix = {arXiv},
       eprint = {2310.18310},
 primaryClass = {astro-ph.GA},
       adsurl = {https://ui.adsabs.harvard.edu/abs/2024MNRAS.527.3494W}
}

@ARTICLE{BLF2020,
       author = {{Benitez-Llambay}, Alejandro and {Frenk}, Carlos},
        title = "{The detailed structure and the onset of galaxy formation in low-mass gaseous dark matter haloes}",
      journal = {\mnras},
         year = 2020,
        month = nov,
       volume = {498},
       number = {4},
        pages = {4887-4900},
          doi = {10.1093/mnras/staa2698},
archivePrefix = {arXiv},
       eprint = {2004.06124},
 primaryClass = {astro-ph.GA},
       adsurl = {https://ui.adsabs.harvard.edu/abs/2020MNRAS.498.4887B}
}

@ARTICLE{HahnAbel2011,
       author = {{Hahn}, Oliver and {Abel}, Tom},
        title = "{Multi-scale initial conditions for cosmological simulations}",
      journal = {\mnras},
         year = 2011,
        month = aug,
       volume = {415},
       number = {3},
        pages = {2101-2121},
          doi = {10.1111/j.1365-2966.2011.18820.x},
       adsurl = {https://ui.adsabs.harvard.edu/abs/2011MNRAS.415.2101H}
}

@ARTICLE{Springel2005,
       author = {{Springel}, Volker},
        title = "{The cosmological simulation code GADGET-2}",
      journal = {\mnras},
         year = 2005,
        month = dec,
       volume = {364},
       number = {4},
        pages = {1105-1134},
          doi = {10.1111/j.1365-2966.2005.09655.x},
       adsurl = {https://ui.adsabs.harvard.edu/abs/2005MNRAS.364.1105S}
}

@INPROCEEDINGS{HaardtMadau2001,
       author = {{Haardt}, F. and {Madau}, P.},
        title = "{Modelling the UV/X-ray cosmic background with CUBA}",
    booktitle = {Clusters of Galaxies and the High Redshift Universe Observed in X-rays},
         year = 2001,
       editor = {{Neumann}, D.~M. and {Tran}, J.~T.~V.},
        month = jan,
          eid = {64},
        pages = {64},
       adsurl = {https://ui.adsabs.harvard.edu/abs/2001cghr.confE..64H}
}

@ARTICLE{Planck2014,
       author = {{Planck Collaboration} and {Ade}, P.~A.~R. and {Aghanim}, N. and {Alves}, M.~I.~R. and {Armitage-Caplan}, C. and {Arnaud}, M. and {Ashdown}, M. and {Atrio-Barandela}, F. and {Aumont}, J. and {Aussel}, H. and {Baccigalupi}, C. and {Banday}, A.~J. and {Barreiro}, R.~B. and {Barrena}, R. and {Bartelmann}, M. and {Bartlett}, J.~G. and {Bartolo}, N. and {Basak}, S. and {Battaner}, E. and {Battye}, R. and {Benabed}, K. and {Beno{\^\i}t}, A. and {Benoit-L{\'e}vy}, A. and {Bernard}, J. -P. and {Bersanelli}, M. and {Bertincourt}, B. and {Bethermin}, M. and {Bielewicz}, P. and {Bikmaev}, I. and {Blanchard}, A. and {Bobin}, J. and {Bock}, J.~J. and {B{\"o}hringer}, H. and {Bonaldi}, A. and {Bonavera}, L. and {Bond}, J.~R. and {Borrill}, J. and {Bouchet}, F.~R. and {Boulanger}, F. and {Bourdin}, H. and {Bowyer}, J.~W. and {Bridges}, M. and {Brown}, M.~L. and {Bucher}, M. and {Burenin}, R. and {Burigana}, C. and {Butler}, R.~C. and {Calabrese}, E. and {Cappellini}, B. and {Cardoso}, J. -F. and {Carr}, R. and {Carvalho}, P. and {Casale}, M. and {Castex}, G. and {Catalano}, A. and {Challinor}, A. and {Chamballu}, A. and {Chary}, R. -R. and {Chen}, X. and {Chiang}, H.~C. and {Chiang}, L. -Y. and {Chon}, G. and {Christensen}, P.~R. and {Churazov}, E. and {Church}, S. and {Clemens}, M. and {Clements}, D.~L. and {Colombi}, S. and {Colombo}, L.~P.~L. and {Combet}, C. and {Comis}, B. and {Couchot}, F. and {Coulais}, A. and {Crill}, B.~P. and {Cruz}, M. and {Curto}, A. and {Cuttaia}, F. and {Da Silva}, A. and {Dahle}, H. and {Danese}, L. and {Davies}, R.~D. and {Davis}, R.~J. and {de Bernardis}, P. and {de Rosa}, A. and {de Zotti}, G. and {D{\'e}chelette}, T. and {Delabrouille}, J. and {Delouis}, J. -M. and {D{\'e}mocl{\`e}s}, J. and {D{\'e}sert}, F. -X. and {Dick}, J. and {Dickinson}, C. and {Diego}, J.~M. and {Dolag}, K. and {Dole}, H. and {Donzelli}, S. and {Dor{\'e}}, O. and {Douspis}, M. and {Ducout}, A. and {Dunkley}, J. and {Dupac}, X. and {Efstathiou}, G. and {Elsner}, F. and {En{\ss}lin}, T.~A. and {Eriksen}, H.~K. and {Fabre}, O. and {Falgarone}, E. and {Falvella}, M.~C. and {Fantaye}, Y. and {Fergusson}, J. and {Filliard}, C. and {Finelli}, F. and {Flores-Cacho}, I. and {Foley}, S. and {Forni}, O. and {Fosalba}, P. and {Frailis}, M. and {Fraisse}, A.~A. and {Franceschi}, E. and {Freschi}, M. and {Fromenteau}, S. and {Frommert}, M. and {Gaier}, T.~C. and {Galeotta}, S. and {Gallegos}, J. and {Galli}, S. and {Gandolfo}, B. and {Ganga}, K. and {Gauthier}, C. and {G{\'e}nova-Santos}, R.~T. and {Ghosh}, T. and {Giard}, M. and {Giardino}, G. and {Gilfanov}, M. and {Girard}, D. and {Giraud-H{\'e}raud}, Y. and {Gjerl{\o}w}, E. and {Gonz{\'a}lez-Nuevo}, J. and {G{\'o}rski}, K.~M. and {Gratton}, S. and {Gregorio}, A. and {Gruppuso}, A. and {Gudmundsson}, J.~E. and {Haissinski}, J. and {Hamann}, J. and {Hansen}, F.~K. and {Hansen}, M. and {Hanson}, D. and {Harrison}, D.~L. and {Heavens}, A. and {Helou}, G. and {Hempel}, A. and {Henrot-Versill{\'e}}, S. and {Hern{\'a}ndez-Monteagudo}, C. and {Herranz}, D. and {Hildebrandt}, S.~R. and {Hivon}, E. and {Ho}, S. and {Hobson}, M. and {Holmes}, W.~A. and {Hornstrup}, A. and {Hou}, Z. and {Hovest}, W. and {Huey}, G. and {Huffenberger}, K.~M. and {Hurier}, G. and {Ili{\'c}}, S. and {Jaffe}, A.~H. and {Jaffe}, T.~R. and {Jasche}, J. and {Jewell}, J. and {Jones}, W.~C. and {Juvela}, M. and {Kalberla}, P. and {Kangaslahti}, P. and {Keih{\"a}nen}, E. and {Kerp}, J. and {Keskitalo}, R. and {Khamitov}, I. and {Kiiveri}, K. and {Kim}, J. and {Kisner}, T.~S. and {Kneissl}, R. and {Knoche}, J. and {Knox}, L. and {Kunz}, M. and {Kurki-Suonio}, H. and {Lacasa}, F. and {Lagache}, G. and {L{\"a}hteenm{\"a}ki}, A. and {Lamarre}, J. -M. and {Langer}, M. and {Lasenby}, A. and {Lattanzi}, M. and {Laureijs}, R.~J. and {Lavabre}, A. and {Lawrence}, C.~R. and {Le Jeune}, M. and {Leach}, S. and {Leahy}, J.~P. and {Leonardi}, R. and {Le{\'o}n-Tavares}, J. and {Leroy}, C. and {Lesgourgues}, J. and {Lewis}, A. and {Li}, C. and {Liddle}, A. and {Liguori}, M. and {Lilje}, P.~B. and {Linden-V{\o}rnle}, M. and {Lindholm}, V. and {L{\'o}pez-Caniego}, M. and {Lowe}, S. and {Lubin}, P.~M. and {Mac{\'\i}as-P{\'e}rez}, J.~F. and {MacTavish}, C.~J. and {Maffei}, B. and {Maggio}, G. and {Maino}, D. and {Mandolesi}, N. and {Mangilli}, A. and {Marcos-Caballero}, A. and {Marinucci}, D. and {Maris}, M. and {Marleau}, F. and {Marshall}, D.~J. and {Martin}, P.~G. and {Mart{\'\i}nez-Gonz{\'a}lez}, E. and {Masi}, S. and {Massardi}, M. and {Matarrese}, S. and {Matsumura}, T. and {Matthai}, F. and {Maurin}, L. and {Mazzotta}, P. and {McDonald}, A. and {McEwen}, J.~D. and {McGehee}, P. and {Mei}, S. and {Meinhold}, P.~R. and {Melchiorri}, A. and {Melin}, J. -B. and {Mendes}, L. and {Menegoni}, E. and {Mennella}, A. and {Migliaccio}, M. and {Mikkelsen}, K. and {Millea}, M. and {Miniscalco}, R. and {Mitra}, S. and {Miville-Desch{\^e}nes}, M. -A. and {Molinari}, D. and {Moneti}, A. and {Montier}, L. and {Morgante}, G. and {Morisset}, N. and {Mortlock}, D. and {Moss}, A. and {Munshi}, D. and {Murphy}, J.~A. and {Naselsky}, P. and {Nati}, F. and {Natoli}, P. and {Negrello}, M. and {Nesvadba}, N.~P.~H. and {Netterfield}, C.~B. and {N{\o}rgaard-Nielsen}, H.~U. and {North}, C. and {Noviello}, F. and {Novikov}, D. and {Novikov}, I. and {O'Dwyer}, I.~J. and {Orieux}, F. and {Osborne}, S. and {O'Sullivan}, C. and {Oxborrow}, C.~A. and {Paci}, F. and {Pagano}, L. and {Pajot}, F. and {Paladini}, R. and {Pandolfi}, S. and {Paoletti}, D. and {Partridge}, B. and {Pasian}, F. and {Patanchon}, G. and {Paykari}, P. and {Pearson}, D. and {Pearson}, T.~J. and {Peel}, M. and {Peiris}, H.~V. and {Perdereau}, O. and {Perotto}, L. and {Perrotta}, F. and {Pettorino}, V. and {Piacentini}, F. and {Piat}, M. and {Pierpaoli}, E. and {Pietrobon}, D. and {Plaszczynski}, S. and {Platania}, P. and {Pogosyan}, D. and {Pointecouteau}, E. and {Polenta}, G. and {Ponthieu}, N. and {Popa}, L. and {Poutanen}, T. and {Pratt}, G.~W. and {Pr{\'e}zeau}, G. and {Prunet}, S. and {Puget}, J. -L. and {Pullen}, A.~R. and {Rachen}, J.~P. and {Racine}, B. and {Rahlin}, A. and {R{\"a}th}, C. and {Reach}, W.~T. and {Rebolo}, R. and {Reinecke}, M. and {Remazeilles}, M. and {Renault}, C. and {Renzi}, A. and {Riazuelo}, A. and {Ricciardi}, S. and {Riller}, T. and {Ringeval}, C. and {Ristorcelli}, I. and {Robbers}, G. and {Rocha}, G. and {Roman}, M. and {Rosset}, C. and {Rossetti}, M. and {Roudier}, G. and {Rowan-Robinson}, M. and {Rubi{\~n}o-Mart{\'\i}n}, J.~A. and {Ruiz-Granados}, B. and {Rusholme}, B. and {Salerno}, E. and {Sandri}, M. and {Sanselme}, L. and {Santos}, D. and {Savelainen}, M. and {Savini}, G. and {Schaefer}, B.~M. and {Schiavon}, F. and {Scott}, D. and {Seiffert}, M.~D. and {Serra}, P. and {Shellard}, E.~P.~S. and {Smith}, K. and {Smoot}, G.~F. and {Souradeep}, T. and {Spencer}, L.~D. and {Starck}, J. -L. and {Stolyarov}, V. and {Stompor}, R. and {Sudiwala}, R. and {Sunyaev}, R. and {Sureau}, F. and {Sutter}, P. and {Sutton}, D. and {Suur-Uski}, A. -S. and {Sygnet}, J. -F. and {Tauber}, J.~A. and {Tavagnacco}, D. and {Taylor}, D. and {Terenzi}, L. and {Texier}, D. and {Toffolatti}, L. and {Tomasi}, M. and {Torre}, J. -P. and {Tristram}, M. and {Tucci}, M. and {Tuovinen}, J. and {T{\"u}rler}, M. and {Tuttlebee}, M. and {Umana}, G. and {Valenziano}, L. and {Valiviita}, J. and {Van Tent}, B. and {Varis}, J. and {Vibert}, L. and {Viel}, M. and {Vielva}, P. and {Villa}, F. and {Vittorio}, N. and {Wade}, L.~A. and {Wandelt}, B.~D. and {Watson}, C. and {Watson}, R. and {Wehus}, I.~K. and {Welikala}, N. and {Weller}, J. and {White}, M. and {White}, S.~D.~M. and {Wilkinson}, A. and {Winkel}, B. and {Xia}, J. -Q. and {Yvon}, D. and {Zacchei}, A. and {Zibin}, J.~P. and {Zonca}, A.},
        title = "{Planck 2013 results. I. Overview of products and scientific results}",
      journal = {\aap},
         year = 2014,
        month = nov,
       volume = {571},
          eid = {A1},
        pages = {A1},
          doi = {10.1051/0004-6361/201321529},
       adsurl = {https://ui.adsabs.harvard.edu/abs/2014A&A...571A...1P}
}

@ARTICLE{Wiersma2009,
       author = {{Wiersma}, Robert P.~C. and {Schaye}, Joop and {Smith}, Britton D.},
        title = "{The effect of photoionization on the cooling rates of enriched, astrophysical plasmas}",
      journal = {\mnras},
         year = 2009,
        month = feb,
       volume = {393},
       number = {1},
        pages = {99-107},
          doi = {10.1111/j.1365-2966.2008.14191.x},
       adsurl = {https://ui.adsabs.harvard.edu/abs/2009MNRAS.393...99W}
}

@ARTICLE{Han2018,
       author = {{Han}, Jiaxin and {Cole}, Shaun and {Frenk}, Carlos S. and {Benitez-Llambay}, Alejandro and {Helly}, John},
        title = "{HBT+: an improved code for finding subhaloes and building merger trees in cosmological simulations}",
      journal = {\mnras},
         year = 2018,
        month = feb,
       volume = {474},
       number = {1},
        pages = {604-617},
          doi = {10.1093/mnras/stx2792},
       adsurl = {https://ui.adsabs.harvard.edu/abs/2018MNRAS.474..604H}
}

@ARTICLE{ManamiRoy2024,
       author = {{Roy}, Manami and {Su}, Kung-Yi and {Tonnesen}, Stephanie and {Fielding}, Drummond B. and {Faucher-Gigu{\`e}re}, Claude-Andr{\'e}},
        title = "{Seeding the CGM: how satellites populate the cold phase of milky way haloes}",
      journal = {\mnras},
         year = 2024,
        month = jan,
       volume = {527},
       number = {1},
        pages = {265-280},
          doi = {10.1093/mnras/stad3142},
archivePrefix = {arXiv},
       eprint = {2310.04404},
 primaryClass = {astro-ph.GA},
       adsurl = {https://ui.adsabs.harvard.edu/abs/2024MNRAS.527..265R}
}

@ARTICLE{TridentMethodsPaper,
author = {{Hummels}, Cameron B. and {Smith}, Britton D. and {Silvia}, Devin W.},
title = "{Trident: A Universal Tool for Generating Synthetic Absorption Spectra from Astrophysical Simulations}",
journal = {\apj},
year = 2017,
month = sep,
volume = {847},
number = {1},
eid = {59},
pages = {59},
doi = {10.3847/1538-4357/aa7e2d},
archivePrefix = {arXiv},
eprint = {1612.03935},
primaryClass = {astro-ph.IM},
adsurl = {https://ui.adsabs.harvard.edu/abs/2017ApJ...847...59H}
}

@ARTICLE{Ehlert2013MetalsM86,
       author = {{Ehlert}, S. and {Werner}, N. and {Simionescu}, A. and {Allen}, S.~W. and {Kenney}, J.~D.~P. and {Million}, E.~T. and {Finoguenov}, A.},
        title = "{Ripping apart at the seams: the network of stripped gas surrounding M86}",
      journal = {\mnras},
         year = 2013,
        month = apr,
       volume = {430},
       number = {3},
        pages = {2401-2410},
          doi = {10.1093/mnras/stt060},
archivePrefix = {arXiv},
       eprint = {1212.3612},
 primaryClass = {astro-ph.CO},
       adsurl = {https://ui.adsabs.harvard.edu/abs/2013MNRAS.430.2401E}
}

@ARTICLE{Gastaldello2021,
       author = {{Gastaldello}, Fabio and {Simionescu}, Aurora and {Mernier}, Francois and {Biffi}, Veronica and {Gaspari}, Massimo and {Sato}, Kosuke and {Matsushita}, Kyoko},
        title = "{The Metal Content of the Hot Atmospheres of Galaxy Groups}",
      journal = {Universe},
         year = 2021,
        month = jun,
       volume = {7},
       number = {7},
          eid = {208},
        pages = {208},
          doi = {10.3390/universe7070208},
archivePrefix = {arXiv},
       eprint = {2106.13258},
 primaryClass = {astro-ph.CO},
       adsurl = {https://ui.adsabs.harvard.edu/abs/2021Univ....7..208G}
}

@ARTICLE{DiCesare2024MetalsAndMergers,
       author = {{Di Cesare}, C. and {Ginolfi}, M. and {Graziani}, L. and {Schneider}, R. and {Romano}, M. and {Popping}, G.},
        title = "{Carbon envelopes around merging galaxies at z {\ensuremath{\sim}} 4.5}",
      journal = {\aap},
         year = 2024,
        month = oct,
       volume = {690},
          eid = {A255},
        pages = {A255},
          doi = {10.1051/0004-6361/202449164},
archivePrefix = {arXiv},
       eprint = {2401.03020},
 primaryClass = {astro-ph.GA},
       adsurl = {https://ui.adsabs.harvard.edu/abs/2024A&A...690A.255D}
}

@ARTICLE{Rahmati2016EagleAGNsAndIons,
       author = {{Rahmati}, Alireza and {Schaye}, Joop and {Crain}, Robert A. and {Oppenheimer}, Benjamin D. and {Schaller}, Matthieu and {Theuns}, Tom},
        title = "{Cosmic distribution of highly ionized metals and their physical conditions in the EAGLE simulations}",
      journal = {\mnras},
         year = 2016,
        month = jun,
       volume = {459},
       number = {1},
        pages = {310-332},
          doi = {10.1093/mnras/stw453},
archivePrefix = {arXiv},
       eprint = {1511.01094},
 primaryClass = {astro-ph.GA},
       adsurl = {https://ui.adsabs.harvard.edu/abs/2016MNRAS.459..310R}
}

@ARTICLE{Segers2017FluctuatingAGNs,
       author = {{Segers}, Marijke C. and {Oppenheimer}, Benjamin D. and {Schaye}, Joop and {Richings}, Alexander J.},
        title = "{Metals in the circumgalactic medium are out of ionization equilibrium due to fluctuating active galactic nuclei}",
      journal = {\mnras},
         year = 2017,
        month = oct,
       volume = {471},
       number = {1},
        pages = {1026-1044},
          doi = {10.1093/mnras/stx1633},
archivePrefix = {arXiv},
       eprint = {1704.05470},
 primaryClass = {astro-ph.GA},
       adsurl = {https://ui.adsabs.harvard.edu/abs/2017MNRAS.471.1026S}
}

@ARTICLE{Man2019AGNandEnvironmentSDSS,
       author = {{Man}, Zhong-yi and {Peng}, Ying-jie and {Kong}, Xu and {Guo}, Ke-xin and {Zhang}, Cheng-peng and {Dou}, Jing},
        title = "{The dependence of AGN activity on environment in SDSS}",
      journal = {\mnras},
         year = 2019,
        month = sep,
       volume = {488},
       number = {1},
        pages = {89-98},
          doi = {10.1093/mnras/stz1706},
archivePrefix = {arXiv},
       eprint = {1907.01563},
 primaryClass = {astro-ph.GA},
       adsurl = {https://ui.adsabs.harvard.edu/abs/2019MNRAS.488...89M}
}

@ARTICLE{Hani2019DiversityOfCGMauriga,
       author = {{Hani}, Maan H. and {Ellison}, Sara L. and {Sparre}, Martin and {Grand}, Robert J.~J. and {Pakmor}, R{\"u}ediger and {Gomez}, Facundo A. and {Springel}, Volker},
        title = "{The diversity of the circumgalactic medium around z = 0 Milky Way-mass galaxies from the Auriga simulations}",
      journal = {\mnras},
         year = 2019,
        month = sep,
       volume = {488},
       number = {1},
        pages = {135-152},
          doi = {10.1093/mnras/stz1708},
archivePrefix = {arXiv},
       eprint = {1907.04336},
 primaryClass = {astro-ph.GA},
       adsurl = {https://ui.adsabs.harvard.edu/abs/2019MNRAS.488..135H}
}

@ARTICLE{RuszkowskiPfrommer2023CosmicRayReview,
       author = {{Ruszkowski}, Mateusz and {Pfrommer}, Christoph},
        title = "{Cosmic ray feedback in galaxies and galaxy clusters}",
      journal = {\aapr},
         year = 2023,
        month = dec,
       volume = {31},
       number = {1},
          eid = {4},
        pages = {4},
          doi = {10.1007/s00159-023-00149-2},
archivePrefix = {arXiv},
       eprint = {2306.03141},
 primaryClass = {astro-ph.HE},
       adsurl = {https://ui.adsabs.harvard.edu/abs/2023A&ARv..31....4R}
}

@ARTICLE{DeFilippis2024CGMandCosmicRays,
       author = {{DeFelippis}, Daniel and {Bournaud}, Fr{\'e}d{\'e}ric and {Bouch{\'e}}, Nicolas and {Tollet}, Edouard and {Farcy}, Marion and {Rey}, Maxime and {Rosdahl}, Joakim and {Blaizot}, J{\'e}r{\'e}my},
        title = "{The effect of cosmic rays on the observational properties of the CGM}",
      journal = {\mnras},
         year = 2024,
        month = may,
       volume = {530},
       number = {1},
        pages = {52-65},
          doi = {10.1093/mnras/stae837},
archivePrefix = {arXiv},
       eprint = {2403.14748},
 primaryClass = {astro-ph.GA},
       adsurl = {https://ui.adsabs.harvard.edu/abs/2024MNRAS.530...52D}
}

@ARTICLE{Salem2016CosmicRaysCGM,
       author = {{Salem}, Munier and {Bryan}, Greg L. and {Corlies}, Lauren},
        title = "{Role of cosmic rays in the circumgalactic medium}",
      journal = {\mnras},
         year = 2016,
        month = feb,
       volume = {456},
       number = {1},
        pages = {582-601},
          doi = {10.1093/mnras/stv2641},
archivePrefix = {arXiv},
       eprint = {1511.05144},
 primaryClass = {astro-ph.GA},
       adsurl = {https://ui.adsabs.harvard.edu/abs/2016MNRAS.456..582S}
}

@ARTICLE{Li2011SNeRates,
       author = {{Li}, Weidong and {Chornock}, Ryan and {Leaman}, Jesse and {Filippenko}, Alexei V. and {Poznanski}, Dovi and {Wang}, Xiaofeng and {Ganeshalingam}, Mohan and {Mannucci}, Filippo},
        title = "{Nearby supernova rates from the Lick Observatory Supernova Search - III. The rate-size relation, and the rates as a function of galaxy Hubble type and colour}",
      journal = {\mnras},
         year = 2011,
        month = apr,
       volume = {412},
       number = {3},
        pages = {1473-1507},
          doi = {10.1111/j.1365-2966.2011.18162.x},
archivePrefix = {arXiv},
       eprint = {1006.4613},
 primaryClass = {astro-ph.SR},
       adsurl = {https://ui.adsabs.harvard.edu/abs/2011MNRAS.412.1473L}
}

@ARTICLE{Ma2025SNeRates,
       author = {{Ma}, Xiaoran and {Wang}, Xiaofeng and {Mo}, Jun and {Andrew Howell}, D. and {Pellegrino}, Craig and {Zhang}, Jujia and {Wu}, Chengyuan and {Yan}, Shengyu and {Liu}, Dongdong and {Arcavi}, Iair and {Chen}, Zhihao and {Farah}, Joseph and {Padilla Gonzalez}, Estefania and {Guo}, Fangzhou and {Hiramatsu}, Daichi and {Li}, Gaici and {Lin}, Han and {Liu}, Jialian and {McCully}, Curtis and {Newsome}, Megan and {Sai}, Hanna and {Terreran}, Giacomo and {Xiang}, Danfeng and {Zhang}, Xinhan},
        title = "{Supernovae at distances <40 Mpc: II. Supernova rate in the local Universe}",
      journal = {\aap},
         year = 2025,
        month = jun,
       volume = {698},
          eid = {A306},
        pages = {A306},
          doi = {10.1051/0004-6361/202452685},
archivePrefix = {arXiv},
       eprint = {2504.04507},
 primaryClass = {astro-ph.HE},
       adsurl = {https://ui.adsabs.harvard.edu/abs/2025A&A...698A.306M}
}

@ARTICLE{VanDeVoort2021MagneticFieldsAndCGM,
       author = {{van de Voort}, Freeke and {Bieri}, Rebekka and {Pakmor}, R{\"u}diger and {G{\'o}mez}, Facundo A. and {Grand}, Robert J.~J. and {Marinacci}, Federico},
        title = "{The effect of magnetic fields on properties of the circumgalactic medium}",
      journal = {\mnras},
         year = 2021,
        month = mar,
       volume = {501},
       number = {4},
        pages = {4888-4902},
          doi = {10.1093/mnras/staa3938},
archivePrefix = {arXiv},
       eprint = {2008.07537},
 primaryClass = {astro-ph.GA},
       adsurl = {https://ui.adsabs.harvard.edu/abs/2021MNRAS.501.4888V}
}

@ARTICLE{Buie2022MagneticFieldsAndCGM,
       author = {{Buie}, Edward and {Scannapieco}, Evan and {Mark Voit}, G.},
        title = "{Modeling Photoionized Turbulent Material in the Circumgalactic Medium. III. Effects of Corotation and Magnetic Fields}",
      journal = {\apj},
         year = 2022,
        month = mar,
       volume = {927},
       number = {1},
          eid = {30},
        pages = {30},
          doi = {10.3847/1538-4357/ac4bc2},
archivePrefix = {arXiv},
       eprint = {2201.09325},
 primaryClass = {astro-ph.GA},
       adsurl = {https://ui.adsabs.harvard.edu/abs/2022ApJ...927...30B}
}

@ARTICLE{Marinacci2015LargeScalePropertiesMagneticFields,
       author = {{Marinacci}, Federico and {Vogelsberger}, Mark and {Mocz}, Philip and {Pakmor}, R{\"u}diger},
        title = "{The large-scale properties of simulated cosmological magnetic fields}",
      journal = {\mnras},
         year = 2015,
        month = nov,
       volume = {453},
       number = {4},
        pages = {3999-4019},
          doi = {10.1093/mnras/stv1692},
archivePrefix = {arXiv},
       eprint = {1506.00005},
 primaryClass = {astro-ph.CO},
       adsurl = {https://ui.adsabs.harvard.edu/abs/2015MNRAS.453.3999M}
}

@ARTICLE{Marinacci2018FirstResultsTNGmagneticfields,
       author = {{Marinacci}, Federico and {Vogelsberger}, Mark and {Pakmor}, R{\"u}diger and {Torrey}, Paul and {Springel}, Volker and {Hernquist}, Lars and {Nelson}, Dylan and {Weinberger}, Rainer and {Pillepich}, Annalisa and {Naiman}, Jill and {Genel}, Shy},
        title = "{First results from the IllustrisTNG simulations: radio haloes and magnetic fields}",
      journal = {\mnras},
         year = 2018,
        month = nov,
       volume = {480},
       number = {4},
        pages = {5113-5139},
          doi = {10.1093/mnras/sty2206},
archivePrefix = {arXiv},
       eprint = {1707.03396},
 primaryClass = {astro-ph.CO},
       adsurl = {https://ui.adsabs.harvard.edu/abs/2018MNRAS.480.5113M}
}

@ARTICLE{McDonough2025MagneticFieldsInSatellites,
       author = {{McDonough}, Bryanne and {Poulin}, Alexander},
        title = "{Magnetic Fields of Satellite Galaxies Stronger than Comparable Centrals in TNG100}",
      journal = {Research Notes of the American Astronomical Society},
         year = 2025,
        month = apr,
       volume = {9},
       number = {4},
          eid = {93},
        pages = {93},
          doi = {10.3847/2515-5172/adce74},
archivePrefix = {arXiv},
       eprint = {2504.07895},
 primaryClass = {astro-ph.GA},
       adsurl = {https://ui.adsabs.harvard.edu/abs/2025RNAAS...9...93M}
}

@ARTICLE{Werhahn2025MagneticFieldsInSatellites,
       author = {{Werhahn}, Maria and {Pakmor}, R{\"u}diger and {Bieri}, Rebekka and {van de Voort}, Freeke and {Talbot}, Rosie Y. and {Springel}, Volker},
        title = "{Environment matters: stronger magnetic fields in satellite galaxies}",
      journal = {\mnras},
         year = 2025,
        month = jul,
       volume = {540},
       number = {4},
        pages = {3431-3440},
          doi = {10.1093/mnras/staf873},
archivePrefix = {arXiv},
       eprint = {2409.17229},
 primaryClass = {astro-ph.GA},
       adsurl = {https://ui.adsabs.harvard.edu/abs/2025MNRAS.540.3431W}
}

@ARTICLE{Rintoul2025MagneticFieldsAndRamPressure,
       author = {{Rintoul}, Thomas A. and {van de Voort}, Freeke and {Hannington}, Andrew T. and {Pakmor}, R{\"u}diger and {Bieri}, Rebekka and {Werhahn}, Maria and {Talbot}, Rosie Y.},
        title = "{The role of magnetic fields in ram pressure stripping of satellite galaxies in the circumgalactic medium around massive galaxies}",
      journal = {\mnras},
         year = 2025,
        month = nov,
       volume = {543},
       number = {4},
        pages = {4321-4334},
          doi = {10.1093/mnras/staf1718},
archivePrefix = {arXiv},
       eprint = {2506.18983},
 primaryClass = {astro-ph.GA},
       adsurl = {https://ui.adsabs.harvard.edu/abs/2025MNRAS.543.4321R}
}

@ARTICLE{Shreeram2025CGMprojectionXray,
       author = {{Shreeram}, Soumya and {Comparat}, Johan and {Merloni}, Andrea and {Zhang}, Yi and {Ponti}, Gabriele and {Nandra}, Kirpal and {ZuHone}, John and {Marini}, Ilaria and {Vladutescu-Zopp}, Stephan and {Popesso}, Paola and {Pakmor}, Ruediger and {Seppi}, Riccardo and {Peroux}, Celine and {Sorini}, Daniele},
        title = "{Quantifying observational projection effects with a simulation-based hot CGM model}",
      journal = {\aap},
         year = 2025,
        month = may,
       volume = {697},
          eid = {A22},
        pages = {A22},
          doi = {10.1051/0004-6361/202452271},
archivePrefix = {arXiv},
       eprint = {2409.10397},
 primaryClass = {astro-ph.GA},
       adsurl = {https://ui.adsabs.harvard.edu/abs/2025A&A...697A..22S}
}

@ARTICLE{Nelson2018DistributionOfHighlyIonizedOxygenInTNG,
       author = {{Nelson}, Dylan and {Kauffmann}, Guinevere and {Pillepich}, Annalisa and {Genel}, Shy and {Springel}, Volker and {Pakmor}, R{\"u}diger and {Hernquist}, Lars and {Weinberger}, Rainer and {Torrey}, Paul and {Vogelsberger}, Mark and {Marinacci}, Federico},
        title = "{The abundance, distribution, and physical nature of highly ionized oxygen O VI, O VII, and O VIII in IllustrisTNG}",
      journal = {\mnras},
         year = 2018,
        month = jun,
       volume = {477},
       number = {1},
        pages = {450-479},
          doi = {10.1093/mnras/sty656},
archivePrefix = {arXiv},
       eprint = {1712.00016},
 primaryClass = {astro-ph.GA},
       adsurl = {https://ui.adsabs.harvard.edu/abs/2018MNRAS.477..450N}
}

@ARTICLE{Nelson2021ColdCircumgalacticMedium,
       author = {{Nelson}, Dylan and {Byrohl}, Chris and {Peroux}, Celine and {Rubin}, Kate H.~R. and {Burchett}, Joseph N.},
        title = "{The cold circumgalactic medium in emission: Mg II haloes in TNG50}",
      journal = {\mnras},
         year = 2021,
        month = nov,
       volume = {507},
       number = {3},
        pages = {4445-4463},
          doi = {10.1093/mnras/stab2177},
archivePrefix = {arXiv},
       eprint = {2106.09023},
 primaryClass = {astro-ph.GA},
       adsurl = {https://ui.adsabs.harvard.edu/abs/2021MNRAS.507.4445N}
}

@INPROCEEDINGS{ChenZahedy2026CGMreview,
       author = {{Chen}, Hsiao-Wen and {Zahedy}, Fakhri S.},
        title = "{The circumgalactic medium}",
    booktitle = {Encyclopedia of Astrophysics, Volume 4},
         year = 2026,
       volume = {4},
        month = jan,
        pages = {370-400},
          doi = {10.1016/B978-0-443-21439-4.00059-6},
archivePrefix = {arXiv},
       eprint = {2412.10579},
 primaryClass = {astro-ph.GA},
       adsurl = {https://ui.adsabs.harvard.edu/abs/2026enap....4..370C}
}

@ARTICLE{Fumagalli2024CGMreview,
       author = {{Fumagalli}, Michele},
        title = "{The multiphase circumgalactic medium and its relation to galaxies: an observational perspective}",
      journal = {arXiv e-prints},
         year = 2024,
        month = aug,
          eid = {arXiv:2409.00174},
        pages = {arXiv:2409.00174},
          doi = {10.48550/arXiv.2409.00174},
archivePrefix = {arXiv},
       eprint = {2409.00174},
 primaryClass = {astro-ph.GA},
       adsurl = {https://ui.adsabs.harvard.edu/abs/2024arXiv240900174F}
}

@ARTICLE{Dressler1980GalaxyMorphology,
       author = {{Dressler}, A.},
        title = "{Galaxy morphology in rich clusters: implications for the formation and evolution of galaxies.}",
      journal = {\apj},
         year = 1980,
        month = mar,
       volume = {236},
        pages = {351-365},
          doi = {10.1086/157753},
       adsurl = {https://ui.adsabs.harvard.edu/abs/1980ApJ...236..351D}
}

@ARTICLE{GunnAndGott1972,
       author = {{Gunn}, James E. and {Gott}, III, J. Richard},
        title = "{On the Infall of Matter Into Clusters of Galaxies and Some Effects on Their Evolution}",
      journal = {\apj},
         year = 1972,
        month = aug,
       volume = {176},
        pages = {1},
          doi = {10.1086/151605},
       adsurl = {https://ui.adsabs.harvard.edu/abs/1972ApJ...176....1G}
}

@ARTICLE{Nielsen2018ApJ,
       author = {{Nielsen}, Nikole M. and {Kacprzak}, Glenn G. and {Pointon}, Stephanie K. and {Churchill}, Christopher W. and {Murphy}, Michael T.},
        title = "{MAGIICAT VI. The Mg II Intragroup Medium Is Kinematically Complex}",
      journal = {\apj},
         year = 2018,
        month = dec,
       volume = {869},
       number = {2},
          eid = {153},
        pages = {153},
          doi = {10.3847/1538-4357/aaedbd},
archivePrefix = {arXiv},
       eprint = {1808.09562},
 primaryClass = {astro-ph.GA},
       adsurl = {https://ui.adsabs.harvard.edu/abs/2018ApJ...869..153N}
}

@ARTICLE{Muzahid2021MUSEQuBES,
       author = {{Muzahid}, Sowgat and {Schaye}, Joop and {Cantalupo}, Sebastiano and {Marino}, Raffaella Anna and {Bouch{\'e}}, Nicolas F. and {Johnson}, Sean and {Maseda}, Michael and {Wendt}, Martin and {Wisotzki}, Lutz and {Zabl}, Johannes},
        title = "{MUSEQuBES: characterizing the circumgalactic medium of redshift {\ensuremath{\approx}}3.3 Ly {\ensuremath{\alpha}} emitters}",
      journal = {\mnras},
         year = 2021,
        month = dec,
       volume = {508},
       number = {4},
        pages = {5612-5637},
          doi = {10.1093/mnras/stab2933},
archivePrefix = {arXiv},
       eprint = {2105.05260},
 primaryClass = {astro-ph.GA},
       adsurl = {https://ui.adsabs.harvard.edu/abs/2021MNRAS.508.5612M}
}

@ARTICLE{Banerjee2025MUSEQuBES,
           author = {{Banerjee}, Eshita and {Muzahid}, Sowgat and {Schaye}, Joop and {Blaizot}, J{\'e}r{\'e}my and {Bouch{\'e}}, Nicolas and {Cantalupo}, Sebastiano and {Johnson}, Sean D. and {Matthee}, Jorryt and {Verhamme}, Anne},
            title = "{MUSEQuBES: Connecting H I Absorption with Ly{\ensuremath{\alpha}} Emitters at z {\ensuremath{\approx}} 3.3}",
          journal = {\apj},
             year = 2025,
            month = feb,
           volume = {980},
           number = {2},
              eid = {171},
            pages = {171},
              doi = {10.3847/1538-4357/ada7e9},
    archivePrefix = {arXiv},
           eprint = {2411.11959},
     primaryClass = {astro-ph.GA},
           adsurl = {https://ui.adsabs.harvard.edu/abs/2025ApJ...980..171B}
    }

@ARTICLE{Banerjee2023MUSEQuBESCIV,
           author = {{Banerjee}, Eshita and {Muzahid}, Sowgat and {Schaye}, Joop and {Johnson}, Sean D. and {Cantalupo}, Sebastiano},
            title = "{MUSEQuBES: the relation between Ly {\ensuremath{\alpha}} emitters and C IV absorbers at z {\ensuremath{\approx}} 3.3}",
          journal = {\mnras},
             year = 2023,
            month = oct,
           volume = {524},
           number = {4},
            pages = {5148-5165},
              doi = {10.1093/mnras/stad2022},
    archivePrefix = {arXiv},
           eprint = {2304.04788},
     primaryClass = {astro-ph.GA},
           adsurl = {https://ui.adsabs.harvard.edu/abs/2023MNRAS.524.5148B}
    }

@ARTICLE{Qu2023CUBS,
       author = {{Qu}, Zhijie and {Chen}, Hsiao-Wen and {Rudie}, Gwen C. and {Johnson}, Sean D. and {Zahedy}, Fakhri S. and {DePalma}, David and {Boettcher}, Erin and {Cantalupo}, Sebastiano and {Chen}, Mandy C. and {Cooksey}, Kathy L. and {Faucher-Gigu{\`e}re}, Claude-Andr{\'e} and {Li}, Jennifer I.-Hsiu and {Lopez}, Sebastian and {Schaye}, Joop and {Simcoe}, Robert A.},
        title = "{The Cosmic Ultraviolet Baryon Survey (CUBS) - VI. Connecting physical properties of the cool circumgalactic medium to galaxies at z {\ensuremath{\approx}} 1}",
      journal = {\mnras},
         year = 2023,
        month = sep,
       volume = {524},
       number = {1},
        pages = {512-528},
          doi = {10.1093/mnras/stad1886},
archivePrefix = {arXiv},
       eprint = {2306.11274},
 primaryClass = {astro-ph.GA},
       adsurl = {https://ui.adsabs.harvard.edu/abs/2023MNRAS.524..512Q}
}

@ARTICLE{FaucherGiguere2011,
       author = {{Faucher-Gigu{\`e}re}, Claude-Andr{\'e} and {Kere{\v{s}}}, Du{\v{s}}an},
        title = "{The small covering factor of cold accretion streams}",
      journal = {\mnras},
         year = 2011,
        month = mar,
       volume = {412},
       number = {1},
        pages = {L118-L122},
          doi = {10.1111/j.1745-3933.2011.01018.x},
archivePrefix = {arXiv},
       eprint = {1011.1693},
 primaryClass = {astro-ph.CO},
       adsurl = {https://ui.adsabs.harvard.edu/abs/2011MNRAS.412L.118F}
}

@ARTICLE{VanDeVoort2012,
       author = {{van de Voort}, Freeke and {Schaye}, Joop and {Altay}, Gabriel and {Theuns}, Tom},
        title = "{Cold accretion flows and the nature of high column density H I absorption at redshift 3}",
      journal = {\mnras},
         year = 2012,
        month = apr,
       volume = {421},
       number = {4},
        pages = {2809-2819},
          doi = {10.1111/j.1365-2966.2012.20487.x},
archivePrefix = {arXiv},
       eprint = {1109.5700},
 primaryClass = {astro-ph.CO},
       adsurl = {https://ui.adsabs.harvard.edu/abs/2012MNRAS.421.2809V}
}

@ARTICLE{Hummels2019CGMandResolution,
       author = {{Hummels}, Cameron B. and {Smith}, Britton D. and {Hopkins}, Philip F. and {O'Shea}, Brian W. and {Silvia}, Devin W. and {Werk}, Jessica K. and {Lehner}, Nicolas and {Wise}, John H. and {Collins}, David C. and {Butsky}, Iryna S.},
        title = "{The Impact of Enhanced Halo Resolution on the Simulated Circumgalactic Medium}",
      journal = {\apj},
         year = 2019,
        month = sep,
       volume = {882},
       number = {2},
          eid = {156},
        pages = {156},
          doi = {10.3847/1538-4357/ab378f},
archivePrefix = {arXiv},
       eprint = {1811.12410},
 primaryClass = {astro-ph.GA},
       adsurl = {https://ui.adsabs.harvard.edu/abs/2019ApJ...882..156H}
}

@ARTICLE{VanDeVoort2019CGMHighResolution,
       author = {{van de Voort}, Freeke and {Springel}, Volker and {Mandelker}, Nir and {van den Bosch}, Frank C. and {Pakmor}, R{\"u}diger},
        title = "{Cosmological simulations of the circumgalactic medium with 1 kpc resolution: enhanced H I column densities}",
      journal = {\mnras},
         year = 2019,
        month = jan,
       volume = {482},
       number = {1},
        pages = {L85-L89},
          doi = {10.1093/mnrasl/sly190},
archivePrefix = {arXiv},
       eprint = {1808.04369},
 primaryClass = {astro-ph.GA},
       adsurl = {https://ui.adsabs.harvard.edu/abs/2019MNRAS.482L..85V}
}

@ARTICLE{Peeples2019Foggie,
       author = {{Peeples}, Molly S. and {Corlies}, Lauren and {Tumlinson}, Jason and {O'Shea}, Brian W. and {Lehner}, Nicolas and {O'Meara}, John M. and {Howk}, J. Christopher and {Earl}, Nicholas and {Smith}, Britton D. and {Wise}, John H. and {Hummels}, Cameron B.},
        title = "{Figuring Out Gas \& Galaxies in Enzo (FOGGIE). I. Resolving Simulated Circumgalactic Absorption at 2 {\ensuremath{\leq}} z {\ensuremath{\leq}} 2.5}",
      journal = {\apj},
         year = 2019,
        month = mar,
       volume = {873},
       number = {2},
          eid = {129},
        pages = {129},
          doi = {10.3847/1538-4357/ab0654},
archivePrefix = {arXiv},
       eprint = {1810.06566},
 primaryClass = {astro-ph.GA},
       adsurl = {https://ui.adsabs.harvard.edu/abs/2019ApJ...873..129P}
}

@ARTICLE{Ramesh2024GIBLESmallScaleGas,
       author = {{Ramesh}, Rahul and {Nelson}, Dylan},
        title = "{Zooming in on the circumgalactic medium with GIBLE: Resolving small-scale gas structure in cosmological simulations}",
      journal = {\mnras},
         year = 2024,
        month = feb,
       volume = {528},
       number = {2},
        pages = {3320-3339},
          doi = {10.1093/mnras/stae237},
archivePrefix = {arXiv},
       eprint = {2307.11143},
 primaryClass = {astro-ph.GA},
       adsurl = {https://ui.adsabs.harvard.edu/abs/2024MNRAS.528.3320R}
}

@ARTICLE{Ramesh2024GIBLEMagneticFields,
       author = {{Ramesh}, Rahul and {Nelson}, Dylan and {Fielding}, Drummond and {Br{\"u}ggen}, Marcus},
        title = "{Zooming in on the circumgalactic medium with GIBLE. The topology and draping of magnetic fields around cold clouds}",
      journal = {\aap},
         year = 2024,
        month = apr,
       volume = {684},
          eid = {L16},
        pages = {L16},
          doi = {10.1051/0004-6361/202348786},
archivePrefix = {arXiv},
       eprint = {2404.01370},
 primaryClass = {astro-ph.GA},
       adsurl = {https://ui.adsabs.harvard.edu/abs/2024A&A...684L..16R}
}

@ARTICLE{Lofthouse2020MAGG_I,
       author = {{Lofthouse}, Emma K. and {Fumagalli}, Michele and {Fossati}, Matteo and {O'Meara}, John M. and {Murphy}, Michael T. and {Christensen}, Lise and {Prochaska}, J. Xavier and {Cantalupo}, Sebastiano and {Bielby}, Richard M. and {Cooke}, Ryan J. and {Lusso}, Elisabeta and {Morris}, Simon L.},
        title = "{MUSE Analysis of Gas around Galaxies (MAGG) - I: Survey design and the environment of a near pristine gas cloud at z {\ensuremath{\approx}} 3.5}",
      journal = {\mnras},
         year = 2020,
        month = jan,
       volume = {491},
       number = {2},
        pages = {2057-2074},
          doi = {10.1093/mnras/stz3066},
archivePrefix = {arXiv},
       eprint = {1910.13458},
 primaryClass = {astro-ph.GA},
       adsurl = {https://ui.adsabs.harvard.edu/abs/2020MNRAS.491.2057L}
}

@ARTICLE{Knobel2009LinkingLength,
       author = {{Knobel}, C. and {Lilly}, S.~J. and {Iovino}, A. and {Porciani}, C. and {Kova{\v{c}}}, K. and {Cucciati}, O. and {Finoguenov}, A. and {Kitzbichler}, M.~G. and {Carollo}, C.~M. and {Contini}, T. and {Kneib}, J.-P. and {Le F{\`e}vre}, O. and {Mainieri}, V. and {Renzini}, A. and {Scodeggio}, M. and {Zamorani}, G. and {Bardelli}, S. and {Bolzonella}, M. and {Bongiorno}, A. and {Caputi}, K. and {Coppa}, G. and {de la Torre}, S. and {de Ravel}, L. and {Franzetti}, P. and {Garilli}, B. and {Kampczyk}, P. and {Lamareille}, F. and {Le Borgne}, J.-F. and {Le Brun}, V. and {Maier}, C. and {Mignoli}, M. and {Pello}, R. and {Peng}, Y. and {Perez Montero}, E. and {Ricciardelli}, E. and {Silverman}, J.~D. and {Tanaka}, M. and {Tasca}, L. and {Tresse}, L. and {Vergani}, D. and {Zucca}, E. and {Abbas}, U. and {Bottini}, D. and {Cappi}, A. and {Cassata}, P. and {Cimatti}, A. and {Fumana}, M. and {Guzzo}, L. and {Koekemoer}, A.~M. and {Leauthaud}, A. and {Maccagni}, D. and {Marinoni}, C. and {McCracken}, H.~J. and {Memeo}, P. and {Meneux}, B. and {Oesch}, P. and {Pozzetti}, L. and {Scaramella}, R.},
        title = "{An Optical Group Catalog to z = 1 from the zCOSMOS 10 k Sample}",
      journal = {\apj},
         year = 2009,
        month = jun,
       volume = {697},
       number = {2},
        pages = {1842-1860},
          doi = {10.1088/0004-637X/697/2/1842},
archivePrefix = {arXiv},
       eprint = {0903.3411},
 primaryClass = {astro-ph.CO},
       adsurl = {https://ui.adsabs.harvard.edu/abs/2009ApJ...697.1842K}
}

@ARTICLE{Diener2013LinkingLength,
       author = {{Diener}, C. and {Lilly}, S.~J. and {Knobel}, C. and {Zamorani}, G. and {Lemson}, G. and {Kampczyk}, P. and {Scoville}, N. and {Carollo}, C.~M. and {Contini}, T. and {Kneib}, J.-P. and {Le Fevre}, O. and {Mainieri}, V. and {Renzini}, A. and {Scodeggio}, M. and {Bardelli}, S. and {Bolzonella}, M. and {Bongiorno}, A. and {Caputi}, K. and {Cucciati}, O. and {de la Torre}, S. and {de Ravel}, L. and {Franzetti}, P. and {Garilli}, B. and {Iovino}, A. and {Kova{\v{c}}}, K. and {Lamareille}, F. and {Le Borgne}, J.-F. and {Le Brun}, V. and {Maier}, C. and {Mignoli}, M. and {Pello}, R. and {Peng}, Y. and {Perez Montero}, E. and {Presotto}, V. and {Silverman}, J. and {Tanaka}, M. and {Tasca}, L. and {Tresse}, L. and {Vergani}, D. and {Zucca}, E. and {Bordoloi}, R. and {Cappi}, A. and {Cimatti}, A. and {Coppa}, G. and {Koekemoer}, A.~M. and {L{\'o}pez-Sanjuan}, C. and {McCracken}, H.~J. and {Moresco}, M. and {Nair}, P. and {Pozzetti}, L. and {Welikala}, N.},
        title = "{Proto-groups at 1.8 < z < 3 in the zCOSMOS-deep Sample}",
      journal = {\apj},
         year = 2013,
        month = mar,
       volume = {765},
       number = {2},
          eid = {109},
        pages = {109},
          doi = {10.1088/0004-637X/765/2/109},
archivePrefix = {arXiv},
       eprint = {1210.2723},
 primaryClass = {astro-ph.CO},
       adsurl = {https://ui.adsabs.harvard.edu/abs/2013ApJ...765..109D}
}

@ARTICLE{Asplund2009,
       author = {{Asplund}, Martin and {Grevesse}, Nicolas and {Sauval}, A. Jacques and {Scott}, Pat},
        title = "{The Chemical Composition of the Sun}",
      journal = {\araa},
         year = 2009,
        month = sep,
       volume = {47},
       number = {1},
        pages = {481-522},
          doi = {10.1146/annurev.astro.46.060407.145222},
archivePrefix = {arXiv},
       eprint = {0909.0948},
 primaryClass = {astro-ph.SR},
       adsurl = {https://ui.adsabs.harvard.edu/abs/2009ARA&A..47..481A}
}

@ARTICLE{Agertz2007,
       author = {{Agertz}, Oscar and {Moore}, Ben and {Stadel}, Joachim and {Potter}, Doug and {Miniati}, Francesco and {Read}, Justin and {Mayer}, Lucio and {Gawryszczak}, Artur and {Kravtsov}, Andrey and {Nordlund}, {\r{A}}ke and {Pearce}, Frazer and {Quilis}, Vicent and {Rudd}, Douglas and {Springel}, Volker and {Stone}, James and {Tasker}, Elizabeth and {Teyssier}, Romain and {Wadsley}, James and {Walder}, Rolf},
        title = "{Fundamental differences between SPH and grid methods}",
      journal = {\mnras},
         year = 2007,
        month = sep,
       volume = {380},
       number = {3},
        pages = {963-978},
          doi = {10.1111/j.1365-2966.2007.12183.x},
archivePrefix = {arXiv},
       eprint = {astro-ph/0610051},
 primaryClass = {astro-ph},
       adsurl = {https://ui.adsabs.harvard.edu/abs/2007MNRAS.380..963A}
}

@ARTICLE{Hani2018MergersAndCGM,
       author = {{Hani}, Maan H. and {Sparre}, Martin and {Ellison}, Sara L. and {Torrey}, Paul and {Vogelsberger}, Mark},
        title = "{Galaxy mergers moulding the circum-galactic medium - I. The impact of a major merger}",
      journal = {\mnras},
         year = 2018,
        month = mar,
       volume = {475},
       number = {1},
        pages = {1160-1176},
          doi = {10.1093/mnras/stx3252},
archivePrefix = {arXiv},
       eprint = {1801.06183},
 primaryClass = {astro-ph.GA},
       adsurl = {https://ui.adsabs.harvard.edu/abs/2018MNRAS.475.1160H}
}

@ARTICLE{Dutta2023MgIIemission,
       author = {{Dutta}, Rajeshwari and {Fossati}, Matteo and {Fumagalli}, Michele and {Revalski}, Mitchell and {Lofthouse}, Emma K. and {Nelson}, Dylan and {Papini}, Giulia and {Rafelski}, Marc and {Cantalupo}, Sebastiano and {Arrigoni Battaia}, Fabrizio and {Dayal}, Pratika and {Longobardi}, Alessia and {P{\'e}roux}, Celine and {Prichard}, Laura J. and {Prochaska}, J. Xavier},
        title = "{Metal line emission from galaxy haloes at z {\ensuremath{\approx}} 1}",
      journal = {\mnras},
         year = 2023,
        month = jun,
       volume = {522},
       number = {1},
        pages = {535-558},
          doi = {10.1093/mnras/stad1002},
archivePrefix = {arXiv},
       eprint = {2302.09087},
 primaryClass = {astro-ph.GA},
       adsurl = {https://ui.adsabs.harvard.edu/abs/2023MNRAS.522..535D}
}

@ARTICLE{Sims2026CAMELSEnvironment,
       author = {{Sims}, Xavier and {Angl{\'e}s-Alc{\'a}zar}, Daniel and {Oh}, Boon-Kiat and {Nagai}, Daisuke and {Mercedes-Feliz}, Jonathan and {Medlock}, Isabel and {Ni}, Yueying and {Lovell}, Christopher C. and {Villaescusa-Navarro}, Francisco},
        title = "{CAMELS Environments: The Impact of Local Neighbours on Galaxy Evolution across the SIMBA, IllustrisTNG, ASTRID, and Swift-EAGLE Simulations}",
      journal = {arXiv e-prints},
         year = 2026,
        month = jan,
          eid = {arXiv:2601.06290},
        pages = {arXiv:2601.06290},
          doi = {10.48550/arXiv.2601.06290},
archivePrefix = {arXiv},
       eprint = {2601.06290},
 primaryClass = {astro-ph.GA},
       adsurl = {https://ui.adsabs.harvard.edu/abs/2026arXiv260106290S}
}

@ARTICLE{Cherrey2024MNRAS.528..481C,
       author = {{Cherrey}, Maxime and {Bouch{\'e}}, Nicolas F. and {Zabl}, Johannes and {Schroetter}, Ilane and {Wendt}, Martin and {Langan}, Ivanna and {Richard}, Johan and {Schaye}, Joop and {Mercier}, Wilfried and {Epinat}, Beno{\^\i}t and {Contini}, Thierry},
        title = "{MusE GAs FLOw and Wind (MEGAFLOW) X. The cool gas and covering fraction of Mg II in galaxy groups}",
      journal = {\mnras},
         year = 2024,
        month = feb,
       volume = {528},
       number = {1},
        pages = {481-498},
          doi = {10.1093/mnras/stad3764},
archivePrefix = {arXiv},
       eprint = {2312.01762},
 primaryClass = {astro-ph.GA},
       adsurl = {https://ui.adsabs.harvard.edu/abs/2024MNRAS.528..481C}
}

@ARTICLE{Ferland2013Cloudy,
       author = {{Ferland}, G.~J. and {Porter}, R.~L. and {van Hoof}, P.~A.~M. and {Williams}, R.~J.~R. and {Abel}, N.~P. and {Lykins}, M.~L. and {Shaw}, G. and {Henney}, W.~J. and {Stancil}, P.~C.},
        title = "{The 2013 Release of Cloudy}",
      journal = {\rmxaa},
         year = 2013,
        month = apr,
       volume = {49},
        pages = {137-163},
          doi = {10.48550/arXiv.1302.4485},
archivePrefix = {arXiv},
       eprint = {1302.4485},
 primaryClass = {astro-ph.GA},
       adsurl = {https://ui.adsabs.harvard.edu/abs/2013RMxAA..49..137F}
}

% %%%%%%%%%%%%%%%%%%%%%%%%%%%%%%%%%%%%%%%%%%%%%%%%%%%%%%%%%%%%%%
% Example below of non-structurated natbib references  
% To use the v8.3 macros with this form of composition of bibliography,
% the option "bibyear" should be added to the command line
% "\documentclass[bibyear]{aa}".
% %%%%%%%%%%%%%%%%%%%%%%%%%%%%%%%%%%%%%%%%%%%%%%%%%%%%%%%%%%%%%%

%%%%%%%%%%%%%%%%%%%%%%%%%%%%%%%%%%%%%%%%%%%%%%%%%%%%%%%%%%%%%%%
% Appendices must be placed after   \end{thebibliography}
% They will be placed automatically on a new page.
%%%%%%%%%%%%%%%%%%%%%%%%%%%%%%%%%%%%%%%%%%%%%%%%%%%%%%%%%%%%%%%
\begin{appendix}
\nolinenumbers
\section{Results for $z=1$}
\label{Appendix:Z1Results}
\begin{figure*}
        \centering
        \includegraphics[width=\textwidth]{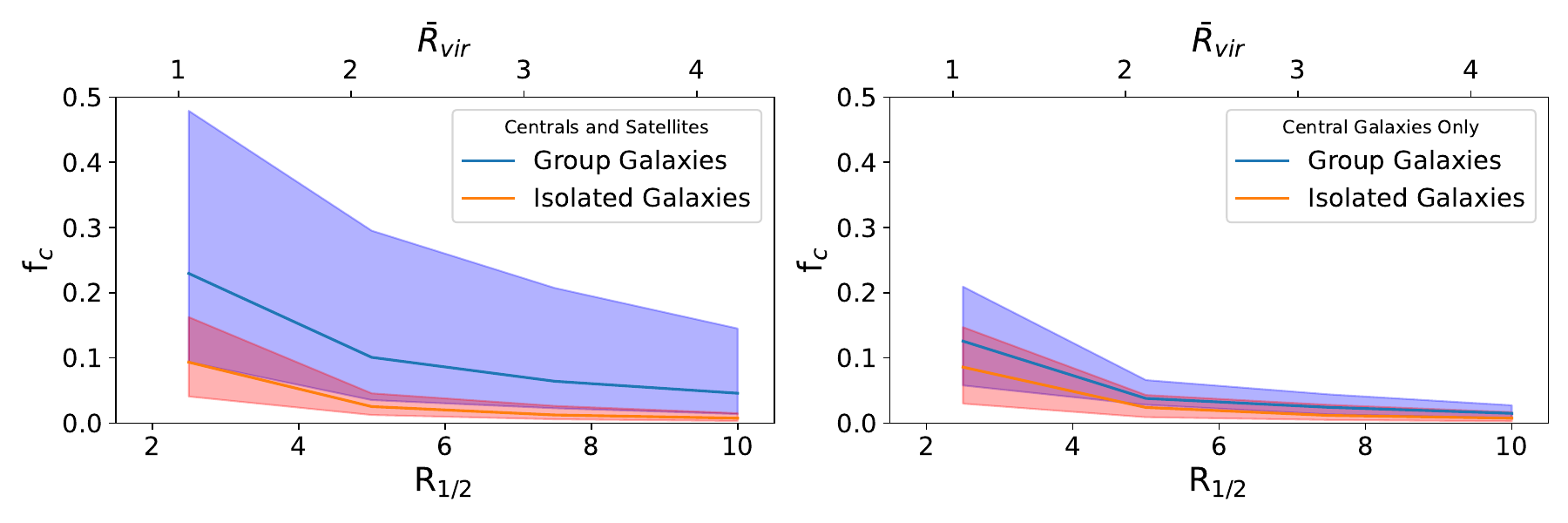}
        \caption{\mgii~at $z=1$: Same as Fig. \ref{Fig:ReproduceTrend} but for $z=1$.
        We obtain similar trends for \mgii~at $z=1$ as at $z=0$.
        }
        \label{Fig:ReproduceTrendMgII_Z1}
\end{figure*}

\begin{figure*}
        \centering
        \includegraphics[width=\textwidth]{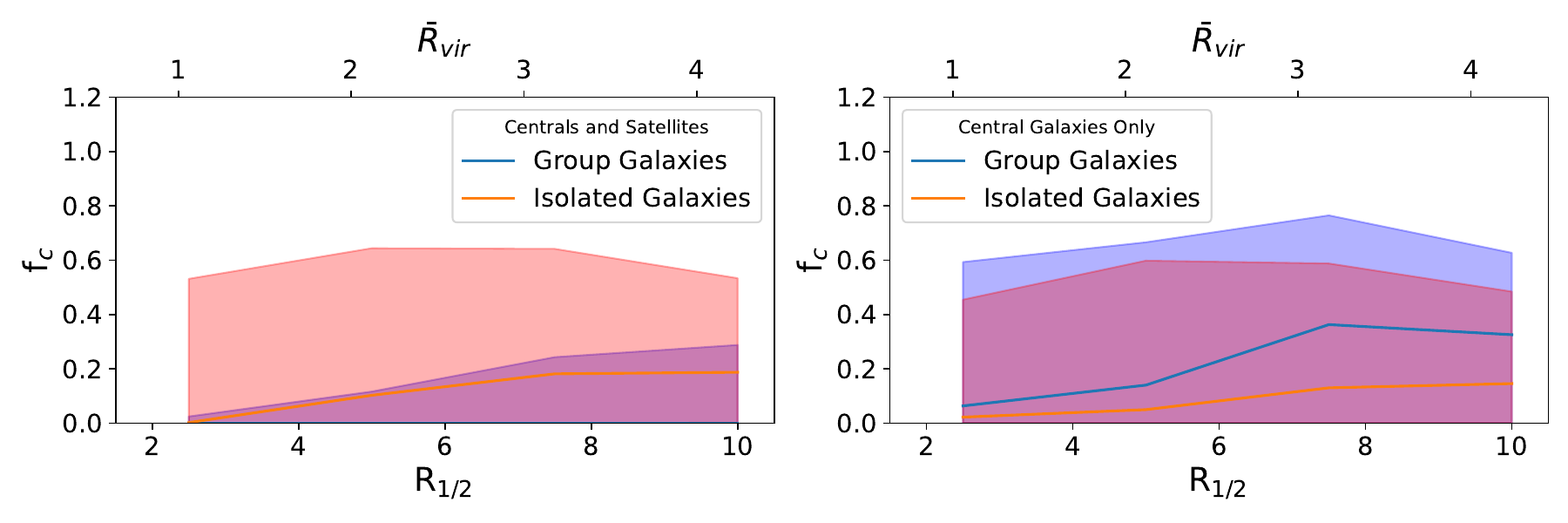}
        \caption{\civ~at $z=1$: Same as Fig. \ref{Fig:ReproduceTrendCIV} but for $z=1$ and a detection limit of $10^{17.5}$ cm$^{-2}$ instead of $10^{18.5}$ cm$^{-2}$.
        We changed the detection limit for this plot since at $z=1$ the detection limit where there is no detection is lower (see. Fig. \ref{Fig:6bTestZ1}).
        We obtain similar trends as at $z=0$.
        }
        \label{Fig:ReproduceTrendCIV_Z1}
\end{figure*}

\begin{figure*}
    \centering    \includegraphics[width=\textwidth]{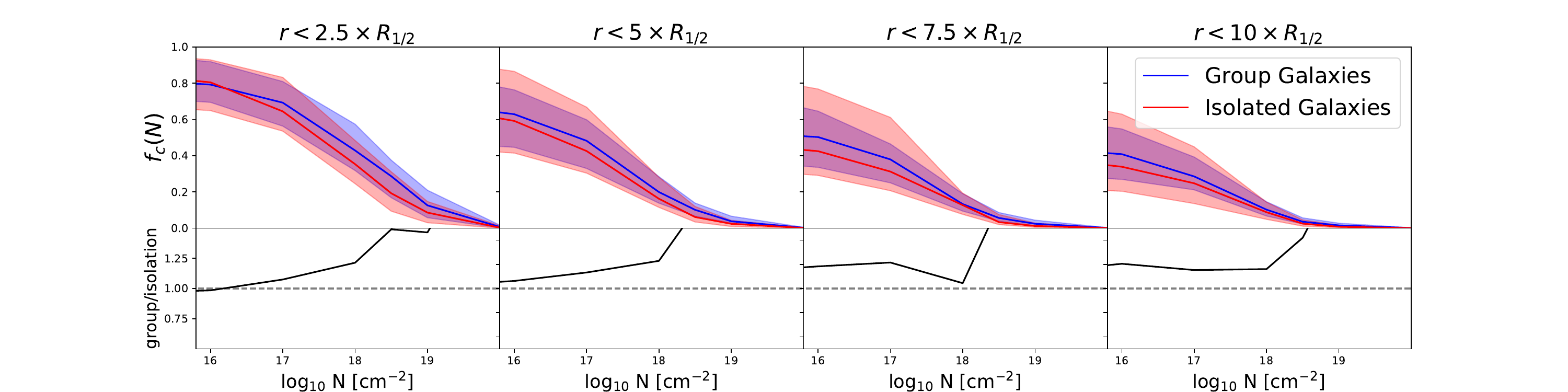}
    \caption{\mgii: Same as Fig. \ref{Fig:6a} but at $z=1$.
    }
    \label{Fig:6aTestZ1}
   \end{figure*}

\begin{figure*}
    
    \centering
    \includegraphics[width=\textwidth]{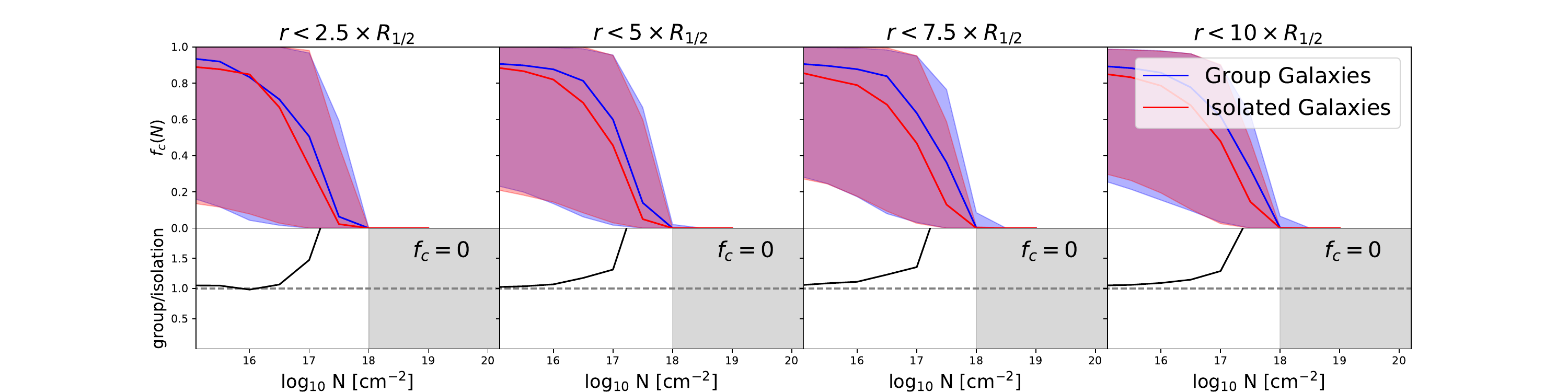}
    \caption{\civ: Same as in Fig. \ref{Fig:6b} but at $z=1$.
    Compared to the case for $z=0$, we reach a covering fraction of $f_c=0$ at a detection limit of $10^{18}$ cm$^{-2}$ instead of $10^{19}$ cm$^{-2}$.
    }
    \label{Fig:6bTestZ1}

\end{figure*}

Here we present the results for $z=1$, which are consistent with the results at $z=0$.
At $z=1$, we have a total of 288 galaxies, out of which 156 are group galaxies (48 centrals) and 132 are isolated galaxies (130 centrals).
The matched samples consist of 41 group and 41 isolated galaxies, if only central galaxies are considered.
If satellites are also included, the matched sample consists of 132 group and 132 isolated galaxies.
We show the results for \mgii~and \civ~with a fixed detection limit in Figs. \ref{Fig:ReproduceTrendMgII_Z1} and \ref{Fig:ReproduceTrendCIV_Z1}.
We show the results for the change of detection limit in Figs. \ref{Fig:6aTestZ1} and \ref{Fig:6bTestZ1}.

\section{Mass-cut in FoF algorithm}
\label{Sec:MassCutFoF}
\begin{figure}
        \centering
        \includegraphics[width=\columnwidth]{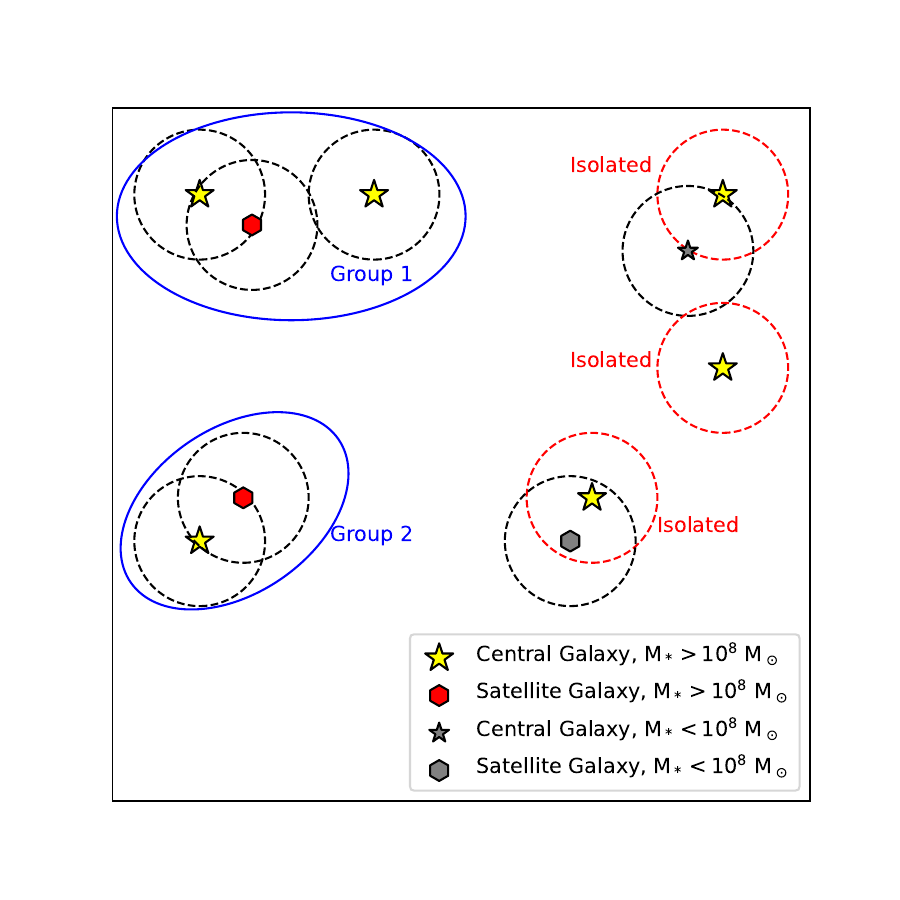}
        \caption{Illustration of the effect of mass-cut on our FoF algorithm.
        In our FoF algorithm we use a radial linking length of 500 kpc to categorize our galaxies as group galaxies or isolated galaxies.
        To be consistent with observed completeness limits, only galaxies with a stellar mass above $10^8$ M$_\odot$ are considered in the FoF algorithm, irrespective of whether they are centrals (yellow stars) or satellites (red diamonds).
        Galaxies with M$_*<10^8$ M$_\odot$ (grey stars and diamonds) are treated as if they do not exist (see text for more information).
        If there is at least one other galaxy with M$_*>10^8$ M$_\odot$ within 500 kpc, which in our illustration is marked by overlapping dashed circles, these galaxies form a group (blue ellipses).
        Otherwise, they are isolated galaxies (red circles).
        We want to stress that this definition of a group is different from the definition of a galaxy group that depends on mass.
        }
        \label{Fig:GroupAndIsolatedGalaxies}
\end{figure}

When splitting our sample into group and isolated galaxies, we disregard all galaxies with a stellar mass below $10^8$ M$_\odot$ as illustrated in Fig. \ref{Fig:GroupAndIsolatedGalaxies}.
Here we show that this mass-cut does not influence our result for the environmental dependence of the CGM.
For that we first rerun the FoF algorithm to identify group and isolated galaxies with three different configurations: i) we disregard galaxies with M$_{cut}<10^7$ M$_\odot$; ii) we disregard galaxies with less than 100 star particles which corresponds to M$_{cut}\lesssim 4.5\times 10^6$ M$_\odot$; iii) we disregard galaxies with M$_{cut}<10^6$ M$_\odot$. We show how the 289 galaxies with $M*\geq 10^8$ M$_\odot$ are distributed into group and isolated galaxies with these new mass cuts in Tab. \ref{tab:MassCuts}.

\begin{table}
\begin{center}
\begin{tabular}{||c|| c c c c c||} 
 \hline
 M$_{cut}$ & N$_{FoF}$ & N$_{g,tot}$ & N$_{i, tot}$ & N$_{g,cent}$ & N$_{i,cent}$ \\ [0.5ex] 
 \hline\hline
 10$^8$ M$_\odot$ & 289 & 158 & 131 & 36 & 127 \\ 
 \hline
 10$^7$ M$_\odot$ & 840 & 220 & 69 & 94 & 69 \\
 \hline
 4.5$\times$ 10$^6$ M$_\odot$ & 1148 & 231 & 58 & 105 & 58 \\
 \hline
 10$^6$ M$_\odot$ & 2015 & 253 & 36 & 127 & 36 \\
 \hline
\end{tabular}
\end{center}
\caption{Table showing the change in group and isolated galaxies for our sample of 289 galaxies with M$_*>10^8$ M$_\odot$ when using different mass cuts in the FoF algorithm. M$_{cut}$ is the mass-cut applied, i.e., all galaxies above this mass are considered in the FoF algorithm. N$_{FoF}$ is the number of galaxies used in the FoF algorithm for each mass-cut. N$_{g,tot}$ is the total number of group galaxies, N$_{i,tot}$ is the total number of isolated galaxies, N$_{g,cent}$ is the number of group central galaxies, and N$_{i,cent}$ is the number of isolated central galaxies.
N$_{g,tot}$ and N$_{i, tot}$ are calculated on the 289 galaxies with M$_*>10^8$ M$_\odot$, while N$_{g,cent}$, and N$_{i,cent}$ are calculated on the 163 central galaxies with M$_*>10^8$ M$_\odot$.}
\label{tab:MassCuts}
\end{table}

\begin{figure*}
        \centering
        \includegraphics[width=\textwidth]{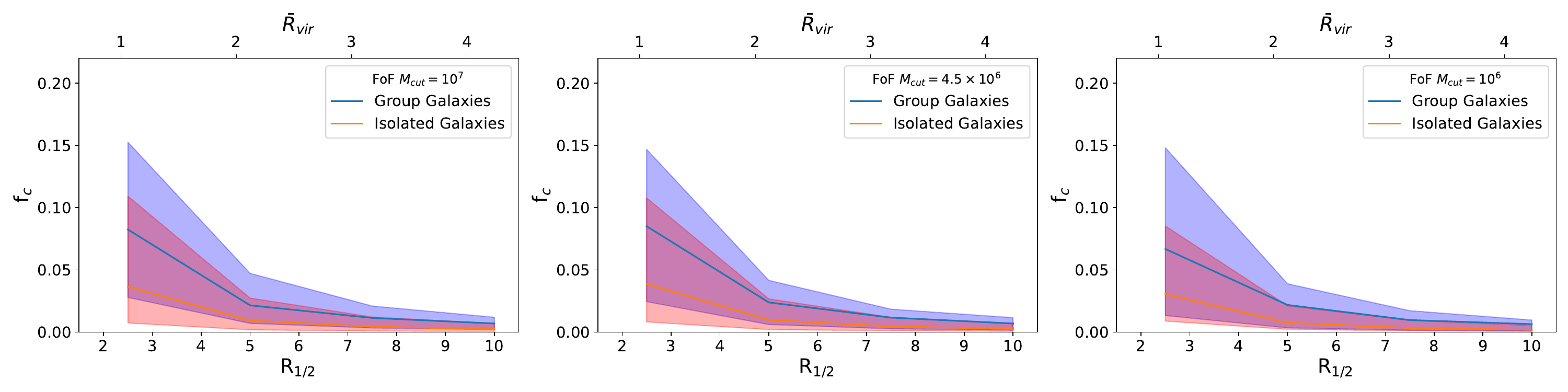}
        \caption{\mgii~at $z=0$ for three different mass-cuts (left panel: $10^7$ M$_\odot$, middle panel: $4.5\times 10^6$ M$_\odot$, right panel: $10^6$ M$_\odot$).
        Blue are group galaxies while red are isolated galaxies.
        The trend is the same as in the right panel of Fig. \ref{Fig:ReproduceTrend} with little variation between the different mass-cuts, i.e., leaving out galaxies with M$_*<10^8$ M$_\odot$ in the FoF algorithm does not influence our results.
        }
        \label{Fig:MgIITrueCatalogue}
\end{figure*}

\begin{figure*}
        \centering
        \includegraphics[width=\textwidth]{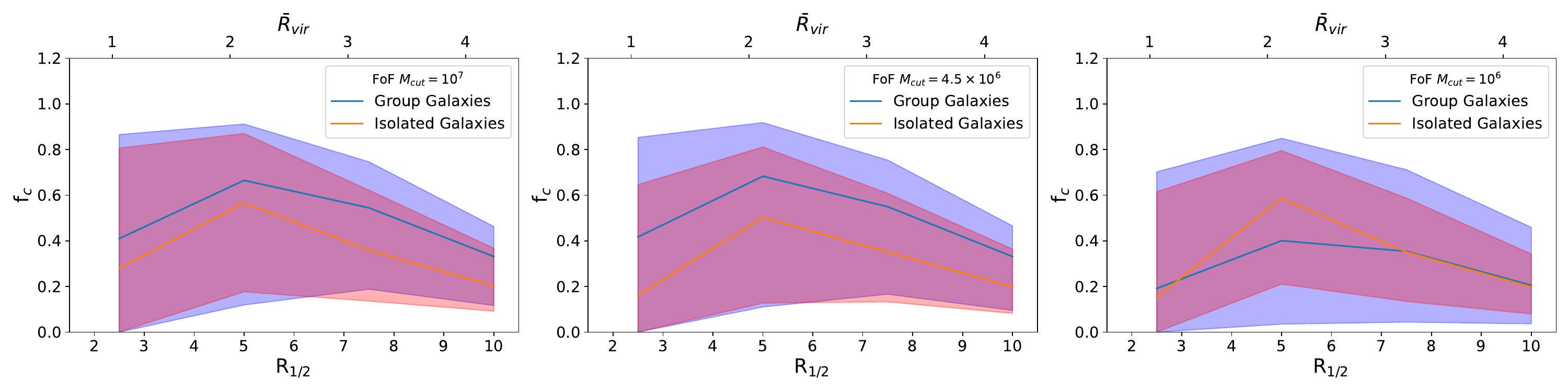}
        \caption{Same as Fig. \ref{Fig:MgIITrueCatalogue} but for \civ.
        The trend is the same as in the right panel of Fig. \ref{Fig:ReproduceTrendCIV}.
        }
        \label{Fig:CIVTrueCatalogue}
\end{figure*}

Including galaxies with lower and lower masses in the FoF algorithm leads to a relabeling of isolated galaxies to group galaxies, i.e., the lower the mass-cut the higher the number of group galaxies.
Thus, galaxies that remain isolated galaxies when going to lower mass-cuts are located in increasingly underdense regions.
By finding matched samples between group central galaxies and isolated central galaxies for each mass-cut and calculating the covering fractions, we can test how the change in mass-cuts influences the environmental dependence of the CGM.
We show the results for \mgii~in Fig. \ref{Fig:MgIITrueCatalogue} and the results for \civ~in Fig. \ref{Fig:CIVTrueCatalogue}.
Going to lower mass-cuts, i.e., shifting galaxies from isolated to group galaxies, does not change our result on the environmental dependence of the CGM.
This is not surprising as we showed in Secs. \ref{Sec:CoveringFractionsInTheCGM} and \ref{Sec:3DStructure} that the CGM of central galaxies is the same, no matter whether they are located in an overdense region or an underdense region, while the CGM of satellite galaxies is intrinsically different from the CGM of central galaxies.

\section{Selecting gas in velocity space}
\label{Sec:GasvLOS}

In our work we selected the gas around the galaxies in a sphere to avoid bias from picking up gas that does not belong to the CGM of the galaxy but nevertheless is within the velocity window along the line-of-sight.
Such additional gas from other galaxies and their CGM along the line-of-sight could contribute to the environmental dependence of the CGM, if it is picked up predominantely in group environments.
Thus, we tested whether selecting gas in velocity space instead of real space introduces an environmental dependence of the CGM.

\begin{figure*}
        \centering
        \includegraphics[width=\textwidth]{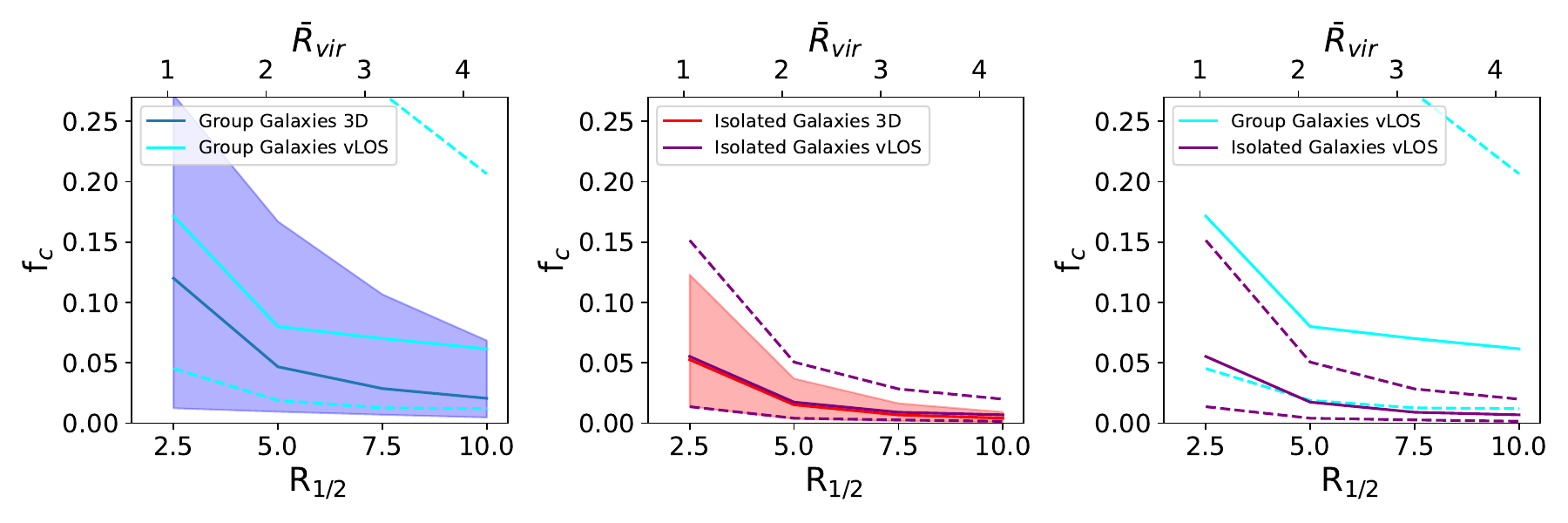}
        \caption{Covering fraction of \mgii~gas at $z=0$ for group and isolated galaxies with satellites included in the sample.
        We compare the difference in selecting gas in a 3D-sphere around the galaxy and selecting it in a window of 500 km\,s$^{-1}$ in velocity space (vLOS).
        In the left panel we show group galaxies with gas selected in 3D (blue) and compare them to the same sample with the gas selected in vLOS (cyan).
        Selecting gas in vLOS increases the covering fraction.
        In the middle panel we show isolated galaxies with gas selected in 3D (red) and gas selected in vLOS (purple).
        Also here, selecting gas in vLOS increases the covering fraction, although the effect is smaller than for group galaxies.
        In the right panel we compare the covering fractions from group (cyan) and isolated galaxies (purple) when the gas is selected in vLOS.
        We see that when satellites are included in the sample and gas is selected in vLOS, the trend for a higher covering fraction in groups becomes even more pronounced.
        }
        \label{Fig:MgIIvLOS_SatIncluded}
\end{figure*}

\begin{figure*}
        \centering
        \includegraphics[width=\textwidth]{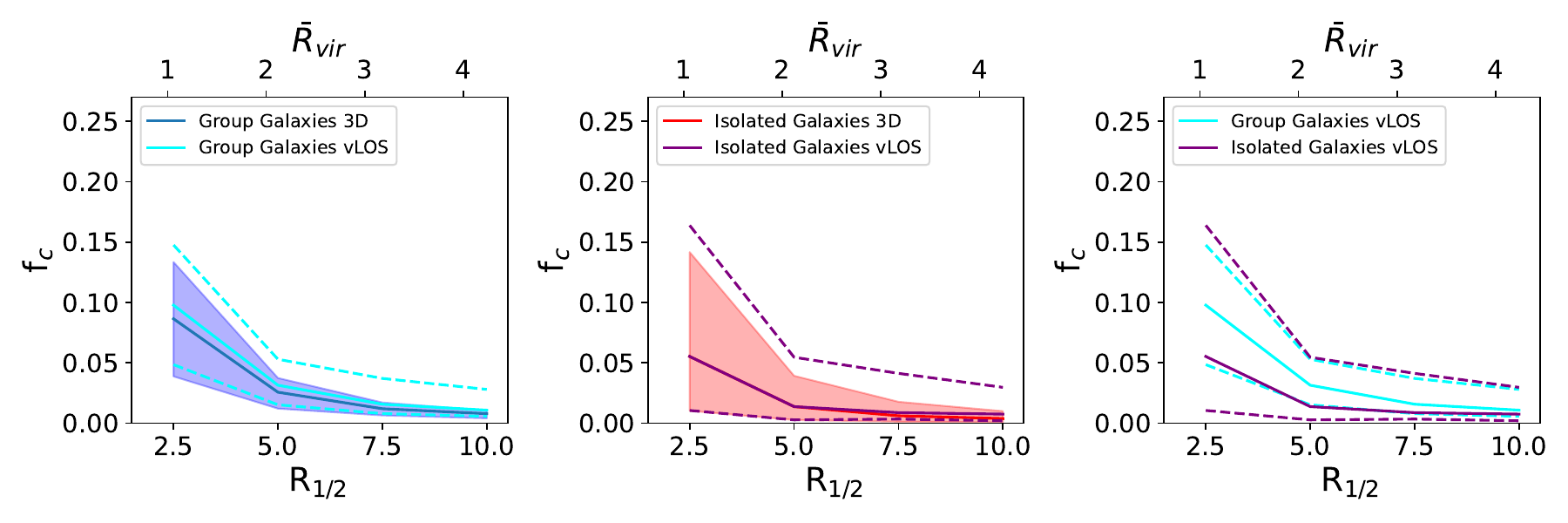}
        \caption{Same as in Fig. \ref{Fig:MgIIvLOS_SatIncluded}, but for a galaxy sample consisting of only central galaxies.
        We see that also in this case the covering fractions of group and isolated galaxies are increased when gas is selected in vLOS.
        However, selecting gas in vLOS does not introduce a difference between group and isolated galaxies when only central galaxies are concerned.
        }
        \label{Fig:MgIIvLOS_onlyCentrals}
\end{figure*}

\begin{figure*}
        \centering
       \includegraphics[width=\textwidth]{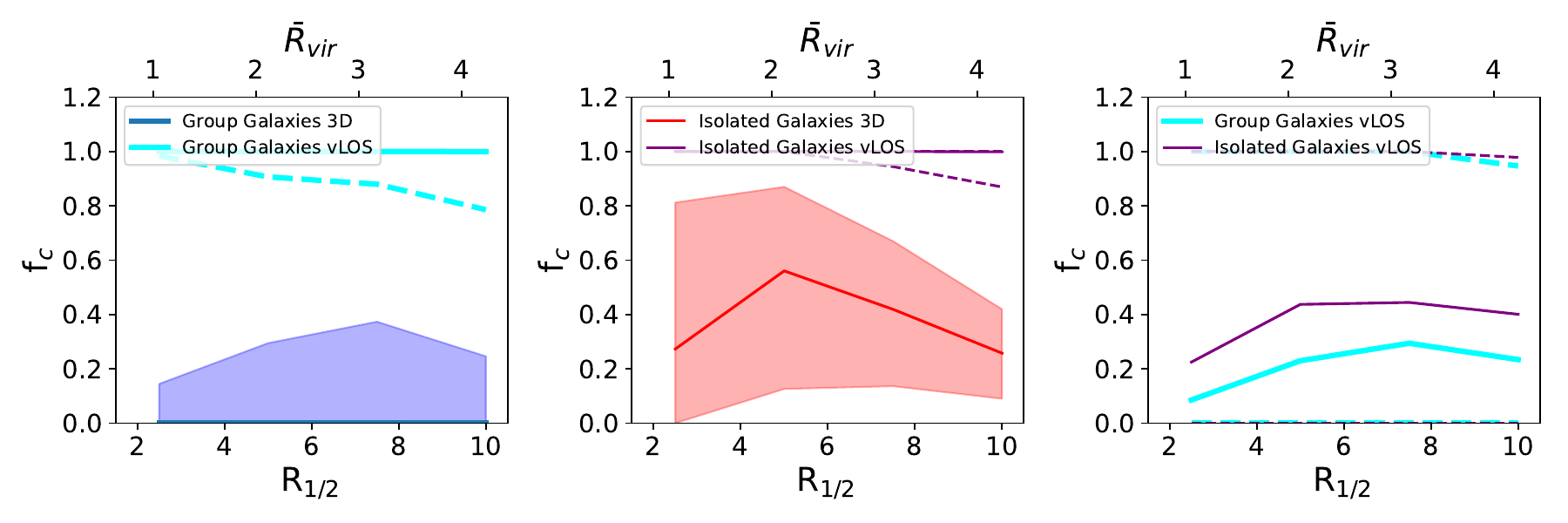}
        \caption{Same as Fig. \ref{Fig:MgIIvLOS_SatIncluded} but for \civ.
        Similar to \mgii, selecting gas in velocity space increases the covering fraction of \civ.
        For \civ, this effect is much stronger as it leads to a median covering fraction of 100\% for both group (left panel) and isolated galaxies (middel panel) when applying a detection limit of $N=10^{18.5}$ cm$^{-2}$.
        However, it does not lead to a difference in covering fractions between group and isolated galaxies as can be seen in the right panel where we compare group and isolated galaxies with gas selected in vLOS and a detection limit of $N=10^{19}$ cm$^{-2}$ to avoid having the covering fractions fully saturated.
        }
        \label{Fig:CIVvLOS_satIncluded}
\end{figure*}

\begin{figure*}
        \centering
       \includegraphics[width=\textwidth]{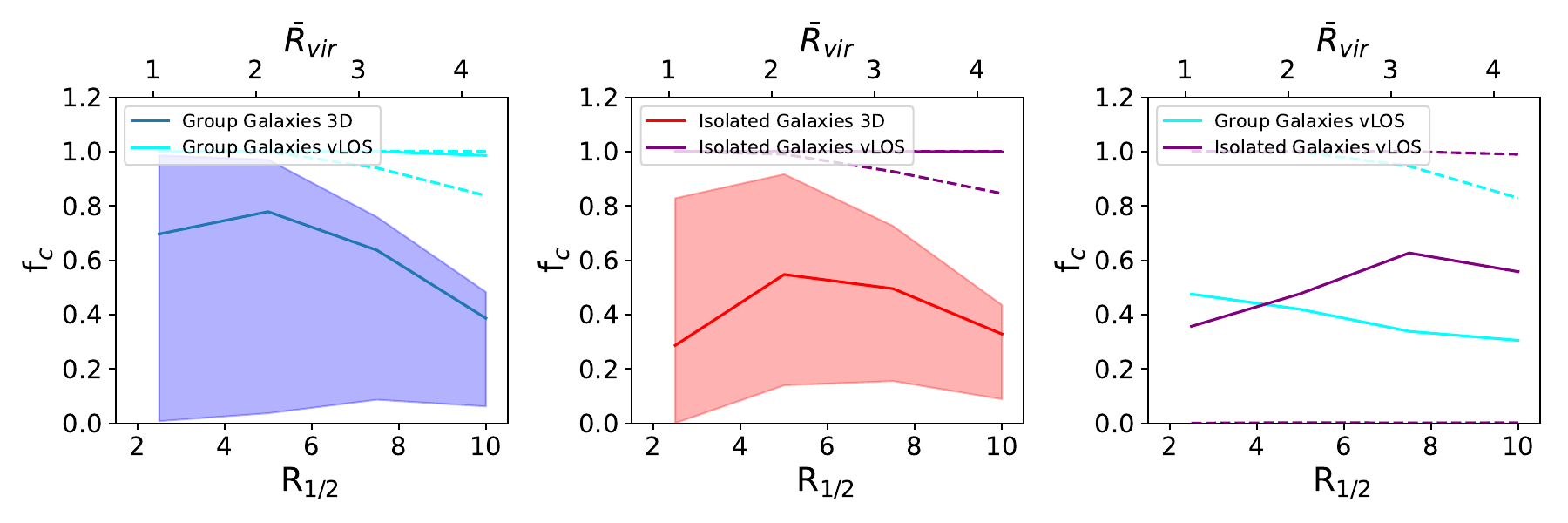}
        \caption{Same as Fig. \ref{Fig:CIVvLOS_satIncluded} but for a sample consisting only of central galaxies.
        Similar to \mgii, selecting gas in velocity space increases the covering fraction of \civ, but does not introduce an environmental dependence of the CGM.
        Also in this plot we used a detection limit of $N=10^{18.5}$ cm$^{-2}$ for the left and middle panel, while for the right panel we use a detection limit of $N=10^{19}$ cm$^{-2}$ to avoid having the covering fractions fully saturated.
        }
        \label{Fig:CIVvLOS_onlyCentrals}
\end{figure*}

For that we repeated the analysis we showed in Figs. \ref{Fig:ReproduceTrend} and \ref{Fig:ReproduceTrendCIV} with gas selected in a window of 500 km\,s$^{-1}$ around the galaxies as typically used in observations \cite[e.g.,][]{Dutta2020,Galbiati+2024}.
We use the same matched samples as in Sec. \ref{Sec:CoveringFractionsInTheCGM}.
We show our results in Figs. \ref{Fig:MgIIvLOS_SatIncluded}, \ref{Fig:MgIIvLOS_onlyCentrals}, \ref{Fig:CIVvLOS_satIncluded}, and \ref{Fig:CIVvLOS_onlyCentrals}.
While the covering fraction of \mgii~and \civ~gas is increased when selecting gas in velocity space (vLOS), our results concerning the environmental dependence of the CGM stay the same, i.e., there is no difference in covering fractions when the galaxies are matched in all relevant aspects that is stellar mass, halo mass, and being the central galaxy of the DM halo.
When including satellites in the sample, the difference between group and isolated galaxies becomes even more pronounced for \mgii.
For \civ~we had to increase the detection limit to $N=10^{19}$ cm$^{-2}$ when probing the difference between group and isolated galaxies and the gas selected in vLOS, since the effect of increased covering fraction is so strong that the median covering fraction is 100\% for both samples when using a detection limit of $N=10^{18.5}$ cm$^{-2}$.
Nevertheless, \civ~also shows the same trend for gas selected in velocity space as it does for gas selected in 3D.
Thus, we conclude that selecting gas in velocity space is not responsible for the environmental dependence of the CGM.

\end{appendix}
\end{document}